\documentclass[aps,prl,twocolumn,showpacs, notitlepage]{revtex4-1}
\usepackage{graphicx}
\usepackage{natbib}
\usepackage[utf8]{inputenc}
\usepackage{epstopdf}
\setcitestyle{super}
\usepackage{color}
\usepackage{amsmath}
\usepackage[normalem]{ulem}
\usepackage{pdfpages}
\usepackage{hyperref}
\makeatletter
\AtBeginDocument{\let\LS@rot\@undefined}
\makeatother

\usepackage{hyphenat}
\begin{document}

\preprint{APS/123-QED}

\title{Formation of short-range magnetic order and avoided ferromagnetic quantum criticality in pressurized LaCrGe$_3$}
\author{Elena Gati$^{1,2}$}
\author{John M. Wilde$^{1,2}$}
\author{Rustem Khasanov$^{3}$}
\author{Li Xiang$^{1,2}$}
\author{Sachith Dissanayake$^{4}$}
\author{Ritu Gupta$^{3}$}
\author{Masaaki Matsuda$^4$}
\author{Feng Ye$^4$}
\author{Bianca Haberl$^4$}
\author{Udhara Kaluarachchi$^{1,2}$}
\author{Robert J. McQueeney$^{1,2}$}
\author{Andreas Kreyssig$^{1,2}$}
\author{Sergey L. Bud'ko$^{1,2}$}
\author{Paul C. Canfield$^{1,2}$}

\address{$^{1}$ Ames Laboratory, US Department of Energy, Iowa State University, Ames, IA, USA}
\address{$^{2}$ Department of Physics and Astronomy, Iowa State University, Ames, IA, USA}
\address{$^{3}$ Laboratory for Muon Spin Spectroscopy, Paul Scherrer Institute, Villigen PSI, Switzerland}
\address{$^{4}$ Neutron Scattering Division, Oak Ridge National Laboratory, Oak Ridge, TN, USA}

\date{\today}

\begin{abstract}
LaCrGe$_3$ has attracted attention as a paradigm example of the avoidance of ferromagnetic (FM) quantum criticality in an itinerant magnet. By combining thermodynamic, transport, x-ray and neutron scattering as well as $\mu$SR measurements, we refined the temperature-pressure phase diagram of LaCrGe$_3$. We provide thermodynamic evidence (i) for the  first-order character of the FM transition when it is suppressed to low temperatures and (ii) for the formation of new phases at high pressures. From our microscopic data, we infer that short-range FM ordered clusters exist in these high-pressure phases. These results suggest that LaCrGe$_3$ is a rare example, which fills the gap between the two extreme limits of avoided FM quantum criticality in clean and strongly disordered metals.
\end{abstract}

\pacs{xxx}

\maketitle

The fluctuations, associated with quantum-critical points (QCP), i.e., second-order phase transitions at zero temperature ($T$), have been considered as crucial \cite{Canfield16b} for the stabilization of intriguing phenomena, such as superconductivity or non-Fermi liquid behavior \cite{Stewart01}. This motivates the search for novel states by tuning a magnetic phase transition\cite{Pfleiderer01,Levy07,Uemera07,Huy07,Westerkamp09,Pfleiderer09,Ubaid10,Saxena00,Aoki01,Cheng15,Ran19} to $T\,=\,0$\,K by external parameters, such as physical pressure, $p$, or chemical substitution. Whereas for antiferromagnetic (AFM) transitions there is a large body of experimental evidence that a QCP can be accessed in metals, e.g., in heavy-fermion systems \cite{Gegenwart08} or in iron-based superconductors \cite{Shibauchi14}, the ferromagnetic (FM) transition in clean, metallic magnets \cite{Brando16} is fundamentally different. Generic considerations \cite{Belitz97,Vojta99,Chubukov04,Belitz05b,Conduit09} suggest that the QCP is avoided when a second-order paramagnetic (PM)-FM transition in a clean, metallic system is suppressed to lower $T$ (with the exception of non-centrosymmetric metals with strong spin-orbit coupling \cite{Kirkpatrick20}). The predicted outcomes are generally either (i) that the PM-FM transition becomes a first-order quantum-phase transition, or (ii) that a new ground state, such as a long-wavelength AFM state (denoted by AFM$_{\textrm{q}}$), intervenes the FM QCP. Experimentally, the first scenario was verified in a variety of systems \cite{Brando16,Uhlarz04,Pfleiderer02,Niklowitz05,Huxley00}, whereas the second scenario has so far been discussed for only a small number of systems. Among those are CeRuPO \cite{Lengyel15}, PrPtAl \cite{Abdul15}, MnP \cite{Cheng15}, Nb$_{1-y}$Fe$_{2+y}$ \cite{Friedemann18,Niklowitz19} and LaCrGe$_3$ \cite{Taufour16,Kaluarachchi17}. \\
For understanding the avoided criticality in clean metallic FM systems, LaCrGe$_3$ \cite{Lin13} turns out to be an important reference system\cite{Taufour16,Kaluarachchi17,Lin13}. First, LaCrGe$_3$ is a simple $3d$ electron system with simple FM structure at ambient $p$. Second, the FM transition can be tuned by $p$ to lower $T$ without changing the level of disorder. Third, earlier studies \cite{Taufour16,Kaluarachchi17} suggested that the FM transition in LaCrGe$_3$ becomes first order at a tricritical point \cite{Taufour10,Taufour16b}, but also indicated the emergence of a new phase above $\,\approx\,1.5\,$GPa. It was argued that the new phase is likely the theoretically predicted AFM$_{\textrm{q}}$ phase. \\
Motivated by identifying the nature of the various phases in LaCrGe$_3$ across the avoided FM QCP region, we present an extensive study of thermodynamic, transport, x-ray diffraction, neutron scattering and muon-spin resonance ($\mu$SR) experiments (see SI\cite{SI} for experimental details). We provide \textit{thermodynamic} evidence that (i) as the FM transition is monotonically suppressed with increasing $p$, the FM transition becomes first order at $p_{\textrm{tr}}\,\approx\,$1.5\,GPa and (ii) two anomalies at $T_1$ and $T_2$, that are very close in $T$, emerge for higher $p$, signaling the occurrence of new phases in the vicinity of the avoided FM QCP. We demonstrate that below $T_1$ the magnetic volume fraction is strongly $T$ dependent. At the same time, our results indicate that even below $T_2\,<\,T_1$ the full-volume magnetism is not long-range ordered and is characterized by a remanent magnetization. These results question the existence of a long-range ordered AFM$_{\textrm{q}}$ phase line emerging near the boundary of the first-order FM transition line in LaCrGe$_3$. Instead, the resulting phase diagram shows features of a subtle interplay of competing magnetic interactions and weak disorder close to the avoided FM QCP. \\
	\begin{figure} 
	\includegraphics[width=0.9\columnwidth]{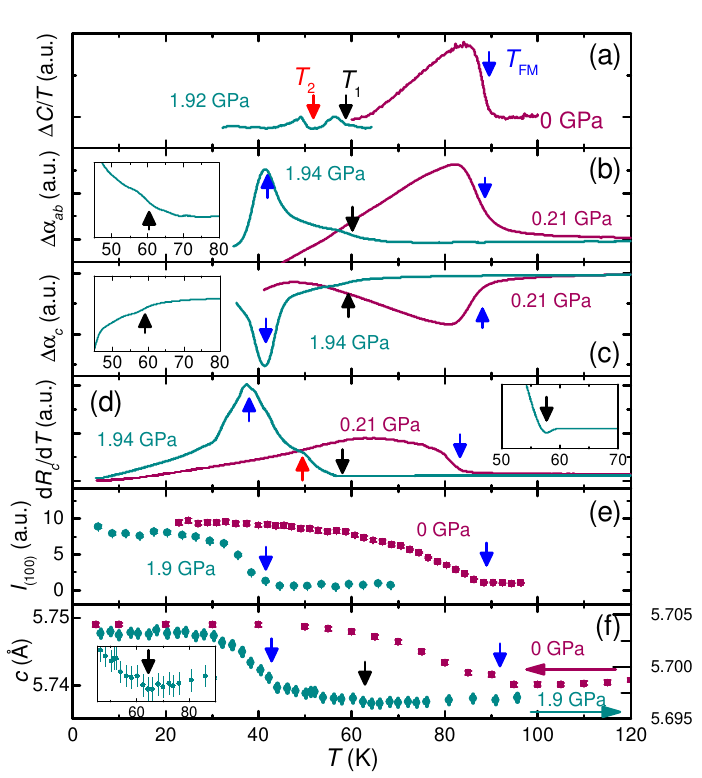}  
	\caption{Thermodynamic, transport and diffraction data of LaCrGe$_3$ for low pressures (close to $p\,\approx\,0$\,GPa) and high pressures ($p\,\approx\,1.9$\,GPa) as a function of temperature, $T$; (a) Anomalous contribution to the specific heat, $\Delta C/T$; (b,c) Anomalous contribution to the thermal expansion coefficient along the $ab$ axes, $\alpha_{ab}$, and the $c$ axis, $\alpha_{c}$; (d) $T$-derivative of the resistance along the $c$ axis, d$R_{c}$/d$T$; (e) Integrated intensity of the (1\,0\,0) neutron-diffraction Bragg peak (nuclear and magnetic contributions); (f) $c$ axis lattice parameters from x-ray (0\,GPa) and neutron (1.9 GPa) diffraction experiments. The arrows indicate the position of various anomalies at $T_{\textrm{FM}}$, $T_1$ and $T_2$. Insets in (b,c,d,f) show the high-$p$ data sets on enlarged scales around $T_1$.}
	\label{fig:anomaly} 
	\end{figure} 
Figure \ref{fig:anomaly} shows representative data sets of the anomalous contribution to specific heat ($\Delta C/T$) (the term ``anomalous'' indicates that data were corrected for a background contribution, see SI\cite{SI}), the anomalous contribution to the thermal expansion coefficient ($\Delta \alpha_{i}$ with $i\,=ab, c$), the $c$ axis resistance ($R_{c}$), the integrated neutron intensity of the (1\,0\,0) Bragg peak ($I_{1\,0\,0}$) and the $c$ lattice parameter for $p\,<\,p_{\textrm{tr}}$ and $p\,>\,p_{\textrm{tr}}$. 

For $p\,\approx\,0-0.21\,\textnormal{GPa}\,<\,p_{\textrm{tr}}$, we find clear anomalies at $T_{\textrm{FM}}\,\simeq\,90\,$K (see blue arrows) that are consistent with FM ordering with moments aligned along the $c$ axis \cite{Lin13}, as suggested by the increase of the $I_{1\,0\,0}$ intensity. The mean-field type thermodynamic signatures are consistent with a second-order phase transition. Notably, the transition is accompanied by sizable lattice changes, as evident from the evolution of $\alpha_{i}$ ($i\,=\,ab, c$) and the $c$ lattice parameter. Specifically, the in-plane $a$ axis (the out-of-plane $c$ axis) decreases (increases) upon entering the FM state.

For $p\,\approx\,1.9\,\textnormal{GPa}\,>\,p_{\textrm{tr}}$, our collection of data show anomalies at three characteristic temperatures. Upon cooling, a clear anomaly occurs in $\Delta C/T$ and d$R_{c}$/d$T$ at $T_1\,\simeq\,60\,$K, together with small, but resolvable changes of the lattice in $a$ and $c$ direction. Interestingly, the anisotropic response of the crystal lattice, $\alpha_{ab}$ and $\alpha_c$, at $T_1$ is similar to the one at $T_{\textrm{FM}}$, albeit much smaller in size, i.e., we find a contraction (expansion) along the $a$ ($c$) axis upon cooling through $T_1$. At $T_2\,\simeq\,50\,$K, another anomaly of similar size in $\Delta C/T$ is clearly resolvable, which however does not have any discernible effect in $\alpha_{ab}$ and $\alpha_c$. Further cooling down to $T_{\textrm{FM}}\,\simeq\,40\,$K results in a strong feature in $\alpha_{i}$ and the $c$ lattice parameter, which, given the increase of $I_{1\,0\,0}$, is associated with the formation of long-range FM order, but does not result in a clear feature in $\Delta C/T$. In contrast to low $p$ though, the symmetric and sharp shape of the anomaly in $\alpha_{i}$ for both directions is strongly reminiscent of a first-order phase transition (cf. also the more step-like change of $c$ and $I_{1\,0\,0}$ at $T_{\textrm{FM}}$). This, together with a sizable thermal hysteresis (see SI\cite{SI}), is clear thermodynamic evidence for the change of the character of the FM transition from second order to first order at $p_{\textrm{tr}}$. \\
	\begin{center}
	\begin{figure}
	\includegraphics[width=0.9\columnwidth]{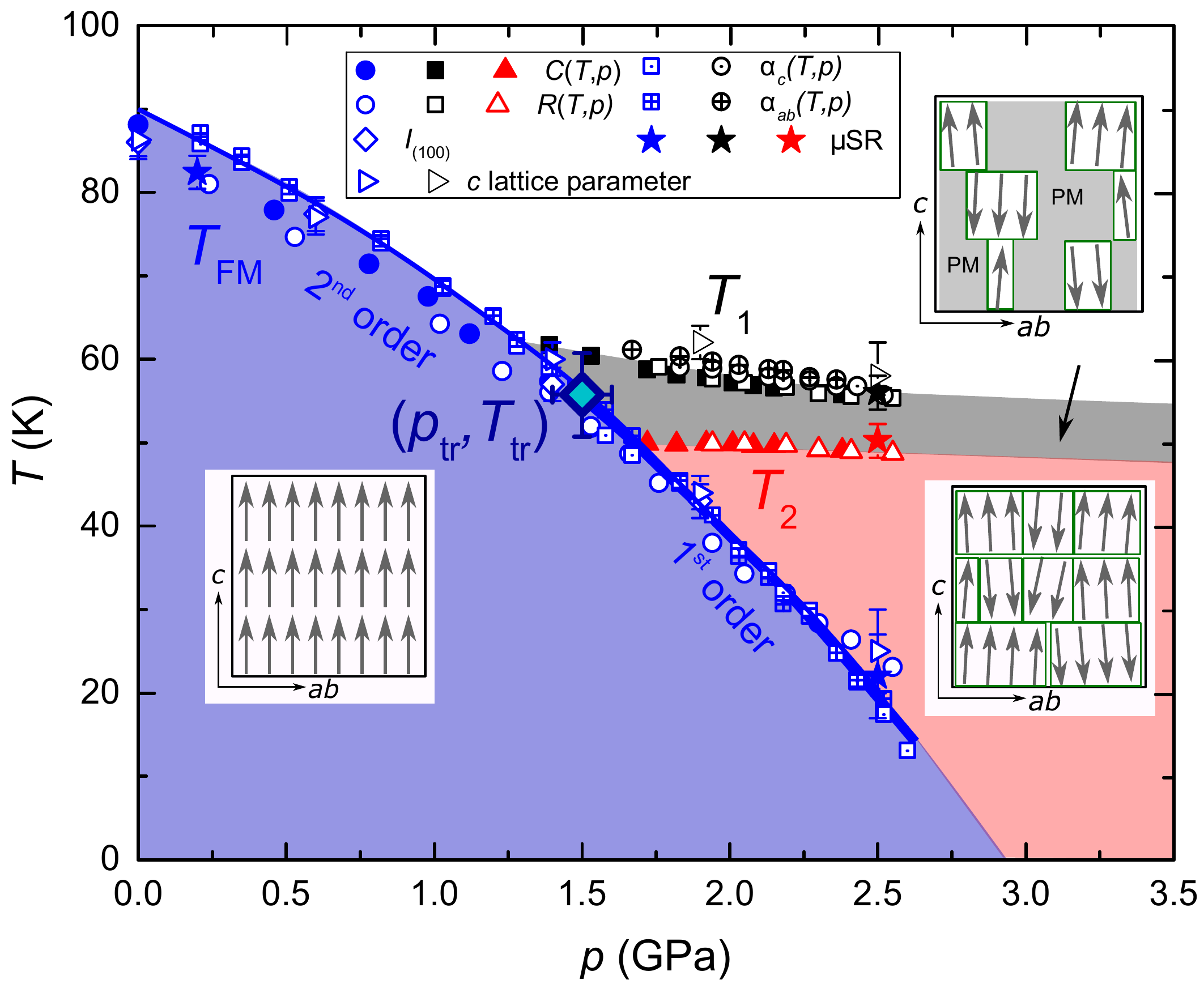} 
	\caption{Temperature-pressure ($T$-$p$) phase diagram of LaCrGe$_3$, constructed from specific heat, thermal expansion, resistance, neutron scattering and $\mu$SR measurements. Lines are a guide to the eye. The blue-shaded region corresponds to the region of ferromagnetic (FM) order, which is schematically depicted in the insets by spins (arrows) pointing along the crystallographic $c$ axis. The rhombus marks the position of the tricritical point at $(p_{\textrm{tr}},T_{\textrm{tr}})$, at which the character of the FM transition changes from second order for low $p$ to first order for high $p$. Black- and red-shaded regions correspond to new phases that occur for $p\,\gtrsim\,1.5\,$GPa. The insets visualize the suggested short-range ordered phases in this $p$ region. For $T_1\,>\,T\,>\,T_2$, small clusters of varying size with FM order are embedded in a paramagnetic (PM) matrix. For $T\,<\,T_2$, these clusters fill the whole sample volume.}
	\label{fig:phasediagram}
	\end{figure}
	\end{center}
The positions of the various anomalies, which we inferred from the full $T$-$p$ data sets up to $\approx\,$2.5\,GPa (see SI\cite{SI}), are compiled in the $T$-$p$ phase diagram in Fig.\,\ref{fig:phasediagram}. Upon suppressing $T_{\textrm{FM}}$ with $p$, the FM transition changes its character from second order to first order at $(p_{\textrm{tr}},T_{\textrm{tr}})\,=\,[1.5(1)\,$GPa, 53(3)\,K] (see SI\cite{SI} for the determination of the position). For $p\,\gtrsim\,p_{\textrm{tr}}$, anomalies at $T_1$ and $T_2$ emerge. (Only the latter phase line was identified in previous studies \cite{Taufour16,Kaluarachchi17}.) The $T_1$ and the $T_2$ lines do not only both emerge in immediate vicinity to $(p_{\textrm{tr}},T_{\textrm{tr}})$, but also closely follow each other in the phase diagram and are suppressed much more slowly by $p$ than $T_{\textrm{FM}}$. Altogether, this phase diagram highlights the complex behavior associated with the avoided FM QCP in LaCrGe$_3$.\\
	\begin{center}
	\begin{figure}
	\includegraphics[width=1\columnwidth]{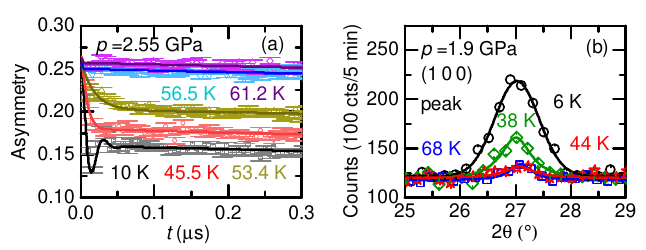} 
	\caption{(a) $\mu$SR spectra of LaCrGe$_3$ in zero field at $p\,=\,2.55\,$GPa. Symbols correspond to the measured data, solid lines correspond to fits by Eq.\,S6 (see SI\cite{SI}); (b) Angle-dependent neutron intensity around the (1\,0\,0) Bragg peak at $p\,=\,1.9$\,GPa. Lines are a guide to the eye.}
	\label{fig:scatteringfigure}
	\end{figure}
	\end{center}
To discuss the nature of the phases below $T_1$ and $T_2$ that are so clearly delineated in Fig.\,\ref{fig:phasediagram} by multiple thermodynamic and transport measurements (see SI\cite{SI}), we turn to $\mu$SR and neutron scattering measurements under pressure. Previous $\mu$SR measurements under $p$ \cite{Taufour16} showed a clear magnetic signal below $\,\approx\,50\,$K at 2.3\,GPa. To confirm this result and to refine the onset temperature, we performed another $\mu$SR study with a finer $T$ data point spacing close to $T_1$ and $T_2$ at 2.55\,GPa. Figure\,\ref{fig:scatteringfigure}\,(a) shows selected zero-field $\mu$SR spectra, that are in full agreement with the notion of some type of local (on the scale of $\mu$SR) magnetic order in the new phases. To discuss this in more detail, we show in Fig.\,\ref{fig:2p5GPa-comparison} the $T$ dependence of the internal field, $B_{\textrm{int}}$, and the transverse relaxation rate, $\lambda_{\textrm{T}}$, as a measure of the width of the field distribution, from zero-field $\mu$SR data. We also include the $T$ dependence of the magnetic asymmetry, $A_{\textrm{mag}}$, as a measure of the magnetic volume fraction, as well as the relaxation rate of the pressure cell, $\lambda_{\textrm{PC}}$, as a measure of the field in the pressure cell that is created by a sample with macroscopic magnetization, from weak-transverse field $\mu$SR data (see SI\cite{SI}). The thermodynamic and transport data for similar $p$ in Fig.\,\ref{fig:2p5GPa-comparison} are used to determine the positions of $T_1\,\approx\,56\,$K and $T_2\,\approx\,49$\,K as well as $T_{\textrm{FM}}\,\approx\,22\,$K. $B_{\textrm{int}}$ sets in between $T_2$ and $T_1$ and increases upon cooling, with low $T$ values similar to the ones in the FM state (see SI\cite{SI}). $\lambda_{\textrm{T}}$ shows a strong increase upon cooling through $T_1$. However, upon further cooling through $T_2$, $\lambda_{\textrm{T}}$ remains at a relatively high, finite value and decreases only slightly below $T_{\textrm{FM}}$. A large $\lambda_{\textrm{T}}$ implies a broad field distribution, characteristic for not well-ordered systems \cite{Yaouanc11}. $A_{\textrm{mag}}$ indicates partial volume fraction for $T_2\,<\,T\,<\,T_1$ and $A_{\textrm{mag}}\,\approx\,0.12$ for $T\,\leq\,T_2$, consistent with full volume fraction (see SI\cite{SI}). Last, $\lambda_{\textrm{PC}}$ is small above $T_1$ and starts to increase just below $T_1$ upon cooling. Below $T_2$, $\lambda_{\textrm{PC}}$ is finite and almost $T$-independent. In addition, we found strong indications for the presence of a remanent field for $T\,=\,35$\,K$\,<\,T_2$ (see SI\cite{SI}). \\

	\begin{center}
	\begin{figure}
	\includegraphics[width=0.9\columnwidth]{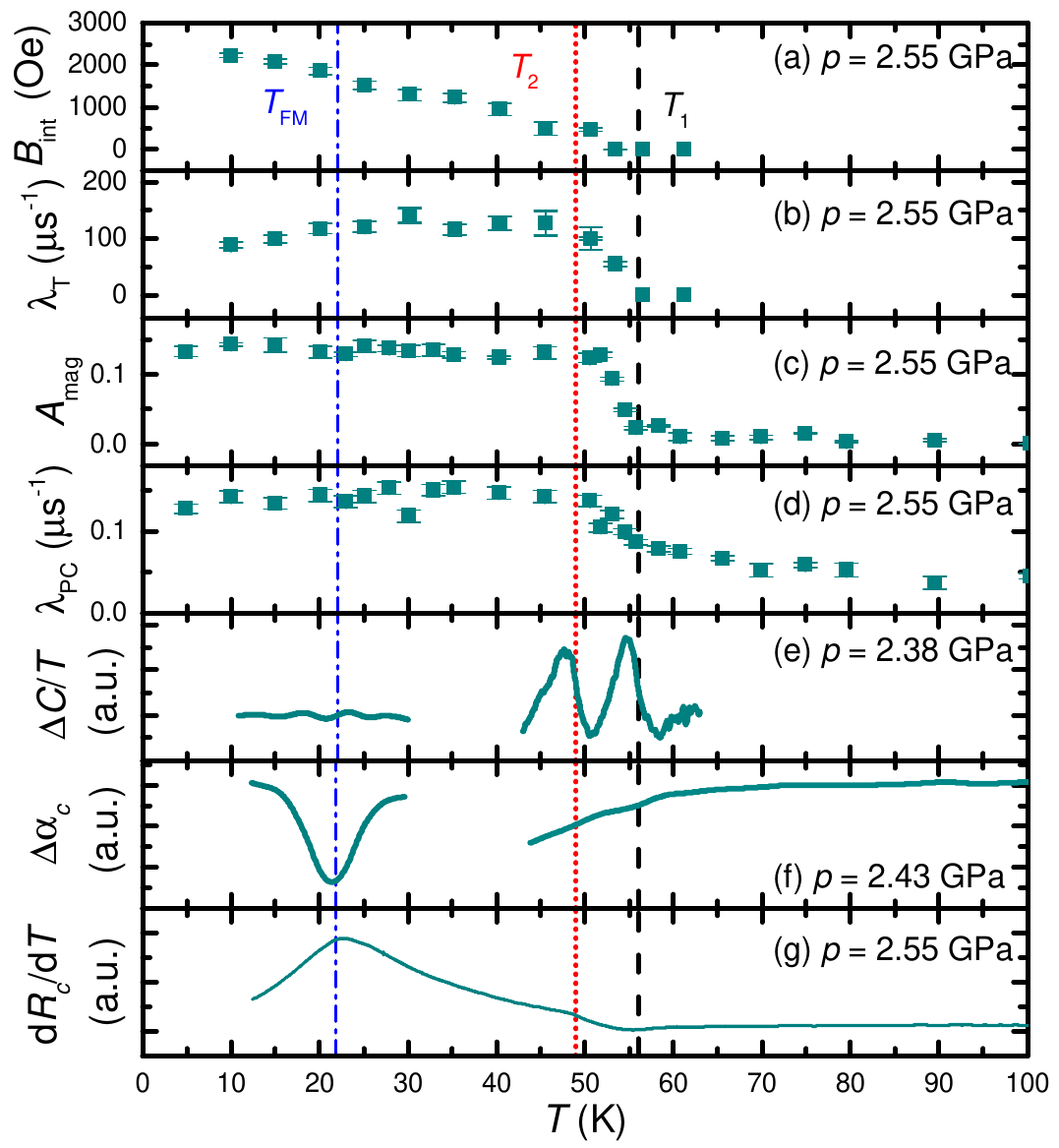} 
	\caption{Comparison of several high-pressure data sets close to a pressure, $p$, of 2.5\,GPa as a function of temperature, $T$. (a) Internal field, $B_{\textrm{int}}$, (b) transverse relaxation rate, $\lambda_{\textrm{T}}$, (c) magnetic asymmetry, $A_{\textrm{mag}}$, and (d) relaxation rate of the pressure cell, $\lambda_{\textrm{PC}}$, from zero-field (a,b) and weak transverse field (wTF) (c,d) $\mu$SR measurements; (e) Anomalous contribution to specific heat, $\Delta C/T$; (f) Anomalous contribution to thermal expansion coefficient along the $c$ axis, $\Delta \alpha_{c}$. The low-$T$ and high-$T$ data are plotted on different scales; (g) $T$-derivative of the $c$ axis resistance, d$R_{c}$/d$T$. Black dashed, red dotted and blue dashed-dotted lines indicate the position of the anomalies at $T_1$, $T_2$ and $T_{\textrm{FM}}$, respectively.}
	\label{fig:2p5GPa-comparison}
	\end{figure}
	\end{center}

In Fig.\,\ref{fig:scatteringfigure}\,(b), we compare the (1\,0\,0) Bragg peak in neutron diffraction for selected $T$ at $p\,=\,1.9\,$GPa. A clear (1\,0\,0) Bragg peak is observed in the PM state ($T\,=\,68$\,K) corresponding to the nuclear contribution, and grows markedly below $T_{\textrm{FM}}\,\simeq\,40\,$K due to the ferromagnetic contribution at $T\,=\,38\,$K and 6\,K. The moment in the FM ground state is $1.4(3)\,\mu_\textrm{B}$, which was determined from $I_{1\,0\,0}$ relative to a set of nuclear Bragg peaks. For $T_{\textrm{FM}}\,<\,T\,<\,T_2$, shown here by the 44\,K data, the (1\,0\,0) Bragg peak is not distinguishable from the data in the PM phase [see also Fig.\,\ref{fig:anomaly}\,(e)]. Furthermore, we cannot resolve any magnetic Bragg peak in the three-dimensional $q$ space in the new phases (see SI\cite{SI}). Overall, we can thus exclude any type of long-range FM or $c$-axis modulated AFM order below $T_1$ and $T_2$, i.e., the previously-suggested AFM$_{\textrm{q}}$ type magnetic order \cite{Taufour16}, with a moment larger than $0.4\,\mu_\textrm{B}$ and $0.7\,\mu_\textrm{B}$, respectively. In addition, we can rule out the formation of a charge-density wave or structural transition at high $p$ from x-ray diffraction studies (see SI\cite{SI}). \\
We now turn to a discussion of the nature of the $p$-induced phases that emerge for $p\,\gtrsim\,p_{\textrm{tr}}$ and $T\,\lesssim\,T_{\textrm{tr}}$. We start by focusing on the range $T\,<\,T_2$, for which the $\mu$SR data suggest $\,\approx\,100\%$ volume fraction, but no magnetic Bragg peak could be resolved in neutron diffraction experiments. Given that the $\mu$SR data indicate a similar $B_{\textrm{int}}$ for the low-$p$ FM state and the new phase below $T_2$ at low $T$, it seems unlikely that the moment of the $T_2$-phase is so low that it falls below our sensitivity in neutron measurements. Following this argument, an obvious scenario, which would reconcile both $\mu$SR and neutron results, would be that the magnetic order below $T_2$ is only short-range. We note that the sizable $\lambda_{\textrm{T}}$ value for $T\,<\,T_2$ is fully consistent with the notion of a short-range ordered state \cite{Yaouanc11}, in which magnetic clusters exist. To discuss the question whether the order within these clusters is FM or AFM, we refer to the observations of a finite $\lambda_{\textrm{PC}}$ and a remanent magnetization below $T_2$ from $\mu$SR. This speaks in favor of FM order in each cluster, whereas the clusters might either align FM or AFM with respect to each other (see inset of Fig.\,\ref{fig:phasediagram} for a schematic picture). We speculate that at least some of the clusters align AFM with respect to each other, since this would explain the small, but finite $\lambda_{\textrm{PC}}$. An estimation of the size of such FM clusters can be inferred from the $\lambda_\textrm{T}$ value as well as the data of the (1\,0\,0) Bragg peak. The large value of $\lambda_\textrm{T}$ between $T_2$ and $T_{\textrm{FM}}$ yields an estimate for the cluster size of 6\,nm \cite{Yaouanc11}. For the neutron data, if we assume a similar moment size as in the FM state, as suggested by a similar $B_{\textrm{int}}$, the absence of a clear magnetic (1\,0\,0) Bragg peak results in an estimate of the average cluster size of less than 12\,nm. This scenario of clusters would also naturally account for a small amount of entropy release upon subsequent cooling through $T_\textrm{FM}$, consistent with the lack of a clear specific heat feature in our experiment (see SI\cite{SI}). Note that moment size and spatial size can change with decreasing $T$, as suggested by a continuous change of $B_{\textrm{int}}$, the $c$ lattice parameter and $\alpha_i$. \\
How is LaCrGe$_3$ for $T_2\,<\,T\,<\,T_1$ then characterized? Our results indicate that in this regime the magnetic volume fraction is strongly $T$-dependent and increases from $\approx\,0$ at $T_1$ upon cooling to $\approx\,100\%$ at $T_2$. We also recall our result of the lattice strain: (i) the lattice response upon cooling through $T_1$ shows the same directional anisotropy as for the FM transition, but only smaller in size, and (ii) there are no pronounced lattice effects at $T_2$. The latter result indicates that no strong modification of the magnetic order occurs at $T_2$, since it would likely result in an additional lattice strain. It thus appears likely, that the magnetic clusters start to form in the range $T_2\,<\,T\,<\,T_1$, and either their number or size is strongly dependent on $T$ (see inset of Fig.\,\ref{fig:phasediagram}). The size and anisotropy of the observed lattice strains are fully consistent with the notion of small FM clusters, in which moments are primarily aligned along the $c$ axis (a small tilt away from the $c$ axis is possible) and in which the partial AFM alignment of the clusters with respect to each other strongly reduces the lattice strain (in contrast to large FM domains in the low-$p$ FM state, resulting in large strains). As an alternative proposal for the nature of the intermediate $T$ phase, we refer to the theoretical idea, that spin-nematic orders can be promoted by quantum fluctuations close to an avoided QCP \cite{Karahasanovic12,Chubukov09}. Experimentally, we cannot rule out this option, which by itself would certainly be exciting. However, if there would be a spin-nematic phase below $T_1$, then it does not couple strongly to the crystalline lattice, since we do not observe any lattice symmetry change across the entire $T$ range for high $p$ (see SI\cite{SI}). \\
Our main results on the avoidance of FM criticality in LaCrGe$_3$ can be summarized as follows. We (i) provided thermodynamic evidence for a change of the transition character from second order to first order, typically considered a hallmark for the avoidance of the QCP in clean metallic FM systems and (ii) argued that short-range magnetic order rather than long-wavelength AFM order \cite{Taufour16} exists for $p\,\geq\,p_{\textrm{tr}}$ between $T_1$ and $T_{\textrm{FM}}$, which is usually associated with the effects of strong disorder \cite{Brando16}. The main question is then what drives the formation of short-range order in LaCrGe$_3$: do the enhanced AFM interactions, that are suggested by theory\cite{Belitz97,Chubukov04,Conduit09,Karahasanovic12,Thomson13,Pedder13}, and the associated frustration between FM and AFM interactions lead to a tendency towards short-range order, or does weak disorder promote short-range order? In fact, an earlier theoretical study \cite{Thomson13} pointed out that the tricritical point can survive up to a critical disorder strength, whereas an amount of disorder smaller than the critical disorder strength can cause a short-range spiral state. So far, this scenario has only been considered to be realized in the stochiometric compound CeFePO \cite{Lausberg12,Brando16}, for which the interpretation is complicated by Kondo physics, and a tuning across the avoided QCP is lacking up to now. Interestingly, CeFePO and LaCrGe$_3$ have a very similar residual resisitivity ratio\cite{Brando16,Taufour16} of $\approx$\,5-10, which is lower than for other clean itinerant FM systems and might indicate a somewhat larger level of disorder. Therefore, our results strongly suggest that LaCrGe$_3$ is a rare example, which fills the gap between two extreme limits of clean and strongly disordered itinerant FM systems. Given that its $p$ tunability allows for accessing the multiple phases without introducing additional disorder, and that its magnetic building block is a 3$d$ element, LaCrGe$_3$ turns out to be a very promising candidate for the comparison to theoretical concepts, that address the effects of weak disorder and modulated AFM orders close to an avoided FM QCP in metals.

\begin{acknowledgments}
We thank A.I. Goldman, V. Taufour and D. Ryan for useful discussions and S. Downing and C. Abel for the growth of single crystals. The authors would like to acknowledge B. Li, D. S. Robinson, C. Kenney-Benson, S. Tkachev, M. Baldini, and S. G. Sinogeikin and D. Popov for their assistance during the x-ray diffraction experiments. We thank C. Tulk, A. M. dos Santos, J. Molaison, and R. Boehler for support of the high-pressure neutron diffraction study, and Y. Uwatoko for providing us the palm cubic pressure cell. Work at the Ames Laboratory was supported by the U.S. Department of Energy, Office of Science, Basic Energy Sciences, Materials Sciences and Engineering Division. The Ames Laboratory is operated for the U.S. Department of Energy by Iowa State University under Contract No. DEAC02-07CH11358. E.G. and L.X. were funded, in part, by the Gordon and Betty Moore Foundation's EPiQS Initiative through Grant No. GBMF4411. In addition, L.X. was funded, in part, by the W. M. Keck Foundation. A portion of this research used resources at the High Flux Isotope Reactor and the Spallation Neutron Source, U.S. DOE Office of Science User Facilities operated by the Oak Ridge National Laboratory. This research used resources of the Advanced Photon Source, a U.S. DOE Office of Science User Facility operated for the US DOE Office of Science by Argonne National Laboratory under Contract No. DE-AC02-06CH11357. We gratefully acknowledge support by HPCAT (Sector 16), Advanced Photon Source (APS), Argonne National Laboratory. HPCAT operations are supported by DOE-NNSA under Grant No. DE-NA0001974, with partial instrumentation funding by NSF. Use of the COMPRES-GSECARS gas loading system was supported by COMPRES under NSF Cooperative Agreement Grant No. EAR-11-57758 and by GSECARS through NSF Grant No. EAR-1128799 and DOE Grant No. DE-FG02-94ER14466. Research of R.G. is supported by the Swiss National Science Foundation (SNF-Grant No. 200021-175935).
\end{acknowledgments}

\bibliographystyle{apsrev}

\begin{thebibliography}{73}
\expandafter\ifx\csname natexlab\endcsname\relax\def\natexlab#1{#1}\fi
\expandafter\ifx\csname bibnamefont\endcsname\relax
  \def\bibnamefont#1{#1}\fi
\expandafter\ifx\csname bibfnamefont\endcsname\relax
  \def\bibfnamefont#1{#1}\fi
\expandafter\ifx\csname citenamefont\endcsname\relax
  \def\citenamefont#1{#1}\fi
\expandafter\ifx\csname url\endcsname\relax
  \def\url#1{\texttt{#1}}\fi
\expandafter\ifx\csname urlprefix\endcsname\relax\def\urlprefix{URL }\fi
\providecommand{\bibinfo}[2]{#2}
\providecommand{\eprint}[2][]{\url{#2}}

\bibitem[{\citenamefont{Canfield and Bud'ko}(2016)}]{Canfield16b}
\bibinfo{author}{\bibfnamefont{P.~C.} \bibnamefont{Canfield}} \bibnamefont{and}
  \bibinfo{author}{\bibfnamefont{S.~L.} \bibnamefont{Bud'ko}},
  \bibinfo{journal}{Rep. Prog. Phys.} \textbf{\bibinfo{volume}{79}},
  \bibinfo{pages}{084506} (\bibinfo{year}{2016}).

\bibitem[{\citenamefont{Stewart}(2001)}]{Stewart01}
\bibinfo{author}{\bibfnamefont{G.~R.} \bibnamefont{Stewart}},
  \bibinfo{journal}{Rev. Mod. Phys.} \textbf{\bibinfo{volume}{73}},
  \bibinfo{pages}{797} (\bibinfo{year}{2001}).

\bibitem[{\citenamefont{Pfleiderer et~al.}(2001)\citenamefont{Pfleiderer,
  Julian, and Lonzarich}}]{Pfleiderer01}
\bibinfo{author}{\bibfnamefont{C.}~\bibnamefont{Pfleiderer}},
  \bibinfo{author}{\bibfnamefont{S.~R.} \bibnamefont{Julian}},
  \bibnamefont{and} \bibinfo{author}{\bibfnamefont{G.~G.}
  \bibnamefont{Lonzarich}}, \bibinfo{journal}{Nature}
  \textbf{\bibinfo{volume}{414}}, \bibinfo{pages}{427} (\bibinfo{year}{2001}).

\bibitem[{\citenamefont{Lévy et~al.}(2007)\citenamefont{Lévy, Sheikin, and
  Huxley}}]{Levy07}
\bibinfo{author}{\bibfnamefont{F.}~\bibnamefont{Lévy}},
  \bibinfo{author}{\bibfnamefont{I.}~\bibnamefont{Sheikin}}, \bibnamefont{and}
  \bibinfo{author}{\bibfnamefont{A.}~\bibnamefont{Huxley}},
  \bibinfo{journal}{Nat. Phys.} \textbf{\bibinfo{volume}{3}},
  \bibinfo{pages}{460} (\bibinfo{year}{2007}).

\bibitem[{\citenamefont{Uemura et~al.}(2007)\citenamefont{Uemura, Goko,
  Gat-Malureanu, Carlo, Russo, Savici, Aczel, MacDougall, Rodriguez, Luke
  et~al.}}]{Uemera07}
\bibinfo{author}{\bibfnamefont{Y.~J.} \bibnamefont{Uemura}},
  \bibinfo{author}{\bibfnamefont{T.}~\bibnamefont{Goko}},
  \bibinfo{author}{\bibfnamefont{I.~M.} \bibnamefont{Gat-Malureanu}},
  \bibinfo{author}{\bibfnamefont{J.~P.} \bibnamefont{Carlo}},
  \bibinfo{author}{\bibfnamefont{P.~L.} \bibnamefont{Russo}},
  \bibinfo{author}{\bibfnamefont{A.~T.} \bibnamefont{Savici}},
  \bibinfo{author}{\bibfnamefont{A.}~\bibnamefont{Aczel}},
  \bibinfo{author}{\bibfnamefont{G.~J.} \bibnamefont{MacDougall}},
  \bibinfo{author}{\bibfnamefont{J.~A.} \bibnamefont{Rodriguez}},
  \bibinfo{author}{\bibfnamefont{G.~M.} \bibnamefont{Luke}},
  \bibnamefont{et~al.}, \bibinfo{journal}{Nat. Phys.}
  \textbf{\bibinfo{volume}{3}}, \bibinfo{pages}{29} (\bibinfo{year}{2007}).

\bibitem[{\citenamefont{Huy et~al.}(2007)\citenamefont{Huy, Gasparini, de~Nijs,
  Huang, Klaasse, Gortenmulder, de~Visser, Hamann, G\"orlach, and
  L\"ohneysen}}]{Huy07}
\bibinfo{author}{\bibfnamefont{N.~T.} \bibnamefont{Huy}},
  \bibinfo{author}{\bibfnamefont{A.}~\bibnamefont{Gasparini}},
  \bibinfo{author}{\bibfnamefont{D.~E.} \bibnamefont{de~Nijs}},
  \bibinfo{author}{\bibfnamefont{Y.}~\bibnamefont{Huang}},
  \bibinfo{author}{\bibfnamefont{J.~C.~P.} \bibnamefont{Klaasse}},
  \bibinfo{author}{\bibfnamefont{T.}~\bibnamefont{Gortenmulder}},
  \bibinfo{author}{\bibfnamefont{A.}~\bibnamefont{de~Visser}},
  \bibinfo{author}{\bibfnamefont{A.}~\bibnamefont{Hamann}},
  \bibinfo{author}{\bibfnamefont{T.}~\bibnamefont{G\"orlach}},
  \bibnamefont{and} \bibinfo{author}{\bibfnamefont{H.~v.}
  \bibnamefont{L\"ohneysen}}, \bibinfo{journal}{Phys. Rev. Lett.}
  \textbf{\bibinfo{volume}{99}}, \bibinfo{pages}{067006}
  (\bibinfo{year}{2007}).

\bibitem[{\citenamefont{Westerkamp et~al.}(2009)\citenamefont{Westerkamp,
  Deppe, K\"uchler, Brando, Geibel, Gegenwart, Pikul, and
  Steglich}}]{Westerkamp09}
\bibinfo{author}{\bibfnamefont{T.}~\bibnamefont{Westerkamp}},
  \bibinfo{author}{\bibfnamefont{M.}~\bibnamefont{Deppe}},
  \bibinfo{author}{\bibfnamefont{R.}~\bibnamefont{K\"uchler}},
  \bibinfo{author}{\bibfnamefont{M.}~\bibnamefont{Brando}},
  \bibinfo{author}{\bibfnamefont{C.}~\bibnamefont{Geibel}},
  \bibinfo{author}{\bibfnamefont{P.}~\bibnamefont{Gegenwart}},
  \bibinfo{author}{\bibfnamefont{A.~P.} \bibnamefont{Pikul}}, \bibnamefont{and}
  \bibinfo{author}{\bibfnamefont{F.}~\bibnamefont{Steglich}},
  \bibinfo{journal}{Phys. Rev. Lett.} \textbf{\bibinfo{volume}{102}},
  \bibinfo{pages}{206404} (\bibinfo{year}{2009}).

\bibitem[{\citenamefont{Pfleiderer}(2009)}]{Pfleiderer09}
\bibinfo{author}{\bibfnamefont{C.}~\bibnamefont{Pfleiderer}},
  \bibinfo{journal}{Rev. Mod. Phys.} \textbf{\bibinfo{volume}{81}},
  \bibinfo{pages}{1551} (\bibinfo{year}{2009}).

\bibitem[{\citenamefont{Ubaid-Kassis et~al.}(2010)\citenamefont{Ubaid-Kassis,
  Vojta, and Schroeder}}]{Ubaid10}
\bibinfo{author}{\bibfnamefont{S.}~\bibnamefont{Ubaid-Kassis}},
  \bibinfo{author}{\bibfnamefont{T.}~\bibnamefont{Vojta}}, \bibnamefont{and}
  \bibinfo{author}{\bibfnamefont{A.}~\bibnamefont{Schroeder}},
  \bibinfo{journal}{Phys. Rev. Lett.} \textbf{\bibinfo{volume}{104}},
  \bibinfo{pages}{066402} (\bibinfo{year}{2010}).

\bibitem[{\citenamefont{Saxena et~al.}(2000)\citenamefont{Saxena, Agarwal,
  Ahilan, Grosche, Haselwimmer, Steiner, Pugh, Walker, Julian, Monthoux
  et~al.}}]{Saxena00}
\bibinfo{author}{\bibfnamefont{S.~S.} \bibnamefont{Saxena}},
  \bibinfo{author}{\bibfnamefont{P.}~\bibnamefont{Agarwal}},
  \bibinfo{author}{\bibfnamefont{K.}~\bibnamefont{Ahilan}},
  \bibinfo{author}{\bibfnamefont{F.~M.} \bibnamefont{Grosche}},
  \bibinfo{author}{\bibfnamefont{R.~K.~W.} \bibnamefont{Haselwimmer}},
  \bibinfo{author}{\bibfnamefont{M.~J.} \bibnamefont{Steiner}},
  \bibinfo{author}{\bibfnamefont{E.}~\bibnamefont{Pugh}},
  \bibinfo{author}{\bibfnamefont{I.~R.} \bibnamefont{Walker}},
  \bibinfo{author}{\bibfnamefont{S.~R.} \bibnamefont{Julian}},
  \bibinfo{author}{\bibfnamefont{P.}~\bibnamefont{Monthoux}},
  \bibnamefont{et~al.}, \bibinfo{journal}{Nature}
  \textbf{\bibinfo{volume}{406}}, \bibinfo{pages}{587} (\bibinfo{year}{2000}).

\bibitem[{\citenamefont{Aoki et~al.}(2001)\citenamefont{Aoki, Huxley,
  Ressouche, Braithwaite, Flouquet, Brison, Lhotel, and Paulsen}}]{Aoki01}
\bibinfo{author}{\bibfnamefont{D.}~\bibnamefont{Aoki}},
  \bibinfo{author}{\bibfnamefont{A.}~\bibnamefont{Huxley}},
  \bibinfo{author}{\bibfnamefont{E.}~\bibnamefont{Ressouche}},
  \bibinfo{author}{\bibfnamefont{D.}~\bibnamefont{Braithwaite}},
  \bibinfo{author}{\bibfnamefont{J.}~\bibnamefont{Flouquet}},
  \bibinfo{author}{\bibfnamefont{J.-P.} \bibnamefont{Brison}},
  \bibinfo{author}{\bibfnamefont{E.}~\bibnamefont{Lhotel}}, \bibnamefont{and}
  \bibinfo{author}{\bibfnamefont{C.}~\bibnamefont{Paulsen}},
  \bibinfo{journal}{Nature} \textbf{\bibinfo{volume}{431}},
  \bibinfo{pages}{613} (\bibinfo{year}{2001}).

\bibitem[{\citenamefont{Cheng et~al.}(2015)\citenamefont{Cheng, Matsubayashi,
  Wu, Sun, Lin, Luo, and Uwatoko}}]{Cheng15}
\bibinfo{author}{\bibfnamefont{J.-G.} \bibnamefont{Cheng}},
  \bibinfo{author}{\bibfnamefont{K.}~\bibnamefont{Matsubayashi}},
  \bibinfo{author}{\bibfnamefont{W.}~\bibnamefont{Wu}},
  \bibinfo{author}{\bibfnamefont{J.~P.} \bibnamefont{Sun}},
  \bibinfo{author}{\bibfnamefont{F.~K.} \bibnamefont{Lin}},
  \bibinfo{author}{\bibfnamefont{J.~L.} \bibnamefont{Luo}}, \bibnamefont{and}
  \bibinfo{author}{\bibfnamefont{Y.}~\bibnamefont{Uwatoko}},
  \bibinfo{journal}{Phys. Rev. Lett.} \textbf{\bibinfo{volume}{114}},
  \bibinfo{pages}{117001} (\bibinfo{year}{2015}).

\bibitem[{\citenamefont{Ran et~al.}(2019)\citenamefont{Ran, Eckberg, Ding,
  Furukawa, Metz, Sahaand, Liu, Zic, Kim, Paglione et~al.}}]{Ran19}
\bibinfo{author}{\bibfnamefont{S.}~\bibnamefont{Ran}},
  \bibinfo{author}{\bibfnamefont{C.}~\bibnamefont{Eckberg}},
  \bibinfo{author}{\bibfnamefont{Q.-P.} \bibnamefont{Ding}},
  \bibinfo{author}{\bibfnamefont{Y.}~\bibnamefont{Furukawa}},
  \bibinfo{author}{\bibfnamefont{T.}~\bibnamefont{Metz}},
  \bibinfo{author}{\bibfnamefont{S.~R.} \bibnamefont{Sahaand}},
  \bibinfo{author}{\bibfnamefont{I.-L.} \bibnamefont{Liu}},
  \bibinfo{author}{\bibfnamefont{M.}~\bibnamefont{Zic}},
  \bibinfo{author}{\bibfnamefont{H.}~\bibnamefont{Kim}},
  \bibinfo{author}{\bibfnamefont{J.}~\bibnamefont{Paglione}},
  \bibnamefont{et~al.}, \bibinfo{journal}{Science}
  \textbf{\bibinfo{volume}{365}}, \bibinfo{pages}{684} (\bibinfo{year}{2019}).

\bibitem[{\citenamefont{Gegenwart et~al.}(2008)\citenamefont{Gegenwart, Si, and
  Steglich}}]{Gegenwart08}
\bibinfo{author}{\bibfnamefont{P.}~\bibnamefont{Gegenwart}},
  \bibinfo{author}{\bibfnamefont{Q.}~\bibnamefont{Si}}, \bibnamefont{and}
  \bibinfo{author}{\bibfnamefont{F.}~\bibnamefont{Steglich}},
  \bibinfo{journal}{Nature Physics} \textbf{\bibinfo{volume}{4}},
  \bibinfo{pages}{186} (\bibinfo{year}{2008}).

\bibitem[{\citenamefont{Shibauchi et~al.}(2014)\citenamefont{Shibauchi,
  Carrington, and Matsuda}}]{Shibauchi14}
\bibinfo{author}{\bibfnamefont{T.}~\bibnamefont{Shibauchi}},
  \bibinfo{author}{\bibfnamefont{A.}~\bibnamefont{Carrington}},
  \bibnamefont{and} \bibinfo{author}{\bibfnamefont{Y.}~\bibnamefont{Matsuda}},
  \bibinfo{journal}{Annual Review of Condensed Matter Physics}
  \textbf{\bibinfo{volume}{5}}, \bibinfo{pages}{113} (\bibinfo{year}{2014}).

\bibitem[{\citenamefont{Brando et~al.}(2016)\citenamefont{Brando, Belitz,
  Grosche, and Kirkpatrick}}]{Brando16}
\bibinfo{author}{\bibfnamefont{M.}~\bibnamefont{Brando}},
  \bibinfo{author}{\bibfnamefont{D.}~\bibnamefont{Belitz}},
  \bibinfo{author}{\bibfnamefont{F.~M.} \bibnamefont{Grosche}},
  \bibnamefont{and} \bibinfo{author}{\bibfnamefont{T.~R.}
  \bibnamefont{Kirkpatrick}}, \bibinfo{journal}{Rev. Mod. Phys.}
  \textbf{\bibinfo{volume}{88}}, \bibinfo{pages}{025006}
  (\bibinfo{year}{2016}).

\bibitem[{\citenamefont{Belitz et~al.}(1997)\citenamefont{Belitz, Kirkpatrick,
  and Vojta}}]{Belitz97}
\bibinfo{author}{\bibfnamefont{D.}~\bibnamefont{Belitz}},
  \bibinfo{author}{\bibfnamefont{T.~R.} \bibnamefont{Kirkpatrick}},
  \bibnamefont{and} \bibinfo{author}{\bibfnamefont{T.}~\bibnamefont{Vojta}},
  \bibinfo{journal}{Phys. Rev. B} \textbf{\bibinfo{volume}{55}},
  \bibinfo{pages}{9452} (\bibinfo{year}{1997}).

\bibitem[{\citenamefont{Vojta et~al.}(1999)\citenamefont{Vojta, Belitz,
  Kirkpatrick, and Narayanan}}]{Vojta99}
\bibinfo{author}{\bibfnamefont{T.}~\bibnamefont{Vojta}},
  \bibinfo{author}{\bibfnamefont{D.}~\bibnamefont{Belitz}},
  \bibinfo{author}{\bibfnamefont{T.}~\bibnamefont{Kirkpatrick}},
  \bibnamefont{and}
  \bibinfo{author}{\bibfnamefont{R.}~\bibnamefont{Narayanan}},
  \bibinfo{journal}{Ann. Phys.} \textbf{\bibinfo{volume}{8}},
  \bibinfo{pages}{593} (\bibinfo{year}{1999}).

\bibitem[{\citenamefont{Chubukov et~al.}(2004)\citenamefont{Chubukov, P\'epin,
  and Rech}}]{Chubukov04}
\bibinfo{author}{\bibfnamefont{A.~V.} \bibnamefont{Chubukov}},
  \bibinfo{author}{\bibfnamefont{C.}~\bibnamefont{P\'epin}}, \bibnamefont{and}
  \bibinfo{author}{\bibfnamefont{J.}~\bibnamefont{Rech}},
  \bibinfo{journal}{Phys. Rev. Lett.} \textbf{\bibinfo{volume}{92}},
  \bibinfo{pages}{147003} (\bibinfo{year}{2004}).

\bibitem[{\citenamefont{Belitz et~al.}(2005)\citenamefont{Belitz, Kirkpatrick,
  and Vojta}}]{Belitz05b}
\bibinfo{author}{\bibfnamefont{D.}~\bibnamefont{Belitz}},
  \bibinfo{author}{\bibfnamefont{T.~R.} \bibnamefont{Kirkpatrick}},
  \bibnamefont{and} \bibinfo{author}{\bibfnamefont{T.}~\bibnamefont{Vojta}},
  \bibinfo{journal}{Rev. Mod. Phys.} \textbf{\bibinfo{volume}{77}},
  \bibinfo{pages}{579} (\bibinfo{year}{2005}).

\bibitem[{\citenamefont{Conduit et~al.}(2009)\citenamefont{Conduit, Green, and
  Simons}}]{Conduit09}
\bibinfo{author}{\bibfnamefont{G.~J.} \bibnamefont{Conduit}},
  \bibinfo{author}{\bibfnamefont{A.~G.} \bibnamefont{Green}}, \bibnamefont{and}
  \bibinfo{author}{\bibfnamefont{B.~D.} \bibnamefont{Simons}},
  \bibinfo{journal}{Phys. Rev. Lett.} \textbf{\bibinfo{volume}{103}},
  \bibinfo{pages}{207201} (\bibinfo{year}{2009}).

\bibitem[{\citenamefont{Kirkpatrick and Belitz}(2020)}]{Kirkpatrick20}
\bibinfo{author}{\bibfnamefont{T.~R.} \bibnamefont{Kirkpatrick}}
  \bibnamefont{and} \bibinfo{author}{\bibfnamefont{D.}~\bibnamefont{Belitz}},
  \bibinfo{journal}{Phys. Rev. Lett.} \textbf{\bibinfo{volume}{124}},
  \bibinfo{pages}{147201} (\bibinfo{year}{2020}).

\bibitem[{\citenamefont{Uhlarz et~al.}(2004)\citenamefont{Uhlarz, Pfleiderer,
  and Hayden}}]{Uhlarz04}
\bibinfo{author}{\bibfnamefont{M.}~\bibnamefont{Uhlarz}},
  \bibinfo{author}{\bibfnamefont{C.}~\bibnamefont{Pfleiderer}},
  \bibnamefont{and} \bibinfo{author}{\bibfnamefont{S.~M.}
  \bibnamefont{Hayden}}, \bibinfo{journal}{Phys. Rev. Lett.}
  \textbf{\bibinfo{volume}{93}}, \bibinfo{pages}{256404}
  (\bibinfo{year}{2004}).

\bibitem[{\citenamefont{Pfleiderer and Huxley}(2002)}]{Pfleiderer02}
\bibinfo{author}{\bibfnamefont{C.}~\bibnamefont{Pfleiderer}} \bibnamefont{and}
  \bibinfo{author}{\bibfnamefont{A.~D.} \bibnamefont{Huxley}},
  \bibinfo{journal}{Phys. Rev. Lett.} \textbf{\bibinfo{volume}{89}},
  \bibinfo{pages}{147005} (\bibinfo{year}{2002}).

\bibitem[{\citenamefont{Niklowitz et~al.}(2005)\citenamefont{Niklowitz,
  Beckers, Lonzarich, Knebel, Salce, Thomasson, Bernhoeft, Braithwaite, and
  Flouquet}}]{Niklowitz05}
\bibinfo{author}{\bibfnamefont{P.~G.} \bibnamefont{Niklowitz}},
  \bibinfo{author}{\bibfnamefont{F.}~\bibnamefont{Beckers}},
  \bibinfo{author}{\bibfnamefont{G.~G.} \bibnamefont{Lonzarich}},
  \bibinfo{author}{\bibfnamefont{G.}~\bibnamefont{Knebel}},
  \bibinfo{author}{\bibfnamefont{B.}~\bibnamefont{Salce}},
  \bibinfo{author}{\bibfnamefont{J.}~\bibnamefont{Thomasson}},
  \bibinfo{author}{\bibfnamefont{N.}~\bibnamefont{Bernhoeft}},
  \bibinfo{author}{\bibfnamefont{D.}~\bibnamefont{Braithwaite}},
  \bibnamefont{and} \bibinfo{author}{\bibfnamefont{J.}~\bibnamefont{Flouquet}},
  \bibinfo{journal}{Phys. Rev. B} \textbf{\bibinfo{volume}{72}},
  \bibinfo{pages}{024424} (\bibinfo{year}{2005}).

\bibitem[{\citenamefont{Huxley et~al.}(2000)\citenamefont{Huxley, Sheikin, and
  Braithwaite}}]{Huxley00}
\bibinfo{author}{\bibfnamefont{A.}~\bibnamefont{Huxley}},
  \bibinfo{author}{\bibfnamefont{I.}~\bibnamefont{Sheikin}}, \bibnamefont{and}
  \bibinfo{author}{\bibfnamefont{D.}~\bibnamefont{Braithwaite}},
  \bibinfo{journal}{Physica B: Condensed Matter}
  \textbf{\bibinfo{volume}{284-288}}, \bibinfo{pages}{1277 }
  (\bibinfo{year}{2000}).

\bibitem[{\citenamefont{Lengyel et~al.}(2015)\citenamefont{Lengyel, Macovei,
  Jesche, Krellner, Geibel, and Nicklas}}]{Lengyel15}
\bibinfo{author}{\bibfnamefont{E.}~\bibnamefont{Lengyel}},
  \bibinfo{author}{\bibfnamefont{M.~E.} \bibnamefont{Macovei}},
  \bibinfo{author}{\bibfnamefont{A.}~\bibnamefont{Jesche}},
  \bibinfo{author}{\bibfnamefont{C.}~\bibnamefont{Krellner}},
  \bibinfo{author}{\bibfnamefont{C.}~\bibnamefont{Geibel}}, \bibnamefont{and}
  \bibinfo{author}{\bibfnamefont{M.}~\bibnamefont{Nicklas}},
  \bibinfo{journal}{Phys. Rev. B} \textbf{\bibinfo{volume}{91}},
  \bibinfo{pages}{035130} (\bibinfo{year}{2015}).

\bibitem[{\citenamefont{Abdul-Jabbar et~al.}(2015)\citenamefont{Abdul-Jabbar,
  Sokolov, O’Neill, Stock, Wermeille, Demmel, Krüger, Green, Lévy-Bertrand,
  Grenier et~al.}}]{Abdul15}
\bibinfo{author}{\bibfnamefont{G.}~\bibnamefont{Abdul-Jabbar}},
  \bibinfo{author}{\bibfnamefont{D.~A.} \bibnamefont{Sokolov}},
  \bibinfo{author}{\bibfnamefont{C.~D.} \bibnamefont{O’Neill}},
  \bibinfo{author}{\bibfnamefont{C.}~\bibnamefont{Stock}},
  \bibinfo{author}{\bibfnamefont{D.}~\bibnamefont{Wermeille}},
  \bibinfo{author}{\bibfnamefont{F.}~\bibnamefont{Demmel}},
  \bibinfo{author}{\bibfnamefont{F.}~\bibnamefont{Krüger}},
  \bibinfo{author}{\bibfnamefont{A.~G.} \bibnamefont{Green}},
  \bibinfo{author}{\bibfnamefont{F.}~\bibnamefont{Lévy-Bertrand}},
  \bibinfo{author}{\bibfnamefont{B.}~\bibnamefont{Grenier}},
  \bibnamefont{et~al.}, \bibinfo{journal}{Nat. Phys.}
  \textbf{\bibinfo{volume}{11}}, \bibinfo{pages}{321–327}
  (\bibinfo{year}{2015}).

\bibitem[{\citenamefont{Friedemann et~al.}(2018)\citenamefont{Friedemann,
  Duncan, Hirschberger, Bauer, Küchler, Neubauer, Brando, Pfleiderer, and
  Grosche}}]{Friedemann18}
\bibinfo{author}{\bibfnamefont{S.}~\bibnamefont{Friedemann}},
  \bibinfo{author}{\bibfnamefont{W.~J.} \bibnamefont{Duncan}},
  \bibinfo{author}{\bibfnamefont{M.}~\bibnamefont{Hirschberger}},
  \bibinfo{author}{\bibfnamefont{T.~W.} \bibnamefont{Bauer}},
  \bibinfo{author}{\bibfnamefont{R.}~\bibnamefont{Küchler}},
  \bibinfo{author}{\bibfnamefont{A.}~\bibnamefont{Neubauer}},
  \bibinfo{author}{\bibfnamefont{M.}~\bibnamefont{Brando}},
  \bibinfo{author}{\bibfnamefont{C.}~\bibnamefont{Pfleiderer}},
  \bibnamefont{and} \bibinfo{author}{\bibfnamefont{F.~M.}
  \bibnamefont{Grosche}}, \bibinfo{journal}{Nat. Phys.}
  \textbf{\bibinfo{volume}{14}}, \bibinfo{pages}{62} (\bibinfo{year}{2018}).

\bibitem[{\citenamefont{Niklowitz et~al.}(2019)\citenamefont{Niklowitz,
  Hirschberger, Lucas, Cermak, Schneidewind, Faulhaber, Mignot, Duncan,
  Neubauer, Pfleiderer et~al.}}]{Niklowitz19}
\bibinfo{author}{\bibfnamefont{P.~G.} \bibnamefont{Niklowitz}},
  \bibinfo{author}{\bibfnamefont{M.}~\bibnamefont{Hirschberger}},
  \bibinfo{author}{\bibfnamefont{M.}~\bibnamefont{Lucas}},
  \bibinfo{author}{\bibfnamefont{P.}~\bibnamefont{Cermak}},
  \bibinfo{author}{\bibfnamefont{A.}~\bibnamefont{Schneidewind}},
  \bibinfo{author}{\bibfnamefont{E.}~\bibnamefont{Faulhaber}},
  \bibinfo{author}{\bibfnamefont{J.-M.} \bibnamefont{Mignot}},
  \bibinfo{author}{\bibfnamefont{W.~J.} \bibnamefont{Duncan}},
  \bibinfo{author}{\bibfnamefont{A.}~\bibnamefont{Neubauer}},
  \bibinfo{author}{\bibfnamefont{C.}~\bibnamefont{Pfleiderer}},
  \bibnamefont{et~al.}, \bibinfo{journal}{Phys. Rev. Lett.}
  \textbf{\bibinfo{volume}{123}}, \bibinfo{pages}{247203}
  (\bibinfo{year}{2019}).

\bibitem[{\citenamefont{Taufour
  et~al.}(2016{\natexlab{a}})\citenamefont{Taufour, Kaluarachchi, Khasanov,
  Nguyen, Guguchia, Biswas, Bonf\`a, De~Renzi, Lin, Kim et~al.}}]{Taufour16}
\bibinfo{author}{\bibfnamefont{V.}~\bibnamefont{Taufour}},
  \bibinfo{author}{\bibfnamefont{U.~S.} \bibnamefont{Kaluarachchi}},
  \bibinfo{author}{\bibfnamefont{R.}~\bibnamefont{Khasanov}},
  \bibinfo{author}{\bibfnamefont{M.~C.} \bibnamefont{Nguyen}},
  \bibinfo{author}{\bibfnamefont{Z.}~\bibnamefont{Guguchia}},
  \bibinfo{author}{\bibfnamefont{P.~K.} \bibnamefont{Biswas}},
  \bibinfo{author}{\bibfnamefont{P.}~\bibnamefont{Bonf\`a}},
  \bibinfo{author}{\bibfnamefont{R.}~\bibnamefont{De~Renzi}},
  \bibinfo{author}{\bibfnamefont{X.}~\bibnamefont{Lin}},
  \bibinfo{author}{\bibfnamefont{S.~K.} \bibnamefont{Kim}},
  \bibnamefont{et~al.}, \bibinfo{journal}{Phys. Rev. Lett.}
  \textbf{\bibinfo{volume}{117}}, \bibinfo{pages}{037207}
  (\bibinfo{year}{2016}{\natexlab{a}}).

\bibitem[{\citenamefont{Kaluarachchi et~al.}(2017)\citenamefont{Kaluarachchi,
  Bud'ko, Canfield, and Taufour}}]{Kaluarachchi17}
\bibinfo{author}{\bibfnamefont{U.~S.} \bibnamefont{Kaluarachchi}},
  \bibinfo{author}{\bibfnamefont{S.~L.} \bibnamefont{Bud'ko}},
  \bibinfo{author}{\bibfnamefont{P.~C.} \bibnamefont{Canfield}},
  \bibnamefont{and} \bibinfo{author}{\bibfnamefont{V.}~\bibnamefont{Taufour}},
  \bibinfo{journal}{Nat. Commun.} \textbf{\bibinfo{volume}{8}},
  \bibinfo{pages}{546} (\bibinfo{year}{2017}).

\bibitem[{\citenamefont{Lin et~al.}(2013)\citenamefont{Lin, Taufour, Bud'ko,
  and Canfield}}]{Lin13}
\bibinfo{author}{\bibfnamefont{X.}~\bibnamefont{Lin}},
  \bibinfo{author}{\bibfnamefont{V.}~\bibnamefont{Taufour}},
  \bibinfo{author}{\bibfnamefont{S.~L.} \bibnamefont{Bud'ko}},
  \bibnamefont{and} \bibinfo{author}{\bibfnamefont{P.~C.}
  \bibnamefont{Canfield}}, \bibinfo{journal}{Phys. Rev. B}
  \textbf{\bibinfo{volume}{88}}, \bibinfo{pages}{094405}
  (\bibinfo{year}{2013}).

\bibitem[{\citenamefont{Taufour et~al.}(2010)\citenamefont{Taufour, Aoki,
  Knebel, and Flouquet}}]{Taufour10}
\bibinfo{author}{\bibfnamefont{V.}~\bibnamefont{Taufour}},
  \bibinfo{author}{\bibfnamefont{D.}~\bibnamefont{Aoki}},
  \bibinfo{author}{\bibfnamefont{G.}~\bibnamefont{Knebel}}, \bibnamefont{and}
  \bibinfo{author}{\bibfnamefont{J.}~\bibnamefont{Flouquet}},
  \bibinfo{journal}{Phys. Rev. Lett.} \textbf{\bibinfo{volume}{105}},
  \bibinfo{pages}{217201} (\bibinfo{year}{2010}).

\bibitem[{\citenamefont{Taufour
  et~al.}(2016{\natexlab{b}})\citenamefont{Taufour, Kaluarachchi, and
  Kogan}}]{Taufour16b}
\bibinfo{author}{\bibfnamefont{V.}~\bibnamefont{Taufour}},
  \bibinfo{author}{\bibfnamefont{U.~S.} \bibnamefont{Kaluarachchi}},
  \bibnamefont{and} \bibinfo{author}{\bibfnamefont{V.~G.} \bibnamefont{Kogan}},
  \bibinfo{journal}{Phys. Rev. B} \textbf{\bibinfo{volume}{94}},
  \bibinfo{pages}{060410} (\bibinfo{year}{2016}{\natexlab{b}}).

\bibitem[{SI()}]{SI}
\bibinfo{howpublished}{See Supplemental Material [url] for experimental details
  and additional thermodynamic, transport and microscopic data, which includes
  Refs.
  \cite{Lin13,Canfield16,Taufour16,Gati19,Budko84,Torikachvili15,Eiling81,Xiang20,Krawitz89,Kabeya11,Sidorov05,Aso06,Dissanayake19,Haberl17,Haberl18,Drozd07,easylab,Khasanov16,Klotz09,Klotz12,Tateiwa09,Smith67,Barron99,Kuechler12,Schmiedeshoff06,Lindbaum02,Kaluarachchi17,Bie07,Cadogan13,Chang04,Renninger68,Ye18,Rodriguez93,Yaouanc11}.}

\bibitem[{\citenamefont{Yaouanc and Dalmas De~R\'{e}otier}(2011)}]{Yaouanc11}
\bibinfo{author}{\bibfnamefont{A.}~\bibnamefont{Yaouanc}} \bibnamefont{and}
  \bibinfo{author}{\bibfnamefont{P.}~\bibnamefont{Dalmas De~R\'{e}otier}},
  \emph{\bibinfo{title}{Muon Spin Rotation, Relaxation and Resonance}}
  (\bibinfo{publisher}{Oxford University Press, Oxford}, \bibinfo{year}{2011}).

\bibitem[{\citenamefont{Karahasanovic et~al.}(2012)\citenamefont{Karahasanovic,
  Kr\"uger, and Green}}]{Karahasanovic12}
\bibinfo{author}{\bibfnamefont{U.}~\bibnamefont{Karahasanovic}},
  \bibinfo{author}{\bibfnamefont{F.}~\bibnamefont{Kr\"uger}}, \bibnamefont{and}
  \bibinfo{author}{\bibfnamefont{A.~G.} \bibnamefont{Green}},
  \bibinfo{journal}{Phys. Rev. B} \textbf{\bibinfo{volume}{85}},
  \bibinfo{pages}{165111} (\bibinfo{year}{2012}).

\bibitem[{\citenamefont{Chubukov and Maslov}(2009)}]{Chubukov09}
\bibinfo{author}{\bibfnamefont{A.~V.} \bibnamefont{Chubukov}} \bibnamefont{and}
  \bibinfo{author}{\bibfnamefont{D.~L.} \bibnamefont{Maslov}},
  \bibinfo{journal}{Phys. Rev. Lett.} \textbf{\bibinfo{volume}{103}},
  \bibinfo{pages}{216401} (\bibinfo{year}{2009}).

\bibitem[{\citenamefont{Thomson et~al.}(2013)\citenamefont{Thomson, Kr\"uger,
  and Green}}]{Thomson13}
\bibinfo{author}{\bibfnamefont{S.~J.} \bibnamefont{Thomson}},
  \bibinfo{author}{\bibfnamefont{F.}~\bibnamefont{Kr\"uger}}, \bibnamefont{and}
  \bibinfo{author}{\bibfnamefont{A.~G.} \bibnamefont{Green}},
  \bibinfo{journal}{Phys. Rev. B} \textbf{\bibinfo{volume}{87}},
  \bibinfo{pages}{224203} (\bibinfo{year}{2013}).

\bibitem[{\citenamefont{Pedder et~al.}(2013)\citenamefont{Pedder, Kr\"uger, and
  Green}}]{Pedder13}
\bibinfo{author}{\bibfnamefont{C.~J.} \bibnamefont{Pedder}},
  \bibinfo{author}{\bibfnamefont{F.}~\bibnamefont{Kr\"uger}}, \bibnamefont{and}
  \bibinfo{author}{\bibfnamefont{A.~G.} \bibnamefont{Green}},
  \bibinfo{journal}{Phys. Rev. B} \textbf{\bibinfo{volume}{88}},
  \bibinfo{pages}{165109} (\bibinfo{year}{2013}).

\bibitem[{\citenamefont{Lausberg et~al.}(2012)\citenamefont{Lausberg, Spehling,
  Steppke, Jesche, Luetkens, Amato, Baines, Krellner, Brando, Geibel
  et~al.}}]{Lausberg12}
\bibinfo{author}{\bibfnamefont{S.}~\bibnamefont{Lausberg}},
  \bibinfo{author}{\bibfnamefont{J.}~\bibnamefont{Spehling}},
  \bibinfo{author}{\bibfnamefont{A.}~\bibnamefont{Steppke}},
  \bibinfo{author}{\bibfnamefont{A.}~\bibnamefont{Jesche}},
  \bibinfo{author}{\bibfnamefont{H.}~\bibnamefont{Luetkens}},
  \bibinfo{author}{\bibfnamefont{A.}~\bibnamefont{Amato}},
  \bibinfo{author}{\bibfnamefont{C.}~\bibnamefont{Baines}},
  \bibinfo{author}{\bibfnamefont{C.}~\bibnamefont{Krellner}},
  \bibinfo{author}{\bibfnamefont{M.}~\bibnamefont{Brando}},
  \bibinfo{author}{\bibfnamefont{C.}~\bibnamefont{Geibel}},
  \bibnamefont{et~al.}, \bibinfo{journal}{Phys. Rev. Lett.}
  \textbf{\bibinfo{volume}{109}}, \bibinfo{pages}{216402}
  (\bibinfo{year}{2012}).

\bibitem[{\citenamefont{Canfield et~al.}(2016)\citenamefont{Canfield, Kong,
  Kaluarachchi, and Jo}}]{Canfield16}
\bibinfo{author}{\bibfnamefont{P.~C.} \bibnamefont{Canfield}},
  \bibinfo{author}{\bibfnamefont{T.}~\bibnamefont{Kong}},
  \bibinfo{author}{\bibfnamefont{U.~S.} \bibnamefont{Kaluarachchi}},
  \bibnamefont{and} \bibinfo{author}{\bibfnamefont{N.~H.} \bibnamefont{Jo}},
  \bibinfo{journal}{Philos. Mag.} \textbf{\bibinfo{volume}{96}},
  \bibinfo{pages}{84} (\bibinfo{year}{2016}).

\bibitem[{\citenamefont{Gati et~al.}(2019)\citenamefont{Gati, Drachuck, Xiang,
  Wang, Bud'ko, and Canfield}}]{Gati19}
\bibinfo{author}{\bibfnamefont{E.}~\bibnamefont{Gati}},
  \bibinfo{author}{\bibfnamefont{G.}~\bibnamefont{Drachuck}},
  \bibinfo{author}{\bibfnamefont{L.}~\bibnamefont{Xiang}},
  \bibinfo{author}{\bibfnamefont{L.-L.} \bibnamefont{Wang}},
  \bibinfo{author}{\bibfnamefont{S.~L.} \bibnamefont{Bud'ko}},
  \bibnamefont{and} \bibinfo{author}{\bibfnamefont{P.~C.}
  \bibnamefont{Canfield}}, \bibinfo{journal}{Rev. Sci. Instrum.}
  \textbf{\bibinfo{volume}{90}}, \bibinfo{pages}{023911}
  (\bibinfo{year}{2019}).

\bibitem[{\citenamefont{Bud'ko et~al.}(1984)\citenamefont{Bud'ko, Voronovskii,
  Gapotchenko, and ltskevich}}]{Budko84}
\bibinfo{author}{\bibfnamefont{S.~L.} \bibnamefont{Bud'ko}},
  \bibinfo{author}{\bibfnamefont{A.~N.} \bibnamefont{Voronovskii}},
  \bibinfo{author}{\bibfnamefont{A.~G.} \bibnamefont{Gapotchenko}},
  \bibnamefont{and} \bibinfo{author}{\bibfnamefont{E.~S.}
  \bibnamefont{ltskevich}}, \bibinfo{journal}{Zh. Eksp. Teor. Fiz.}
  \textbf{\bibinfo{volume}{86}}, \bibinfo{pages}{778} (\bibinfo{year}{1984}).

\bibitem[{\citenamefont{Torikachvili et~al.}(2015)\citenamefont{Torikachvili,
  Kim, Colombier, Bud'ko, and Canfield}}]{Torikachvili15}
\bibinfo{author}{\bibfnamefont{M.~S.} \bibnamefont{Torikachvili}},
  \bibinfo{author}{\bibfnamefont{S.~K.} \bibnamefont{Kim}},
  \bibinfo{author}{\bibfnamefont{E.}~\bibnamefont{Colombier}},
  \bibinfo{author}{\bibfnamefont{S.~L.} \bibnamefont{Bud'ko}},
  \bibnamefont{and} \bibinfo{author}{\bibfnamefont{P.~C.}
  \bibnamefont{Canfield}}, \bibinfo{journal}{Rev. Sci. Instrum.}
  \textbf{\bibinfo{volume}{86}}, \bibinfo{pages}{123904}
  (\bibinfo{year}{2015}).

\bibitem[{\citenamefont{Eiling and Schilling}(1981)}]{Eiling81}
\bibinfo{author}{\bibfnamefont{A.}~\bibnamefont{Eiling}} \bibnamefont{and}
  \bibinfo{author}{\bibfnamefont{J.~S.} \bibnamefont{Schilling}},
  \bibinfo{journal}{Journal of Physics F: Metal Physics}
  \textbf{\bibinfo{volume}{11}}, \bibinfo{pages}{623} (\bibinfo{year}{1981}).

\bibitem[{\citenamefont{Xiang et~al.}(2020)\citenamefont{Xiang, Gati, Bud’ko,
  Ribeiro, Ata, Tutsch, Lang, and Canfield}}]{Xiang20}
\bibinfo{author}{\bibfnamefont{L.}~\bibnamefont{Xiang}},
  \bibinfo{author}{\bibfnamefont{E.}~\bibnamefont{Gati}},
  \bibinfo{author}{\bibfnamefont{S.~L.} \bibnamefont{Bud’ko}},
  \bibinfo{author}{\bibfnamefont{R.~A.} \bibnamefont{Ribeiro}},
  \bibinfo{author}{\bibfnamefont{A.}~\bibnamefont{Ata}},
  \bibinfo{author}{\bibfnamefont{U.}~\bibnamefont{Tutsch}},
  \bibinfo{author}{\bibfnamefont{M.}~\bibnamefont{Lang}}, \bibnamefont{and}
  \bibinfo{author}{\bibfnamefont{P.~C.} \bibnamefont{Canfield}},
  \bibinfo{journal}{Rev. Sci. Instrum.} \textbf{\bibinfo{volume}{91}},
  \bibinfo{pages}{095103} (\bibinfo{year}{2020}).

\bibitem[{\citenamefont{Krawitz et~al.}(1989)\citenamefont{Krawitz, Reichel,
  and Hitterman}}]{Krawitz89}
\bibinfo{author}{\bibfnamefont{A.~D.} \bibnamefont{Krawitz}},
  \bibinfo{author}{\bibfnamefont{D.~G.} \bibnamefont{Reichel}},
  \bibnamefont{and}
  \bibinfo{author}{\bibfnamefont{R.}~\bibnamefont{Hitterman}},
  \bibinfo{journal}{J. Am. Ceram. Soc.} \textbf{\bibinfo{volume}{72}},
  \bibinfo{pages}{515} (\bibinfo{year}{1989}).

\bibitem[{\citenamefont{Kabeya et~al.}(2011)\citenamefont{Kabeya, Imura,
  Deguchi, and K.~Sato}}]{Kabeya11}
\bibinfo{author}{\bibfnamefont{N.}~\bibnamefont{Kabeya}},
  \bibinfo{author}{\bibfnamefont{K.}~\bibnamefont{Imura}},
  \bibinfo{author}{\bibfnamefont{K.}~\bibnamefont{Deguchi}}, \bibnamefont{and}
  \bibinfo{author}{\bibfnamefont{N.}~\bibnamefont{K.~Sato}},
  \bibinfo{journal}{Journal of the Physical Society of Japan}
  \textbf{\bibinfo{volume}{80}}, \bibinfo{pages}{SA098} (\bibinfo{year}{2011}).

\bibitem[{\citenamefont{Sidorov and Sadykov}(2005)}]{Sidorov05}
\bibinfo{author}{\bibfnamefont{V.~A.} \bibnamefont{Sidorov}} \bibnamefont{and}
  \bibinfo{author}{\bibfnamefont{R.~A.} \bibnamefont{Sadykov}},
  \bibinfo{journal}{Journal of Physics: Condensed Matter}
  \textbf{\bibinfo{volume}{17}}, \bibinfo{pages}{S3005} (\bibinfo{year}{2005}).

\bibitem[{\citenamefont{Aso et~al.}(2006)\citenamefont{Aso, Uwatoko, Fujiwara,
  Motoyama, Ban, Homma, Shiokawa, Hirota, and Sato}}]{Aso06}
\bibinfo{author}{\bibfnamefont{N.}~\bibnamefont{Aso}},
  \bibinfo{author}{\bibfnamefont{Y.}~\bibnamefont{Uwatoko}},
  \bibinfo{author}{\bibfnamefont{T.}~\bibnamefont{Fujiwara}},
  \bibinfo{author}{\bibfnamefont{G.}~\bibnamefont{Motoyama}},
  \bibinfo{author}{\bibfnamefont{S.}~\bibnamefont{Ban}},
  \bibinfo{author}{\bibfnamefont{Y.}~\bibnamefont{Homma}},
  \bibinfo{author}{\bibfnamefont{Y.}~\bibnamefont{Shiokawa}},
  \bibinfo{author}{\bibfnamefont{K.}~\bibnamefont{Hirota}}, \bibnamefont{and}
  \bibinfo{author}{\bibfnamefont{N.~K.} \bibnamefont{Sato}},
  \bibinfo{journal}{AIP Conference Proceedings} \textbf{\bibinfo{volume}{850}},
  \bibinfo{pages}{705} (\bibinfo{year}{2006}).

\bibitem[{\citenamefont{Dissanayake et~al.}(2019)\citenamefont{Dissanayake,
  Matsuda, Munakata, Kagi, Gouchi, and Uwatoko}}]{Dissanayake19}
\bibinfo{author}{\bibfnamefont{S.}~\bibnamefont{Dissanayake}},
  \bibinfo{author}{\bibfnamefont{M.}~\bibnamefont{Matsuda}},
  \bibinfo{author}{\bibfnamefont{K.}~\bibnamefont{Munakata}},
  \bibinfo{author}{\bibfnamefont{H.}~\bibnamefont{Kagi}},
  \bibinfo{author}{\bibfnamefont{J.}~\bibnamefont{Gouchi}}, \bibnamefont{and}
  \bibinfo{author}{\bibfnamefont{Y.}~\bibnamefont{Uwatoko}},
  \bibinfo{journal}{Journal of Physics: Condensed Matter}
  \textbf{\bibinfo{volume}{31}}, \bibinfo{pages}{384001}
  (\bibinfo{year}{2019}).

\bibitem[{\citenamefont{Haberl et~al.}(2017)\citenamefont{Haberl, Dissanayake,
  Ye, Daemen, Cheng, Li, Ramirez-Cuesta, Matsuda, Molaison, and
  Boehler}}]{Haberl17}
\bibinfo{author}{\bibfnamefont{B.}~\bibnamefont{Haberl}},
  \bibinfo{author}{\bibfnamefont{S.}~\bibnamefont{Dissanayake}},
  \bibinfo{author}{\bibfnamefont{F.}~\bibnamefont{Ye}},
  \bibinfo{author}{\bibfnamefont{L.~L.} \bibnamefont{Daemen}},
  \bibinfo{author}{\bibfnamefont{Y.}~\bibnamefont{Cheng}},
  \bibinfo{author}{\bibfnamefont{C.~W.} \bibnamefont{Li}},
  \bibinfo{author}{\bibfnamefont{A.-J.~T.} \bibnamefont{Ramirez-Cuesta}},
  \bibinfo{author}{\bibfnamefont{M.}~\bibnamefont{Matsuda}},
  \bibinfo{author}{\bibfnamefont{J.~J.} \bibnamefont{Molaison}},
  \bibnamefont{and} \bibinfo{author}{\bibfnamefont{R.}~\bibnamefont{Boehler}},
  \bibinfo{journal}{High Pressure Research} \textbf{\bibinfo{volume}{37}},
  \bibinfo{pages}{495} (\bibinfo{year}{2017}).

\bibitem[{\citenamefont{Haberl et~al.}(2018)\citenamefont{Haberl, Dissanayake,
  Wu, Myles, dos Santos, Loguillo, Rucker, Armitage, Cochran, Andrews
  et~al.}}]{Haberl18}
\bibinfo{author}{\bibfnamefont{B.}~\bibnamefont{Haberl}},
  \bibinfo{author}{\bibfnamefont{S.}~\bibnamefont{Dissanayake}},
  \bibinfo{author}{\bibfnamefont{Y.}~\bibnamefont{Wu}},
  \bibinfo{author}{\bibfnamefont{D.~A.} \bibnamefont{Myles}},
  \bibinfo{author}{\bibfnamefont{A.~M.} \bibnamefont{dos Santos}},
  \bibinfo{author}{\bibfnamefont{M.}~\bibnamefont{Loguillo}},
  \bibinfo{author}{\bibfnamefont{G.~M.} \bibnamefont{Rucker}},
  \bibinfo{author}{\bibfnamefont{D.~P.} \bibnamefont{Armitage}},
  \bibinfo{author}{\bibfnamefont{M.}~\bibnamefont{Cochran}},
  \bibinfo{author}{\bibfnamefont{K.~M.} \bibnamefont{Andrews}},
  \bibnamefont{et~al.}, \bibinfo{journal}{Rev. Sci. Instrum.}
  \textbf{\bibinfo{volume}{89}}, \bibinfo{pages}{092902}
  (\bibinfo{year}{2018}).

\bibitem[{\citenamefont{Drozd-Rzoska et~al.}(2007)\citenamefont{Drozd-Rzoska,
  Rzoska, Paluch, Imre, and Roland}}]{Drozd07}
\bibinfo{author}{\bibfnamefont{A.}~\bibnamefont{Drozd-Rzoska}},
  \bibinfo{author}{\bibfnamefont{S.~J.} \bibnamefont{Rzoska}},
  \bibinfo{author}{\bibfnamefont{M.}~\bibnamefont{Paluch}},
  \bibinfo{author}{\bibfnamefont{A.~R.} \bibnamefont{Imre}}, \bibnamefont{and}
  \bibinfo{author}{\bibfnamefont{C.~M.} \bibnamefont{Roland}},
  \bibinfo{journal}{The Journal of Chemical Physics}
  \textbf{\bibinfo{volume}{126}}, \bibinfo{pages}{164504}
  (\bibinfo{year}{2007}).

\bibitem[{eas()}]{easylab}
\bibinfo{note}{\url{https://www.almax-easylab.com/ProductDetails.aspx?PID=50\&IID=10029}}.

\bibitem[{\citenamefont{Khasanov et~al.}(2016)\citenamefont{Khasanov, Guguchia,
  Maisuradze, Andreica, Elender, Raselli, Shermadini, Goko, Knecht, Morenzoni
  et~al.}}]{Khasanov16}
\bibinfo{author}{\bibfnamefont{R.}~\bibnamefont{Khasanov}},
  \bibinfo{author}{\bibfnamefont{Z.}~\bibnamefont{Guguchia}},
  \bibinfo{author}{\bibfnamefont{A.}~\bibnamefont{Maisuradze}},
  \bibinfo{author}{\bibfnamefont{D.}~\bibnamefont{Andreica}},
  \bibinfo{author}{\bibfnamefont{M.}~\bibnamefont{Elender}},
  \bibinfo{author}{\bibfnamefont{A.}~\bibnamefont{Raselli}},
  \bibinfo{author}{\bibfnamefont{Z.}~\bibnamefont{Shermadini}},
  \bibinfo{author}{\bibfnamefont{T.}~\bibnamefont{Goko}},
  \bibinfo{author}{\bibfnamefont{F.}~\bibnamefont{Knecht}},
  \bibinfo{author}{\bibfnamefont{E.}~\bibnamefont{Morenzoni}},
  \bibnamefont{et~al.}, \bibinfo{journal}{High Pressure Research}
  \textbf{\bibinfo{volume}{36}}, \bibinfo{pages}{140} (\bibinfo{year}{2016}).

\bibitem[{\citenamefont{Klotz et~al.}(2009)\citenamefont{Klotz, Chervin,
  Munsch, and Marchand}}]{Klotz09}
\bibinfo{author}{\bibfnamefont{S.}~\bibnamefont{Klotz}},
  \bibinfo{author}{\bibfnamefont{J.-C.} \bibnamefont{Chervin}},
  \bibinfo{author}{\bibfnamefont{P.}~\bibnamefont{Munsch}}, \bibnamefont{and}
  \bibinfo{author}{\bibfnamefont{G.~L.} \bibnamefont{Marchand}},
  \bibinfo{journal}{Journal of Physics D: Applied Physics}
  \textbf{\bibinfo{volume}{42}}, \bibinfo{pages}{075413}
  (\bibinfo{year}{2009}).

\bibitem[{\citenamefont{Klotz et~al.}(2012)\citenamefont{Klotz, Takemura,
  Strässle, and Hansen}}]{Klotz12}
\bibinfo{author}{\bibfnamefont{S.}~\bibnamefont{Klotz}},
  \bibinfo{author}{\bibfnamefont{K.}~\bibnamefont{Takemura}},
  \bibinfo{author}{\bibfnamefont{T.}~\bibnamefont{Strässle}},
  \bibnamefont{and} \bibinfo{author}{\bibfnamefont{T.}~\bibnamefont{Hansen}},
  \bibinfo{journal}{Journal of Physics: Condensed Matter}
  \textbf{\bibinfo{volume}{24}}, \bibinfo{pages}{325103}
  (\bibinfo{year}{2012}).

\bibitem[{\citenamefont{Tateiwa and Haga}(2009)}]{Tateiwa09}
\bibinfo{author}{\bibfnamefont{N.}~\bibnamefont{Tateiwa}} \bibnamefont{and}
  \bibinfo{author}{\bibfnamefont{Y.}~\bibnamefont{Haga}},
  \bibinfo{journal}{Rev. Sci. Instrum.} \textbf{\bibinfo{volume}{80}},
  \bibinfo{pages}{123901} (\bibinfo{year}{2009}).

\bibitem[{\citenamefont{Smith and Chu}(1967)}]{Smith67}
\bibinfo{author}{\bibfnamefont{T.~F.} \bibnamefont{Smith}} \bibnamefont{and}
  \bibinfo{author}{\bibfnamefont{C.~W.} \bibnamefont{Chu}},
  \bibinfo{journal}{Phys. Rev.} \textbf{\bibinfo{volume}{159}},
  \bibinfo{pages}{353} (\bibinfo{year}{1967}).

\bibitem[{\citenamefont{Barron and White}(1999)}]{Barron99}
\bibinfo{author}{\bibfnamefont{T.}~\bibnamefont{Barron}} \bibnamefont{and}
  \bibinfo{author}{\bibfnamefont{G.}~\bibnamefont{White}},
  \emph{\bibinfo{title}{Heat Capacity and Thermal Expansion at Low
  Temperatures}} (\bibinfo{publisher}{Springer US}, \bibinfo{year}{1999}).

\bibitem[{\citenamefont{Küchler et~al.}(2012)\citenamefont{Küchler, Bauer,
  Brando, and Steglich}}]{Kuechler12}
\bibinfo{author}{\bibfnamefont{R.}~\bibnamefont{Küchler}},
  \bibinfo{author}{\bibfnamefont{T.}~\bibnamefont{Bauer}},
  \bibinfo{author}{\bibfnamefont{M.}~\bibnamefont{Brando}}, \bibnamefont{and}
  \bibinfo{author}{\bibfnamefont{F.}~\bibnamefont{Steglich}},
  \bibinfo{journal}{Rev. Sci. Instrum.} \textbf{\bibinfo{volume}{83}},
  \bibinfo{pages}{095102} (\bibinfo{year}{2012}).

\bibitem[{\citenamefont{Schmiedeshoff et~al.}(2006)\citenamefont{Schmiedeshoff,
  Lounsbury, Luna, Tracy, Schramm, Tozer, Correa, Hannahs, Murphy, Palm
  et~al.}}]{Schmiedeshoff06}
\bibinfo{author}{\bibfnamefont{G.~M.} \bibnamefont{Schmiedeshoff}},
  \bibinfo{author}{\bibfnamefont{A.~W.} \bibnamefont{Lounsbury}},
  \bibinfo{author}{\bibfnamefont{D.~J.} \bibnamefont{Luna}},
  \bibinfo{author}{\bibfnamefont{S.~J.} \bibnamefont{Tracy}},
  \bibinfo{author}{\bibfnamefont{A.~J.} \bibnamefont{Schramm}},
  \bibinfo{author}{\bibfnamefont{S.~W.} \bibnamefont{Tozer}},
  \bibinfo{author}{\bibfnamefont{V.~F.} \bibnamefont{Correa}},
  \bibinfo{author}{\bibfnamefont{S.~T.} \bibnamefont{Hannahs}},
  \bibinfo{author}{\bibfnamefont{T.~P.} \bibnamefont{Murphy}},
  \bibinfo{author}{\bibfnamefont{E.~C.} \bibnamefont{Palm}},
  \bibnamefont{et~al.}, \bibinfo{journal}{Rev. Sci. Instrum.}
  \textbf{\bibinfo{volume}{77}}, \bibinfo{pages}{123907}
  (\bibinfo{year}{2006}).

\bibitem[{\citenamefont{Lindbaum and Rotter}(2002)}]{Lindbaum02}
\bibinfo{author}{\bibfnamefont{A.}~\bibnamefont{Lindbaum}} \bibnamefont{and}
  \bibinfo{author}{\bibfnamefont{M.}~\bibnamefont{Rotter}},
  \emph{\bibinfo{title}{Handbook of Magnetic Materials}}
  (\bibinfo{publisher}{North Holland}, \bibinfo{year}{2002}), chap.
  \bibinfo{chapter}{Spontaneous magnetoelastic effects in gadolinium
  compounds}.

\bibitem[{\citenamefont{Bie et~al.}(2007)\citenamefont{Bie, Zelinska, Tkachuk,
  and Mar}}]{Bie07}
\bibinfo{author}{\bibfnamefont{H.}~\bibnamefont{Bie}},
  \bibinfo{author}{\bibfnamefont{O.~Y.} \bibnamefont{Zelinska}},
  \bibinfo{author}{\bibfnamefont{A.~V.} \bibnamefont{Tkachuk}},
  \bibnamefont{and} \bibinfo{author}{\bibfnamefont{A.}~\bibnamefont{Mar}},
  \bibinfo{journal}{Chemistry of Materials} \textbf{\bibinfo{volume}{19}},
  \bibinfo{pages}{4613} (\bibinfo{year}{2007}).

\bibitem[{\citenamefont{Cadogan et~al.}(2013)\citenamefont{Cadogan, Lemoine,
  Slater, Mar, and Avdeev}}]{Cadogan13}
\bibinfo{author}{\bibfnamefont{J.}~\bibnamefont{Cadogan}},
  \bibinfo{author}{\bibfnamefont{P.}~\bibnamefont{Lemoine}},
  \bibinfo{author}{\bibfnamefont{B.~R.} \bibnamefont{Slater}},
  \bibinfo{author}{\bibfnamefont{A.}~\bibnamefont{Mar}}, \bibnamefont{and}
  \bibinfo{author}{\bibfnamefont{M.}~\bibnamefont{Avdeev}}, in
  \emph{\bibinfo{booktitle}{Solid Compounds of Transition Elements II}}
  (\bibinfo{publisher}{Trans Tech Publications Ltd}, \bibinfo{year}{2013}),
  vol. \bibinfo{volume}{194} of \emph{\bibinfo{series}{Solid State Phenomena}},
  pp. \bibinfo{pages}{71--74}.

\bibitem[{\citenamefont{Chang}(2004)}]{Chang04}
\bibinfo{author}{\bibfnamefont{S.-L.} \bibnamefont{Chang}},
  \emph{\bibinfo{title}{X-ray Multiple-Wave Diffraction}}
  (\bibinfo{publisher}{Springer-Verlag Berlin Heidelberg New York},
  \bibinfo{year}{2004}).

\bibitem[{\citenamefont{Renninger}(1968)}]{Renninger68}
\bibinfo{author}{\bibfnamefont{M.}~\bibnamefont{Renninger}},
  \bibinfo{journal}{Acta Cryst. A} \textbf{\bibinfo{volume}{24}},
  \bibinfo{pages}{143} (\bibinfo{year}{1968}).

\bibitem[{\citenamefont{Ye et~al.}(2018)\citenamefont{Ye, Liu, Whitfield,
  Osborn, and Rosenkranz}}]{Ye18}
\bibinfo{author}{\bibfnamefont{F.}~\bibnamefont{Ye}},
  \bibinfo{author}{\bibfnamefont{Y.}~\bibnamefont{Liu}},
  \bibinfo{author}{\bibfnamefont{R.}~\bibnamefont{Whitfield}},
  \bibinfo{author}{\bibfnamefont{R.}~\bibnamefont{Osborn}}, \bibnamefont{and}
  \bibinfo{author}{\bibfnamefont{S.}~\bibnamefont{Rosenkranz}},
  \bibinfo{journal}{J. Appl. Crystallogr.} \textbf{\bibinfo{volume}{51}},
  \bibinfo{pages}{315} (\bibinfo{year}{2018}).

\bibitem[{\citenamefont{Rodríguez-Carvajal}(1993)}]{Rodriguez93}
\bibinfo{author}{\bibfnamefont{J.}~\bibnamefont{Rodríguez-Carvajal}},
  \bibinfo{journal}{Physica B: Condensed Matter}
  \textbf{\bibinfo{volume}{192}}, \bibinfo{pages}{55 } (\bibinfo{year}{1993}).

\end{thebibliography}

\clearpage


\section*{Supplementary material}

# Supplementary Information: Formation of short-range magnetic order and avoided ferromagnetic quantum criticality in pressurized LaCrGe₃

Elena Gati[1,2], John M. Wilde[1,2], Rustem Khasanov[3], Li Xiang[1,2], Sachith Dissanayake[4],
Ritu Gupta[3], Masaaki Matsuda[4], Feng Ye[4], Bianca Haberl[4], Udhara Kaluarachchi[1,2],
Robert J. McQueeney[1,2], Andreas Kreyssig[1,2], Sergey L. Bud'ko[1,2], and Paul C. Canfield[1,2]

[1] *Ames Laboratory, US Department of Energy, Iowa State University, Ames, Iowa 50011, USA*
[2] *Department of Physics and Astronomy, Iowa State University, Ames, Iowa 50011, USA*
[3] *Laboratory for Muon Spin Spectroscopy, Paul Scherrer Institute, 5232 Villigen PSI, Switzerland and*
[4] *Neutron Scattering Division, Oak Ridge National Laboratory, Oak Ridge, TN, USA*
(Dated: October 20, 2020)

## EXPERIMENTAL METHODS

*Single crystal synthesis* - Single crystals of LaCrGe₃ were synthesized by the flux-growth technique, as described in Ref. [1]. To this end, high-purity elements (> 3N) were premixed in a molar ratio of 13:13:74 by arc-melting. Afterwards, the ingot was placed in a 2 ml fritted alumina crucible[2] and sealed in a fused silica ampoule under argon atmosphere. Subsequently, the ampoule (with the growth material inside) was heated up to 1100°C over 3 h and held there for 5 h. The growth was then cooled to 800°C over 125 h. The excess liquid was decanted at 800°C using a centrifuge in a final step. The obtained single crystals of rod-like shape were characterized by means of x-ray diffraction, resistance and magnetization at ambient pressure prior to all measurements at finite pressures. These results were well consistent with previous reports[1,3,4] in terms of the Curie temperature $T_{FM}$ as well as the residual resistivity ratio $RRR$.

*Specific heat measurements under pressure* - Specific heat under pressure was measured using the AC calorimetry technique, as described in detail in Refs. [5, 6]. To this end, a single crystal of LaCrGe₃ was placed between a heater and a thermometer. The heater was supplied with an oscillating voltage, and the resulting temperature oscillation of the sample, which is related to the specific heat of the sample, was recorded. Given the non-adiabatic conditions of the pressure-cell environment, absolute values of the specific heat cannot be obtained with high accuracy; nonetheless, the technique of AC calorimetry allows for a decoupling of the sample from the bath (i.e., the pressure medium and the cell), to a good approximation, by choosing the appropriate measurement frequency (see Ref. [5] for details on the procedure of the determination of the measurement frequency). Thus, changes of the specific heat with pressure can be obtained reliably. Our implementation of this technique[5] has proven to be particularly sensitive for the detection of specific heat anomalies of varying size, resulting from different amounts of removed entropy, and over a wide range of phase transition temperatures. This is highly beneficial for the present study, where the pronounced specific heat anomaly at high temperatures close

to 90 K at ambient pressure becomes suppressed to very low temperatures and strongly reduced in size.

The cryogenic environment was provided by a closed-cycle cryostat (Janis SHI-950 with a base temperature of ≈ 3.5 K). Pressure was generated in a piston-cylinder double-wall pressure cell with the outer cylinder made out of Grade 5 titanium alloy (Ti 6Al-4V) and the inner cylinder out of Ni-Cr-Al alloy (see Ref. [7] for a very similar design). A mixture of 4:6 light mineral oil:n-pentane was used as a pressure-transmitting medium. It solidifies at $p \approx 3 - 4$ GPa at room temperature[8], thus ensuring hydrostatic pressure application over the available pressure range. Pressure at low temperatures was determined from the shift of the superconducting transition temperature of elemental lead (Pb)[9], which was determined in resistance measurements. The error in the determination of the low-temperature pressure is estimated to be 0.01 GPa, and pressure changes in this particular cell[10] by less than 0.04 GPa by increasing temperature up to 100 K.

*Thermal expansion measurements under pressure* - Thermal expansion, i.e., the macroscopic length change of a crystal of LaCrGe₃ along a particular crystallographic axis as a function of temperature, was measured using strain gauges, which are sensors whose resistance, $R$, changes upon compression or tension. For our measurements, strain gauges (type FLG-02-23, Tokyo Sokki Kenkyujo Co., Ltd. with $R \approx 120 \Omega$) were fixed rigidly to the sample by using Devcon 5 minute epoxy (No. 14250), and the resulting resistance changes of the strain gauges were recorded and converted into length changes using the known gauge factor. In total, two strain gauges were fixed orthogonally on the same sample to measure the expansion along the $ab$ axes and the $c$ axis simultaneously. Since the strain gauge resistance varies not only due to the expansion of the crystal with temperature, but also due to the intrinsic resistance change of the strain gauge wire material, another set of strain gauges was mounted on a sample of tungsten carbide. Given that tungsten carbide is a very hard material and has a comparatively small expansion coefficient over a wide temperature range, and in particular no anomalous behavior,[11] the resistances of the strain gauges mounted on tung-

sten carbide are used to subtract the intrinsic resistance change of the strain gauge from the measured resistance data on LaCrGe$_3$. This subtraction was performed *in situ* by using two Wheatstone bridges (see, e.g., Ref. [12] for similar designs). To this end, in each bridge, one strain gauge on the sample, one strain gauge on tungsten carbide inside the cell and two thin-film resistors with similar and almost temperature-independent absolute resistances of $\approx 120\,\Omega$, which were placed outside of the cell in the low-$T$ environment, were used. The current for the bridge was supplied by a LakeShore 370 Resistance Bridge, which was also used to measure the voltage across each bridge. The cryogenic environment, pressure cell, pressure medium and manometer were identical to the one for specific heat measurements, see above.

*Resistance measurements under pressure* - Resistance under pressure was measured in a four-point configuration with current directed along the crystallographic $c$ axis (Note that previously-published data[3] were obtained with current in the $ab$ plane). Contacts were made using Epo-tek H20E silver epoxy. The AC resistance was measured by a LakeShore 370 Resistance Bridge. The cryogenic environment, pressure cell, pressure medium and manometer were identical to the one for specific heat measurements, see above.

*High-energy x-ray diffraction measurements on single-crystals under pressure* - High-energy (100 keV) x-ray diffraction measurements were performed on single crystals at station 6-ID-D of the Advanced Photon Source, Argonne National Laboratory. The samples were pressurized in diamond anvil cells (part of the Diacell Bragg Series, Almax easylab[13]) using He-gas as a pressure-transmitting medium. We used diamond anvils with $600\,\mu m$ culets and stainless-steel gaskets preindented to thicknesses of $\approx 60\,\mu m$, with laser-drilled holes of diameter $\approx 310\,\mu m$. The wavelength of a fluorescence line of ruby was used for room-temperature pressure calibration. By measuring the lattice parameter of polycrystalline silver, we determined pressure *in situ* at all temperatures and pressures with an accuracy of 0.1 GPa. Large regions of the $(H\,H\,L)$ plane and the powder diffraction pattern of silver were recorded by a MAR345 image plate positioned 1.249 m behind the DAC while the DAC was rocked 2.4° along two independent axes perpendicular to the incident x-ray beam. At ambient pressure, other planes of high-symmetry were also recorded outside of a DAC. In addition, at ambient pressure high-resolution measurements were taken of the Bragg peaks $(16\,0\,0)$ and $(0\,0\,16)$ with a Pixirad-1 detector positioned 1.210 m behind the sample while rocking around one axis perpendicular to the incident x-ray beam.

*Powder x-ray scattering measurements under pressure* - Powder x-ray diffraction measurements were performed under pressure with 30 keV x-rays at station 16-BM-D of the Advanced Photon Source, Argonne National Laboratory. The powder was made by crushing single crystals of LaCrGe$_3$ and only powder of less than a micron size was loaded into the DAC (Diacell Bragg Series, Almax easylab[13]). The DAC was configured identically as for the high-energy x-ray experiment described above with He-gas as pressure-transmitting medium, but the wavelength of a ruby fluoresence line was used to measure pressure at all temperatures and pressures with an accuracy of 0.1 GPa. Large regions of reciprocal space were recorded on a MAR345 image plate positioned 0.412 m behind the DAC. Individual crystallites of LaCrGe$_3$ still had very sharp peaks and so the sample was rocked along one axis perpendicular to the beam to obtain a better powder average.

*Neutron diffraction measurements at HB1 on single crystals at ambient and finite pressure* - Neutron diffraction measurements were performed on single crystals using the HB1 diffractometer at the High Flux Isotope Reactor, Oak Ridge National Laboratory. For measurements taken at ambient pressure a single crystal was sealed in an Al can containing He exchange gas, which was then attached to the head of a He closed-cycle refrigerator (CCR). We refer to this experiment as N0. The beam collimators placed before the monochromator, between the monochromator and the sample, between the sample and analyzer, and between the analyzer and detector were 48'-80'-80'-240', respectively. HB1 operates at a fixed incident energy of 13.5 meV and contamination from higher harmonics in the incident beam was eliminated using Pyrolytic Graphite (PG) filters.

For measurements with $p < 2\,\text{GPa}$, a piston-cylinder cell was used, which is similar in design to the one in Ref. [14]. We refer to these experiments throughout the text as N1. It was loaded with the single crystal placed within a teflon capsule with Daphne 7373 as the pressure-transmitting medium, which solidifies at $2.5\,\text{GPa}$[8] at room temperature. For the 1.9 GPa measurement, the sample was loaded together with a NaCl single crystal, which was used to measure the pressure within the cell based on the lattice parameter changes before and after applying pressure at room temperature with an accuracy of 0.1 GPa. For all other pressures, the pressure was determined with an accuracy of 0.2 GPa based on a calibrated pressure-load curve measured at room temperature for that specific cell and was corrected for temperature-induced reduction of pressure via previous calibration measurements. The cell was then attached directly to the head of a CCR.

For measurements with $p > 2\,\text{GPa}$, a clamp-type palm cubic anvil cell (PCAC) was used with ZrO$_2$ anvils and a gasket made out of an Al-based alloy[15]. We refer to these experiments throughout the text as N2. A single crystal with volume of $0.9 \times 0.9 \times 1.5$ mm$^3$ was attached to the bottom of a teflon capsule together with a 1:1 mixture of Fluorinet FC70 and FC77 as the pressure-transmitting medium. Although this medium solidifies close to 1.1 GPa at room temperature, previous studies

have shown that the PCAC applies pressure almost hydrostatically up to much higher pressures than the solidification pressure due to the three-dimensional anvil design[15] that allows to compress the medium simultaneously along three orthogonal directions. The pressure was determined with an accuracy of 0.3 GPa at room temperature based on a calibrated pressure-load curve for that specific cell and was corrected for temperature-induced reduction of pressure via previous calibration measurements. After applying pressure the cell was then loaded into a high-capacity CCR with a base temperature of approximately 3 K.

*Neutron elastic scattering measurements at CORELLI under pressure* - Elastic scattering measurements using a diamond anvil cell (DAC)[16] were performed using the time-of-flight diffractometer CORELLI at the Spallation Neutron Source, Oak Ridge National Laboratory. We refer to these experiments throughout the text as N3. CORELLI allows for the simultaneous measurement of large sections of the three-dimensional reciprocal space by utilizing a white-beam Laue technique with energy discrimination by modulating the incident beam with a statistical chopper[17]. This allows CORELLI to efficiently separate the elastic and inelastic channel of the diffuse scattering signal, thus identifying whether the observed correlation is static or not. By applying pressure in a DAC at CORELLI we were able to reach pressures from 0.8 GPa to 3.2 GPa at base temperature of $T \approx 3$ K. DAC's with single-crystal diamond anvils are heavily used for diffraction at the synchrotron, but in order to be used for neutron diffraction the DAC at CORELLI utilizes polycrystalline Versimax diamond anvils[18]. Measurements of MnP have shown that CORELLI is capable of measuring moments as low as $0.25 \, \mu_B / Å^3$ within a DAC[18]. The sample (sample thickness of $\approx 210 \, \mu m$ and sample crossectional area of $0.7 \times 0.7 \, mm^2$) was loaded onto one polycrystalline anvil with the PH15-5 steel gasket (500 $\mu m$ height, 3 mm culet size, 1.3 mm initial gasket hole) in place with deuterated glycerin as a pressure medium, which solidifies at $\approx 5$ GPa at room temperature[19], but remains soft providing close to hydrostatic conditions up to 9 GPa[20]. This was then sealed and pressurized at room temperature with a press. Note that an initial experiment was 4:1 methanol:ethanol as pressure-transmitting medium did not succeed, because the pressure medium evaporated too quickly during sealing. In contrast, glycerin does not readily evaporate on air which ensures that the pressure-transmitting medium is contained. Pressure was assigned to the one obtained from a calibrated pressure-load curve for that specific cell and anvil/gasket set-up at room temperature with an accuracy of 0.5 GPa. After applying pressure the cell was then loaded onto the head of a CCR. We took measurements with the well-focused incident beam passing through the steel gasket from the side. Both gasket and Versimax anvils only yield powder diffraction rings which

can be readily distinguished from the single crystal sample peaks.

*$\mu SR$ measurements under pressure* - Approximately 100 small single crystals of LaCrGe$_3$ (in total $\approx 2$ g) from three batches were placed inside a double-wall piston-cylinder cell. Care was taken to ensure an as high filling factor of the sample space as possible, as well as to ensure a random orientation of the small single crystals. Both, the inner and the outer cylinder of the pressure cell, which was specially designed for the use in $\mu SR$ experiments, are made out of MP35N alloy[21]. The maximum pressure of this cell is $\approx 2.5$ GPa. Daphne 7373 oil was used as a pressure-transmitting medium, which solidifies at room temperature close to 2.5 GPa[8]. The pressure at low temperatures was determined from the shift of the superconducting transition of elemental indium[22], which was also placed in the pressure cell. We estimate the error in this low-temperature pressure to be $\approx 0.05$ GPa. The superconducting transition temperature of indium was determined in AC susceptibility measurements prior to the $\mu SR$ experiments in a separate $^4$He cryostat. In total, $\mu SR$ measurements were performed in zero magnetic field and in several transverse magnetic fields (up to 6000 Oe) at 0.2 GPa as well as at the maximum pressure of 2.5 GPa. The measurements were performed in a $^4$He cryostat with base temperature of 2.2 K at the $\mu E1$ beamline at the Paul-Scherrer-Institute in Villigen, Switzerland, by using the GPD spectrometer. Typically, $5 - 10 \cdot 10^6$ positron events were counted for each data point.

*Pressure media and homogeneity* - For all experiments under finite pressure, the pressure media and cell design was chosen such to ensure hydrostatic or close to hydrostatic pressure conditions in the pressure range of interest[8,15,18,20,23,24].

## SPECIFIC HEAT DATA UNDER PRESSURE

*Specific heat data sets under pressure and procedure to obtain anomalous specific heat contributions* - Figure S1 shows selected data sets of specific heat divided by temperature, $C/T$, which were taken during this study, covering the pressure range 0.46 GPa $\leq p \leq$ 2.38 GPa and temperature range 5 K $\leq T \leq$ 100 K. These data sets were used to extract the anomalous contributions to the specific heat, i.e., the specific heat data corrected for an estimate of the background contribution, resulting from, e.g., phonons. A previous work[1] demonstrated by comparison to the specific heat of the non-magnetic analogue LaVGe$_3$ that the specific heat of LaCrGe$_3$ is dominated by non-magnetic contributions at low temperatures and at temperatures higher than the ferromagnetic transition at ambient pressure. Only close to the ferromagnetic

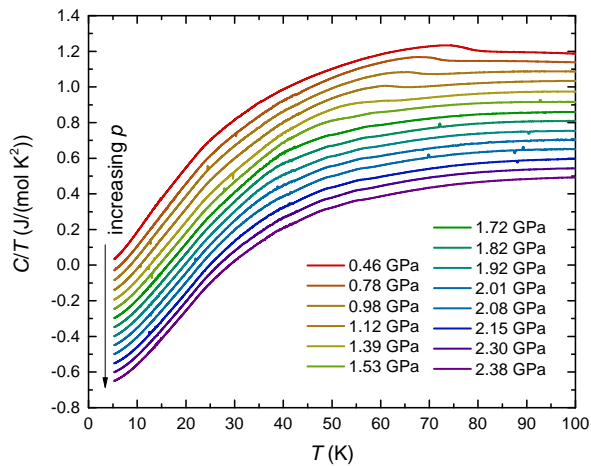

FIG. S1. Specific heat divided by temperature, $C/T$, of LaCrGe$_3$ vs. $T$ over a wide temperature range ($5\,\mathrm{K} \leq T \leq 100\,\mathrm{K}$) for finite pressures ($0.46\,\mathrm{GPa} \leq p \leq 2.38\,\mathrm{GPa}$). Data have been shifted vertically by 0.1 J/(mol K$^2$) for clarity.

phase transition, a substantial magnetic contribution to specific heat was observed. Given that the AC specific heat technique used here does not allow to determine specific heat values to a very high accuracy, we do not refer to specific heat measurements of LaVGe$_3$ for the background subtraction for the LaCrGe$_3$ data under pressure (shown in Fig. S1), and instead follow the procedure, which is illustrated in Fig. S2 for a subset of the data. Following the knowledge of the ambient-pressure study, we approximate the non-magnetic background contribution by fitting the $C/T$ data by a polynomial function of the order of three across a wide temperature range except in the immediate vicinity of any phase transition temperature $T_p$ (i.e., either $T_{\mathrm{FM}}$, $T_1$ or $T_2$). Typically, the range $T_p - 10\,\mathrm{K} \leq T \leq T_p + 5\,\mathrm{K}$ was excluded from the fit, and overall, the fit was typically performed down to $T_p - 20\,\mathrm{K}$ and up to $T_p + 20\,\mathrm{K}$. The so-obtained background curves manifest a shoulder in $C/T$ at $T \approx 80\,\mathrm{K}$, which can also be seen in the ambient-pressure specific heat data on LaVGe$_3$ when replotted as $C/T$ vs. $T^1$. Note that this procedure of background subtraction leads to significant uncertainties in estimating the absolute size of specific heat (and thus, entropy) that is associated with each phase transition. However, the conclusions, which are presented in the main text, are solely based on the analysis of the positions of anomalies in $\Delta C/T$, which should not be affected by the background subtraction procedure.

*Position of anomalies in specific heat data and criteria to determine transition temperatures* - The so-obtained anomalous specific heat contributions, $\Delta C/T$, as a function of temperature, $T$, are shown in Fig. S3 (a), together with the temperature-derivative of the same data in (b),

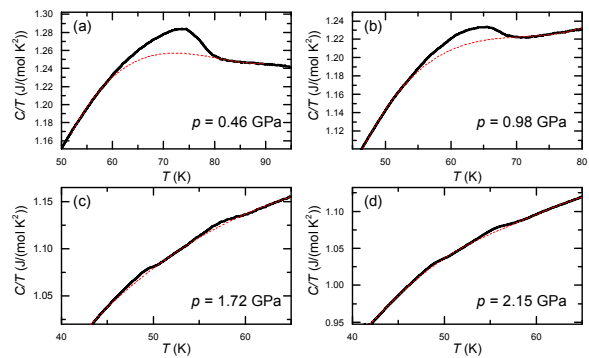

FIG. S2. Illustration of the procedure to obtain the anomalous specific heat contributions, that are associated with various phase transitions in LaCrGe$_3$, for selected pressures (a-d). The background (red dashed line) was obtained by fitting the specific heat data (black line) well below and well above the phase transitions simultaneously with a polynomial of the order three (for details, see text).

for $0\,\mathrm{GPa} \leq p \leq 2.38\,\mathrm{GPa}$. The second-order ferromagnetic (FM) transition at $T_{\mathrm{FM}}$ (indicated by the blue arrow) manifests itself in an almost mean-field-like jump in the specific heat, the size of which becomes progressively reduced with increasing pressure (Note that a discussion of the specific heat signature of the first-order FM transition for $p \geq 1.53\,\mathrm{GPa}$ will be given below). At $p = 1.39\,\mathrm{GPa}$, a second, more subtle anomaly occurs on the high-temperature side of the ferromagnetic specific heat anomaly for the first time. This result suggests the presence of a new phase transition, which was denoted by $T_1$ in the main text. Upon increasing the pressure slightly to $1.53\,\mathrm{GPa}$, these two specific heat anomalies at $T_{\mathrm{FM}}$ and $T_1$, respectively, become more separated in temperature and thus clearly distinguishable. For even higher pressures, the anomaly, which we associate with $T_{\mathrm{FM}}$, continues to drop (see below), but we also observe two specific heat anomalies, which are separated by only $\approx 10\,\mathrm{K}$ in temperature and almost similar in size. The positions of both of these anomalies are almost unchanged in temperature upon increasing pressure (compared to the strong suppression of the $T_{\mathrm{FM}}$-line with pressure). We thus assign the lower-temperature specific heat anomaly for $p \geq 1.72\,\mathrm{GPa}$ to another phase transition at $T_2$, which is distinct from the ferromagnetic transition. Note that the $T_2$-line has previously been reported in literature[3], based on electrical transport measurements, and was assigned to a new magnetic phase transition of likely modulated AFM$_q$. In the main text and also here in the SI, we present neutron and new $\mu$SR data for high pressures, which strongly suggest a new interpretation of the magnetic state of the phase below $T_2$.

To determine the transition temperatures $T_{\mathrm{FM}}$, $T_1$ and $T_2$ from the presented specific heat data, the position of

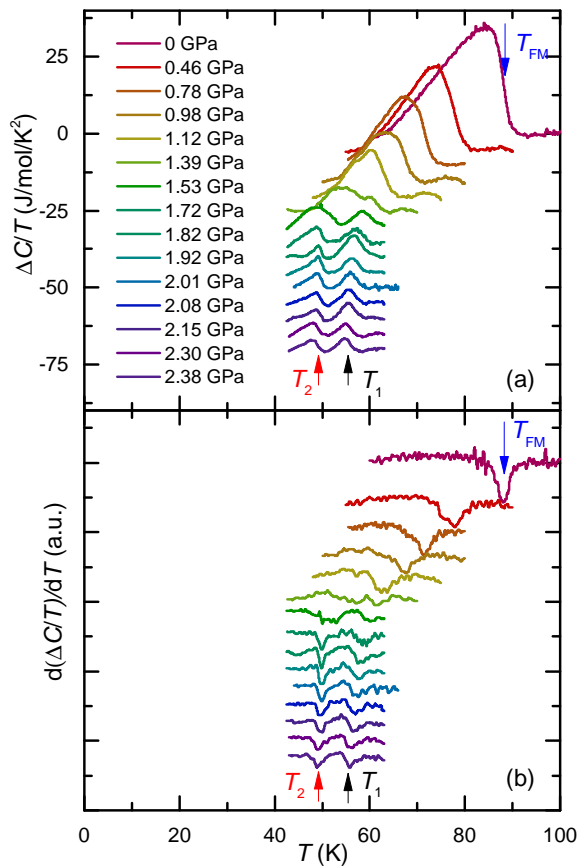

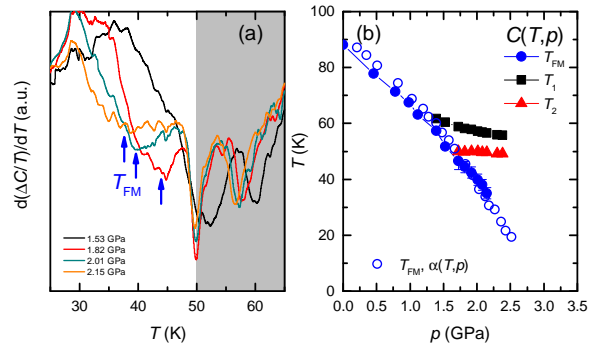

FIG. S4. (a) Derivative of the anomalous specific heat contribution, $d(\Delta C/T)/dT$, vs. $T$ for LaCrGe$_3$ for selected pressures $1.53\,\text{GPa} \leq p \leq 2.15\,\text{GPa}$. ($\Delta C/T$) was obtained by subtracting the specific heat data set at 0.98 GPa, which serves as a proxy for the background contribution for $T \lesssim 50\,\text{K}$ from the measured data. The two large peaks for $T \gtrsim 50\,\text{K}$ (grey area) were identified as the anomalies at $T_1$ and $T_2$ in the main text. In addition, tiny and very broad anomalies, the positions of which are indicated by the arrows, can be identified at lower temperatures and are likely associated with the first-order ferromagnetic transition at $T_{\text{FM}}$ for $p \gtrsim 1.5\,\text{GPa}$; (b) Temperature-phase diagram of LaCrGe$_3$, as constructed from specific heat measurements under pressure. Blue solid circles, black solid squares and red solid triangles correspond to $T_{\text{FM}}$, $T_1$ and $T_2$, respectively. Open blue symbols indicate the position of the clear anomalies at $T_{\text{FM}}$ in the thermal expansion coefficient data (shown below).

FIG. S3. Anomalous contribution to the specific heat, $\Delta C/T$, (a) and temperature-derivative of this data, $d(\Delta C/T)/dT$, (b) vs. temperature, $T$, for LaCrGe$_3$ for finite pressures in the range $0\,\text{GPa} \leq p \leq 2.38\,\text{GPa}$. Blue, black and red arrows in each panel exemplarily indicate the position of anomalies at $T_{\text{FM}}$, $T_1$ and $T_2$, respectively. In all panels, data were shifted vertically for clarity.

the minimum in $d(\Delta C/T)/dT$ was chosen [see position of the arrows, which are exemplarily shown for $p = 0\,\text{GPa}$ and $2.38\,\text{GPa}$ in Fig. S3 (b)]. The application of this criterion yields transition temperatures that are very close to that obtained by iso-entropic construction.

*Signature of ferromagnetic transition in specific heat measurements at and beyond 1.53 GPa* - In the discussion above as well as in the main text, we do not show any specific heat signatures of the ferromagnetic (FM) transition for $p \geq 1.53\,\text{GPa}$. The main reason for this is that we were not able to resolve a clear and sharp feature in the specific heat for these pressures (in contrast to the clear, huge and sharp features in the thermal expansion coefficient, shown in the main text and below). We believe that this is due to a sizable amount of entropy being released upon cooling through $T_1$ and $T_2$, which likely results in only a small amount of entropy being released

on the subsequent cooling through $T_{\text{FM}}$. In turn, then we expect that the specific heat feature becomes very small, likely below the limit below which we can separate it from the background. In addition, the transition changes its character close to 1.5 GPa (see discussion below), and the absolute value of $dT_{\text{FM}}/dp$ gets larger upon increasing $p$. It is therefore reasonable to expect that any specific heat feature above 1.5 GPa might be different in shape and broadened in temperature, making it hard to separate the feature from the (unknown) background contribution. Nonetheless, even if no clear feature can be observed in $C/T$ (or $\Delta C/T$), a potential feature might show up more clearly in the temperature-derivative of these data. For this reason, we show in Fig. S4 (a), the temperature-derivative of $\Delta C/T$ data across a wider temperature range. For this plot, $\Delta C/T$ was obtained by $\Delta C/T = (C(T, p) - C(T, p = 0.98\,\text{GPa}))/T$, with $C(T, p)$ being the temperature-dependent specific heat data at the pressure of interest, and $C(T, p = 0.98\,\text{GPa})$ the temperature-dependent specific heat data at 0.98 GPa. This analysis is needed, since a simple polynomial fit (with order three) is not sufficient to describe the background over a very wide temperature range. In our approach, we assume that the $C(T, p = 0.98\,\text{GPa})$, for which only a FM transition for high temperatures $T > 60\,\text{K}$ occurs, can be used as a good proxy for the background contribution for $T \lesssim 50\,\text{K}$, where we expect

the FM transition to occur for $p \gtrsim 1.53\,\text{GPa}$. The justificiation for this ansatz is based on the comparison of specific heat data on LaCrGe$_3$ and the non-magnetic analogue LaVGe$_3$[1], which showed that the specific heat at temperatures well below the FM transition, is dominated by non-magnetic contributions. Indeed, as a result of our analysis, very subtle and progressively broader minima can be observed in the so-obtained $\mathrm{d}(\Delta C/T)/\mathrm{d}T$ data, the position of which (see arrows) coincide well [see Fig. S4 (b)] with the positions of the sharp and clear anomalies in the thermal expansion coefficient (see main text and Fig. S6). Thus, we assume that this subtle feature in the specific heat is indeed related to the ferromagnetic ordering, in accordance with our hypothesis of smaller entropy associated with the ordering and/or additional broadening of the feature. Nonetheless, from the specific heat data alone, it would not be possible to infer the $T_{\text{FM}}$-line for $p \geq 1.53\,\text{GPa}$ reliably.

## LATTICE PARAMETERS UNDER PRESSURE

*Definition of physical quantities* - Since we discuss and compare measurements from various techniques, which all give insight into the change of the crystalline lattice with pressure and temperature, we first want to define some of the measurement quantities here and elaborate which of the different experimental techniques yields insight into which quantity.

The temperature-dependent relative length change (or alternatively, thermal expansion) of a macroscopic crystal along a crystallographic axis $i$, $(\Delta L/L)_i$ (with $i = ab, c$ for a hexagonal system, such as LaCrGe$_3$) is defined as

$$
\begin{aligned}
&(\Delta L/L)_i(T, p = \text{const.}) \\
&= \frac{L_i(T, p = \text{const.}) - L_i(T_{\text{ref}}, p = \text{const.})}{L_i(T_{\text{ref}}, p = \text{const.})},
\end{aligned} \tag{S1}
$$

with $L_i(T, p = \text{const.})$ being the absolute length of a crystal in $i$-direction at any given temperature, $T$, and $T_{\text{ref}}$, being any reference temperature. The thermal expansion coefficient along a crystallographic axis $i$, $\alpha_i$, is then defined as

$$
\alpha_i = \frac{1}{L_i(T, p = \text{const.})} \frac{\partial L_i(T, p = \text{const.})}{\partial T}, \tag{S2}
$$

and is experimentally often determined to a very good approximation (since $\Delta L_i \ll L_i$) by

$$
\alpha_i = \frac{1}{L_i(300\,\text{K}, p = \text{const.})} \frac{\mathrm{d}\Delta L_i(T, p = \text{const.})}{\mathrm{d}T}, \tag{S3}
$$

with $\Delta L_i = L_i(T, p = \text{const.}) - L_i(T_{\text{ref}}, p = \text{const.})$, and $L_i(300\,\text{K}, p = \text{const.})$ being the length of the crystal at room temperature, which can be determined in an independent measurement. Note that due to the freedom in the choice of $T_{\text{ref}}$, $\Delta L_i$ can only be determined up to a constant. Since $\alpha_i$, as defined above in Eq. S3, is directly proportional to the temperature-derivative of $\Delta L_i$, the size of $\alpha_i$ is independent of the choice of $T_{\text{ref}}$. Note that in Eq. S3, we set the normalization length in the denominator to $L_i(300\,\text{K}, p = \text{const.})$, since $\Delta L_i \ll L_i$.

Experimentally, the relative length change, $(\Delta L/L)_i$, and the thermal expansion coefficient, $\alpha_i$, can be determined from, e.g. capacitive dilatometry at ambient pressure or the strain-gauge technique for finite pressures. From neutron and x-ray diffraction measurements at ambient and finite pressures, the crystallographic lattice parameters $a = b$ and $c$ (for a hexagonal crystal system, such as LaCrGe$_3$) can be inferred at any measured temperature and pressure. Each of these measurement quantities are related by simple equations, and we will do this explicit comparison for our data collection of LaCrGe$_3$ under pressure at the end of this section.

*Functionality of the strain-gauge technique for the determination of the thermal expansion and the thermal expansion coefficient* - Prior to the discussion of our various data sets, taken under finite pressures, we first want to demonstrate the functionality of our strain-gauge-based setup by comparing the relative length change, $(\Delta L/L)_i$, and the thermal expansion coefficient, $\alpha_i$, obtained via the strain-gauge technique (at a relatively low applied pressure) to the data obtained by the technique of capacitive dilatometry at ambient pressure (see Fig. S5). Capacitive dilatometry is a well-established technique for the determination of the thermal expansion of solids and known for its extremely high sensitivity[25,26]. The capacitive dilatometry data, presented in Fig. S5, were obtained by using a dilatometer, which was described earlier in Ref. [27], in a Quantum Design PPMS, which provided the low-temperature environment.

Figures S5 (a) and (b) show the temperature ($T$) dependence of $(\Delta L/L)_i$ and $\alpha_i$ for $i = ab, c$ at ambient pressure, as obtained from using the technique of capacitive dilatometry. (We use the notion of $ab$, since the $a$ and $b$ direction are equivalent in a hexagonal crystal system.) Upon cooling from 150 K, the crystal shrinks along both crystallographic inequivalent directions, as can be seen from a reduction of $(\Delta L/L)_i$, corresponding to positive values of $\alpha_i$ (with small directional anisotropy). Below 90 K, an anomalous behavior of $(\Delta L/L)_i$ and $\alpha_i$ can be observed, which is a result of the well-known ferromagnetic ordering at $T_{\text{FM}} \simeq 89\,\text{K}$[1]. In more detail, upon cooling through this ferromagnetic transition, the length along the $ab$ axes shrinks rapidly, whereas the

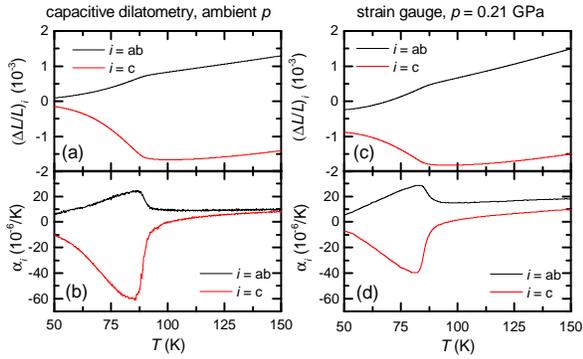

FIG. S5. Comparison of thermal expansion data on LaCrGe$_3$, obtained by capacitive dilatometry (a,b) and a strain-gauge-based method (c,d); (a,b) Relative length change, $(\Delta L/L)_i$ (a), and thermal expansion coefficient, $\alpha_i$ (b), vs. temperature, $T$, along the crystallographic $ab$ and $c$ direction, obtained by utilizing a capacitive dilatometer at ambient pressure; (c,d) Relative length change, $(\Delta L/L)_i$ (c), and thermal expansion coefficient, $\alpha_i$ (d), vs. temperature, $T$, along the crystallographic $ab$ and $c$ direction, obtained by utilizing a strain-gauge-based method at $p = 0.21$ GPa inside the pressure cell. Due to the freedom of choice in $T_{\text{ref}}$, which causes that the relative length change can be only determined up to a constant, the $(\Delta L/L)_i$ values at 150 K were matched to the 150 K values from the capacitive dilatometry data.

length along the $c$ axis shows a very pronounced increase. This response of the crystal lattice to the ferromagnetic order is consistent with a picture of magnetoelastic effects resulting from dipolar coupling[28] of ferromagnetically aligned spins with moments aligned along the crystallographic $c$ direction. The described relative length changes yield a positive anomaly in $\Delta\alpha_{ab}(T)$ and a negative anomaly in $\Delta\alpha_c(T)$, with $|\Delta\alpha_c(T)| \simeq 3|\Delta\alpha_{ab}(T)|$ ($\Delta\alpha_i(T)$ corresponds to the anomalous contribution to the thermal expansion coefficients after subtraction of non-magnetic background contributions, not shown in Fig. S6). Given that the temperature dependence of $\alpha_i(T)$ is closely related to the temperature dependence of the specific heat, $C(T)$, via the (uniaxial) Grüneisen parameter, it can be expected that the anomalous contributions, $\Delta\alpha_i(T)$, are similar in shape to anomalous contributions to the specific heat, $\Delta C(T)$. Indeed, similar to the specific heat measurements, the thermal expansion coefficients, $\alpha_i(T)$, display almost mean-field like changes at the phase transition temperature $T_{\text{FM}}$. For reasons of consistency with the specific heat data and the chosen criteria, the positions of the extrema in $d\alpha_i/dT$ (i.e., the minimum in $d\alpha_{ab}/dT$ and the maximum in $d\alpha_c/dT$) were chosen to determine $T_{\text{FM}} = 89.5$ K at ambient pressure.

For the comparison, Figs. S5 (c) and (d) show the temperature dependence of $(\Delta L/L)_i$ and $\alpha_i$ for $i = ab, c$, obtained from the strain-gauge technique, as described in the section on experimental methods. The presented data were taken inside the pressure cell, which was closed

hand-tight prior to the experiment without applying load to the piston. This procedure caused that the lowest-pressure measurements were actually performed already at a finite pressure of 0.21 GPa, as determined from the low-temperature Pb manometer. Whereas this small pressure leads to a small, but measurable shift of the transition temperature, it does not compromise our comparison, since LaCrGe$_3$ still undergoes a ferromagnetic transition with very similar responses of the crystalline lattice (as demonstrated by our x-ray and neutron diffraction data under pressure, see below). Again, upon cooling from high temperature, we find a decrease of the length along both inequivalent directions (i.e., positive $\alpha_i$ values). Consistent with our dilatometry data, we find a strong decrease (large increase) of the length along the $ab$ axes ($c$ axis). The anomalies in $\alpha_i(T)$ with $i = ab, c$ are also almost mean-field like. Applying the same criterion for the determination of $T_{\text{FM}}$ as above yields $T_{\text{FM}}(0.21$ GPa$) = 86$ K (see Fig. S6 for the temperature-derivatives of these data).

This suppression of $T_{\text{FM}}$ with modest pressures is fully consistent with our analysis of the phase diagram from specific heat measurements. In terms of the absolute $\alpha_i$ [and $(\Delta L/L)_i$], we find that the maximum value of $\alpha_{ab}$, determined from the strain-gauge technique, is similar to the one of the capacitive dilatometry data, whereas the value of $\alpha_c$ is smaller by about 1/3. Reasons for this discrepancy can be manifold. First, the strain gauges are rigidly glued to the samples. However, the glue will not transmit the strain perfectly, thus naturally leading to the observations of slightly smaller length changes in the strain-gauge measurements. If this was the case, then the fact, that the $\alpha_{ab}$ values match better, suggests that the strain transmission of the glue for the strain gauge of the $ab$ axes was higher. Second, another option is related to the expansion of the tungsten-carbide samples, which we use for the subtraction of the intrinsic strain-gauge response. Strictly speaking, in our strain-gauge technique, we measure the length change of our sample relative to the one of the tungsten carbide pieces. However, tungsten carbide is known for its small expansivity[11], and thus, this scenario is highly unlikely.

*Anomalies in the thermal expansion coefficient under pressure and criteria for inferring phase transition temperatures* - Figures S6 (a) and (b) show the anomalous contributions to the thermal expansion coefficients, $\Delta\alpha_i$ with $i = ab, c$, of LaCrGe$_3$ for finite pressures up to 2.43 GPa. These $\Delta\alpha_i$ data were obtained by subtracting a background contribution, which was obtained by fitting a data set at 2.60 GPa, for which the ferromagnetic transition $T_{\text{FM}}$ is suppressed to $T < 10$ K. We find that the above-described pronounced thermal expansion anomalies at $T_{\text{FM}}$, i.e., the positive anomaly in $\alpha_{ab}$ and

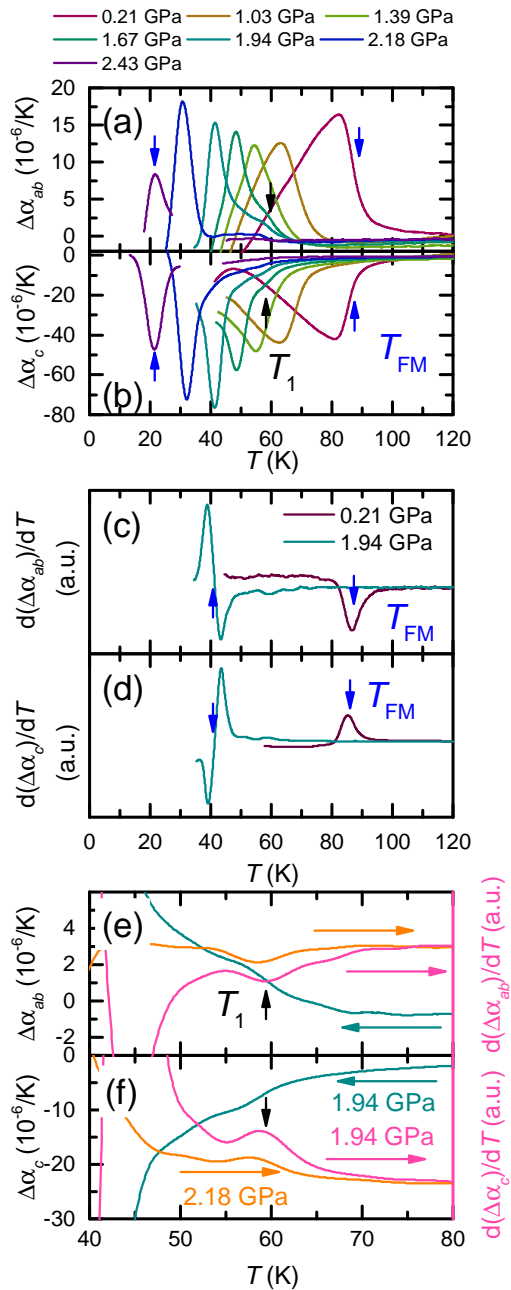

FIG. S6. Analysis of the thermal expansion anomalies; (a,b) Anomalous contribution to the thermal expansion coefficient along the $ab$ axes, $\Delta\alpha_{ab}$ (a), and along the $c$ axis, $\Delta\alpha_c$ (b), vs. temperature, $T$, of LaCrGe$_3$ for pressures $0.21\,\text{GPa} \leq p \leq 2.43\,\text{GPa}$. Blue (black) arrows indicate the criteria for the determination of $T_{\text{FM}}$ ($T_1$). Note that the change in criterion for $T_{\text{FM}}$ is related to the change of the character of the transition from second-order to first-order at $p_{\text{tr}} \simeq 1.5\,\text{GPa}$ (see text for details). Data between 30 K and 40 K are omitted due to an anomaly in the strain-gauge response that is not intrinsic to LaCrGe$_3$; (c,d) Derivative of the anomalous thermal expansion coefficients along the $ab$ axes, $\text{d}(\Delta\alpha_{ab})/\text{d}T$, (c) and along the $c$ axis, $\text{d}(\Delta\alpha_c)/\text{d}T$ (d) for $p = 0.21\,\text{GPa}$ and $1.94\,\text{GPa}$. The blue arrows indicate the criteria to determine $T_{\text{FM}}$ from these data sets; (e,f) Enlarged view on $\Delta\alpha_i$ (left axis) and $\text{d}\Delta\alpha_i/\text{d}T$ (right axis) for $i = ab$ (e) and $i = c$ (f) at $p = 1.94\,\text{GPa}$ around the phase transition temperature $T_1$. The criterion, which was chosen to determine $T_1$, is indicated by the black arrows. In each panel, $\text{d}(\Delta\alpha_i)/\text{d}T$ at $p = 2.18\,\text{GPa}$ is included to demonstrate that, if present, any feature at $T_2 \simeq 49\,\text{K}$ is distinctly smaller than the one at $T_1$.

the negative anomaly in $\alpha_c$, are shifted to lower temperatures with increasing $p$. Importantly, in contrast to the signature of the ferromagnetic transition in specific heat measurements, the feature in the thermal expansion remains clear and measurable over the full, investigated pressure range, thus allowing us to reliably determine the $T_{\text{FM}}$-line across wide ranges of the phase diagram. At the same time, we find that the shape of the expansion anomalies along both directions changes its shape upon increasing pressure. Specifically, the almost mean-field-type $\Delta\alpha_i$, with $i = ab, c$, anomalies for low pressures change into symmetric and sharp peaks for higher pressures. These changes of the anomaly shape strongly suggest a change of the character of the phase transition from second-order to first-order upon increasing pressure, consistent with previous reports[3] as well as the generic avoidance of ferromagnetic criticality in itinerant ferromagnets. The detailed determination of the associated tricritical point at $(p_{\text{tr}}, T_{\text{tr}})$ from an analysis of the asymmetry and the width of the thermal expansion coefficient feature will be discussed below. Here, we would now like to discuss the implications for the choice of criterion to determine $T_{\text{FM}}$ from the present thermal expansion coefficient data. For low pressures, the mean-field-type anomaly gives rise to a pronounced minimum in $\text{d}(\Delta\alpha_{ab})/\text{d}T$ (maximum in $\text{d}(\Delta\alpha_c)/\text{d}T$, as exemplarily shown in Figs. S6 (c) and (d) for $p = 0.21\,\text{GPa}$. We chose the positions of these extrema to determine $T_{\text{FM}}$ for low pressures. In contrast, the sharp anomaly in the thermal expansion coefficient for high pressures gives rise to an anomaly in $\text{d}(\Delta\alpha_i)/\text{d}T$ ($i = ab, c$) with pronounced over- and undershoots on the low- and high-temperature side, see, e.g., the $p = 1.94\,\text{GPa}$ data sets in Figs. S6 (c) and (d). Correspondingly, we chose the mid-point in $\text{d}(\Delta\alpha_i)/\text{d}T$ between the maximum and minimum values of the anomaly to determine $T_{\text{FM}}$ for high pressures (see blue arrows).

In addition to the FM anomaly, a closer look on the $\Delta\alpha_i(T)$ ($i = ab, c$) data reveal a smaller, but nonetheless clear anomaly at $T_1$. To show this, we present in Figs. S6 (e) and (f) the $\Delta\alpha_i(T)$ (left axis) as well as the $\text{d}(\Delta\alpha_i)/\text{d}T$ (right axis) for $p = 1.94\,\text{GPa}$ on enlarged scales around $T_1$. The anomalies in $\Delta\alpha_i(T)$ can be seen with bare eyes, but become very obvious in $\text{d}(\Delta\alpha_i)/\text{d}T$, where we observe a minimum in $\text{d}(\Delta\alpha_{ab})/\text{d}T$ and a maximum in $\text{d}(\Delta\alpha_c)/\text{d}T$. As already pointed out in the main text, this result implies that the lattice responds in the same way to the phase transition at $T_1$ as to the ferromagnetic order, albeit smaller in size, i.e., the crystal shrinks in the $ab$ plane and expands along the $c$ axis upon cooling. The positions of the extrema in $\text{d}(\Delta\alpha_i)/\text{d}T$ were used to infer $T_1$.

However, the phase transition at $T_2$, which gives rise to a clear specific heat feature, does not result in a pronounced feature close to 50 K in the thermal expansion coefficient. To demonstrate this, we also added the

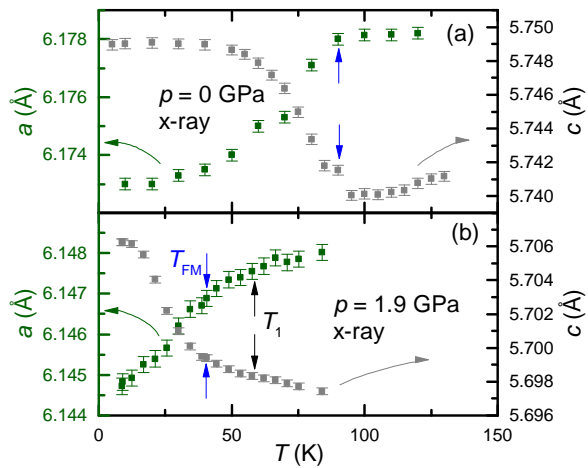

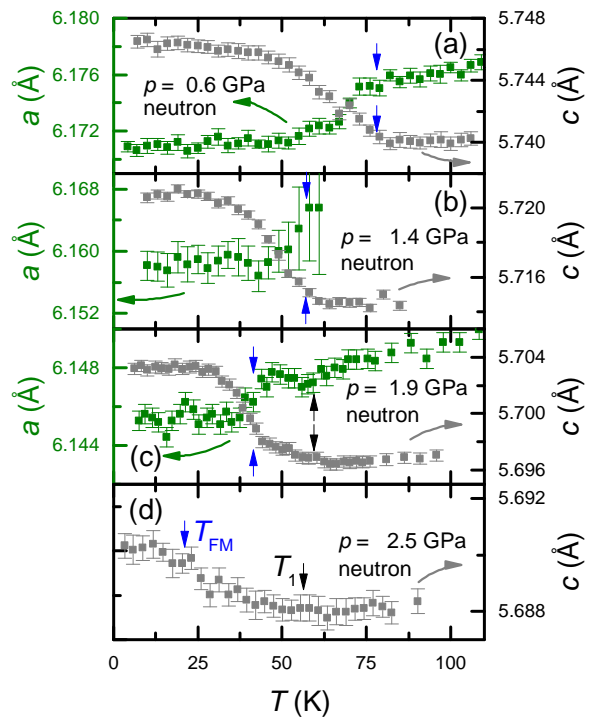

FIG. S7. Lattice parameters $a$ (left axis) and $c$ (right axis) at ambient pressure (a) and at $p = 1.9$ GPa (b), as determined from x-ray diffraction experiments. The positions of the arrows correspond to the transition temperatures,determined from our thermodynamic measurements and approximately coincide with the points where the behavior of the lattice parameters deviates from the extrapolated high-temperature behavior.

data sets of $\Delta\alpha_i(T)$ ($i = ab, c$) for $p = 2.18$ GPa in Figs. S6 (e) and (f), since the ferromagnetic transition is suppressed well below 50 K for this pressure. No clear feature is discernible in either the $ab$ axes or the $c$ axis data. Thus, we can conclude that the anomalous lattice effects are distinctly larger at $T_1$ than at $T_2$, however, both are distinctly smaller than the one induced by long-range FM ordering.

*Lattice parameters under pressure from x-ray and neutron diffraction* - In Fig. S7, we show the temperature dependence of the crystallographic lattice parameters, $a$ and $c$, determined from x-ray diffraction measurements on single crystals, at ambient pressure (a) and at $p = 1.9$ GPa (b). The $c$ lattice parameters were determined by measuring the position of the $(0\,0\,16)$ Bragg peaks with the Pixirad-1 detector at ambient pressure, and the $(0\,0\,4)$ Bragg peaks with the MAR345 detector while under applied pressure in the DAC. The $a$ lattice parameters were determined the same way using the position of the $(16\,0\,0)$ and $(2\,2\,0)$ Bragg peaks for ambient pressure and with applied pressure, respectively. At ambient pressure, the change of the lattice parameters is very consistent with the behavior found in measurements of the thermal expansion via capacitive dilatometry or the strain-gauge technique, which were discussed earlier in this section, i.e., the in-plane $a$ lattice parameter shrinks and the $c$ parameter increases upon cooling through the ferromagnetic (FM) transition $T_{FM} \simeq 90$ K. The positions of the arrows correspond to

the transition temperatures determined from our thermodynamic measurements and approximately coincide with the points where the behavior of the lattice parameters deviates from the extrapolated high-temperature behavior. The lattice parameters show a typical temperature dependence, consistent with the second-order nature of the phase transition. At 1.9 GPa, the FM transition is suppressed to much lower temperatures ($T_{FM} \simeq 40$ K) and still results in a strong increase of the $c$ lattice parameter upon cooling, whereas the $a$ lattice parameter shows a discernible decrease. In addition, a very subtle, kink-like feature might be discernible at much higher temperatures at $T \approx 60$ K at 1.9 GPa in the $a$ and $c$ lattice parameters, respectively (see black arrow). If indeed present, this feature coincides with $T_1$. A clearer feature around $T_1$ on the basis of neutron diffraction data will be presented below.

Figure S8 shows the temperature dependence of the $a$

FIG. S8. Lattice parameters $a$ (left axis) and $c$ (right axis) at $p = 0.6$ GPa (a), at $p = 1.4$GPa (b), at $p = 1.9$GPa (c), and at $p = 2.5$GPa (d), as determined from neutron diffraction experiments [N1 (a-c), N2 (d)]. The positions of the arrows correspond to the transition temperatures determined from our thermodynamic measurements and approximately coincide with the points where the behavior of the lattice parameters deviates from the extrapolated high-temperature behavior.

and $c$ lattice parameters, inferred from neutron diffraction measurements on single crystals at HB1, for various pressures. Again, similar to the discussed other low-pressure data sets, the $a$ ($c$) lattice parameters at $p = 0.6$ GPa show a decrease (an increase) upon cooling through $T_{FM}(p)$, which follow an order-parameter type of behavior, and thus, are consistent with the notion of a second-order phase transition. The $c$ lattice parameter at 1.4 GPa also shows an order-parameter type decrease at $T_{FM}(p)$. At 1.9 GPa, the FM transition is shifted to even lower temperatures in our neutron and x-ray diffraction data. In addition, the $c$ lattice parameter shows a kink at much higher temperatures $T \approx 60$ K, which is thus likely associated with the phase transition at $T_1$ (This aspect becomes much clearer from a direct comparison of the lattice parameter data and the thermal expansion data, obtained by the strain-gauge data, which will be presented in the upcoming section). This kink-like feature in the $c$ lattice parameter is accompanied by a very subtle change of the slope of the $a$ lattice parameter. Increasing pressure even further to 2.5 GPa, we still find a kink-like feature in the $c$ lattice parameter close to $T_1$ (as identified in our thermodynamic and transport data), and upon further cooling $c$ shows a increase in the low-temperature region. This increase is likely associated with the (broadened) first-order FM transition, as it gets suppressed closer to zero. Again, we will provide further evidence for the underlying phase transition in the next section, when we compare the different lattice parameter and length change data sets.

*Comparison of lattice parameter data from x-ray and neutron diffraction measurements under pressure to relative length change data, obtained from the strain-gauge-based technique under pressure* - After the presentation of the measured data of relative length change and the lattice parameters under pressure, we want to turn to the explicit comparison of the various data sets, taken at very similar pressures. The result of this comparison is shown in Fig. S9. [Note that for each panel the axes are scaled such that they correspond to the same relative length changes, and thus, the overlap of different data sets demonstrate the agreement of the data even on a quantitative level.] For the majority of the data sets, in particular for all taken at finite pressures $p \geq 0.6$ GPa, we find a good agreement of the lattice parameter data from neutron and x-ray diffraction experiments with the relative length change, determined from the strain-gauge technique. Only for the data at/close to ambient pressure, we find small discrepancies in the absolute values of the relative length change, as well as minor differences in the position of the FM anomaly. The latter can be assigned to the difference in pressure, at which the two data sets were taken. The reasons for the differences in absolute values only for this pressure are largely unknown and

potential reasons were discussed in detail above, when we compared the low-pressure strain gauge data to the data from capacitive dilatometry at ambient pressure.

Nonetheless, the good agreement of the diffraction and relative length change data provides strong support for statements made above. In particular, the broadened increase of the $c$ lattice parameter at very low temperatures indeed coincides with the increase of $(\Delta L/L)_c$ data, which can clearly be associated with a feature of the FM ordering. Thus, the $c$ axis increase at low temperatures is also likely a result of the ferromagnetic order, which is suppressed to very low temperatures.

## ISOSTRUCTURAL NATURE OF ALL SALIENT PHASE TRANSITIONS

In this section, we show additional x-ray diffraction data, which demonstrate that all salient phase transitions, which were observed in our thermodynamic and diffraction data, only result in a change of the crystallographic lattice parameters, but are not accompanied by any symmetry changes of the crystallographic structure. This allows us to exclude any type of charge-density wave or structural phase transition for the high-pressure phases in LaCrGe$_3$.

*High-energy x-ray diffraction of single-crystals* - High-energy x-ray diffraction data were taken on single crystals of LaCrGe$_3$ to search for any structural anomalies. These data are shown as two-dimensional images of the $(H H L)$ plane in Fig. S11, and as longitudinal cuts through the Bragg peak $(2\,2\,0)$ in Fig. S12 for $p = 1.5$ GPa, 1.9 GPa, and 4 GPa for several temperatures. We note that the previous study of LaCrGe$_3$[3,4] suggested that at $p = 4$ GPa and base temperature, LaCrGe$_3$ is in the new magnetic phase (see Fig. S10). Therefore we expect to probe the properties of the new phase at 4 GPa, well separated from the low-$p$ ferromagnetism. From the data in Fig. S11 we see no indication of additional Bragg peaks from LaCrGe$_3$ within a dynamical range of $10^5$, which indicates that there is no superstructure and no charge-density wave at any pressure. In addition, we show in Fig. S12 that we did not observe splitting or broadening of the Bragg peaks which would indicate a symmetry-lowering lattice distortion.

At ambient pressure, LaCrGe$_3$ was reported to adopt a hexagonal perovskite structure with space group $P6_3/mmc$[29,30]. The single-crystal x-ray data indicates that the crystal structure remains hexagonal or trigonal and with a $c$-glide plane parallel to the uniaxial $c$-direction through all salient phase transitions at all pressures. Given these constraints and the parent space group of LaCrGe$_3$, $P6_3/mmc$[29,30], several structural phases can occur, (i) $P6_3/mmc$, (iia) $P\bar{6}2c$ with a

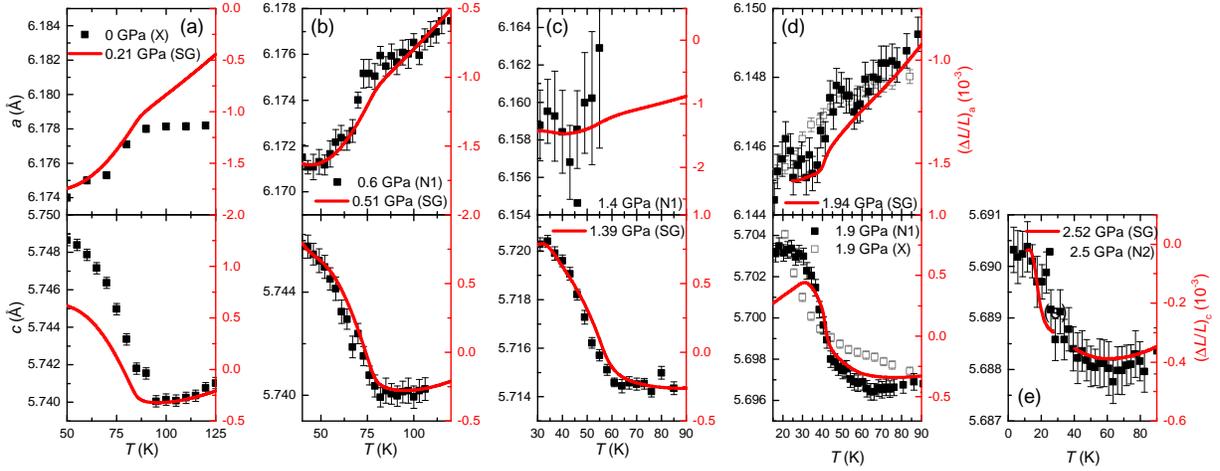

FIG. S9. Comparison of $a$ and $c$ lattice parameters of LaCrGe$_3$, determined from x-ray and neutron diffraction experiments, with the relative length change $(\Delta L/L)_a$ and $(\Delta L/L)_c$ of LaCrGe$_3$, determined via the strain-gauge technique, as a function of temperature, $T$, for ambient and finite pressures. In each panel, the top shows the comparison of the $a$ lattice parameter (left axis, symbols) with the relative length change $(\Delta L/L)_a$ (right axis, solid line), and the bottom shows the comparison of the $c$ lattice parameter (left axis, symbols) with the relative length change $(\Delta L/L)_c$ (right axis, solid line). Note that for each panel the left and right axes are scaled such that they correspond to the same relative length changes. X-ray data (abbreviated as X) were taken at ambient pressure (a) and 1.9 GPa (d), neutron data (in cells N1 and N2) at pressures $p = 0.6$ GPa (b), 1.4 GPa (c), 1.9 GPa (d) and 2.5 GPa (e). Relative length change data (abbreviated as SG) were taken at $p = 0.21$ GPa (a), 0.51 GPa (b), 1.39 GPa (c), 1.94 GPa (d) and 2.52 GPa (e). The larger error bar in (c) of the lattice parameter data is related to shorter counting time.

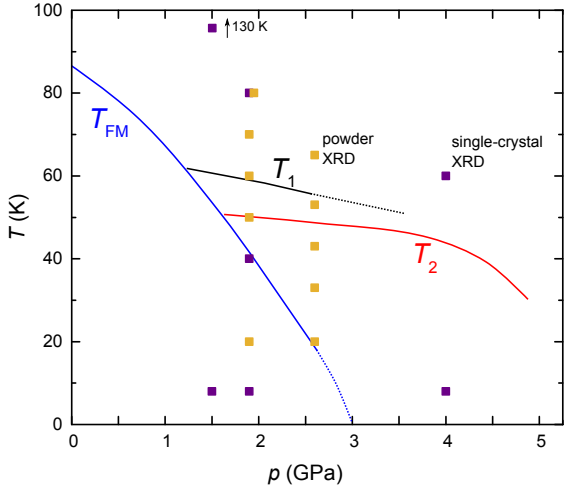

FIG. S10. Temperature-pressure phase diagram of LaCrGe$_3$, as obtained from the present study, showing the phase transition lines at $T_{FM}$, $T_1$ and $T_2$. High-pressure data for $T_1$ is taken from previous works[3,4]. Purple (dark yellow) symbols indicate the temperature/pressure combinations, for which single-crystal (powder) x-ray diffraction data were taken with high statistics and which are shown in Figs. S11-S13. Dotted lines correspond to extrapolation and not actual data.

rotation in the $ab$-plane of the triangularly arranged Ge atoms as a new degree of freedom, (iib) $P6_3mc$ with a shift of the La planes in the $c$-direction relative to the Cr and Ge planes, and (iii) $P31c$ which is a combina-

tion of both. Phase transitions from the parent space group, $P6_3/mmc$, to $P\bar{6}2c$, $P6_3mc$ or $P31c$ would leave the lattice symmetry and the reflection conditions unchanged, but could be differentiated from analysis of the Bragg peak intensities. To investigate whether there is a change of Bragg peak intensities, we performed an x-ray diffraction study on powder samples under pressure, the results of which will be discussed in the following.

*Powder x-ray diffraction in a DAC* - To measure the largest number of Bragg peak intensities possible at once within a DAC we recorded x-ray diffraction data on powder samples of LaCrGe$_3$ down to 20 K. To minimize the effect of the small sample size and create a better polycrystalline average, individual scans at each temperature were measured on a MAR345 with the DAC at different angles along an axis perpendicular to the incident beam while rocking the sample along another axis perpendicular to the incident beam. Each scan was azimuthally integrated and combined at each temperature and pressure measured. To reduce the impact of comparatively large single-crystal grains in the sample, very strong individual Bragg peaks were excluded from the azimuthal integration. The powder x-ray diffraction data processed this way is shown in Fig. S13, and shows virtually no change in the Bragg peak intensity as a function of temperature. The single-crystal and powder x-ray diffraction data strongly indicate that all the phase transitions for LaCrGe$_3$ are isostructural in nature.

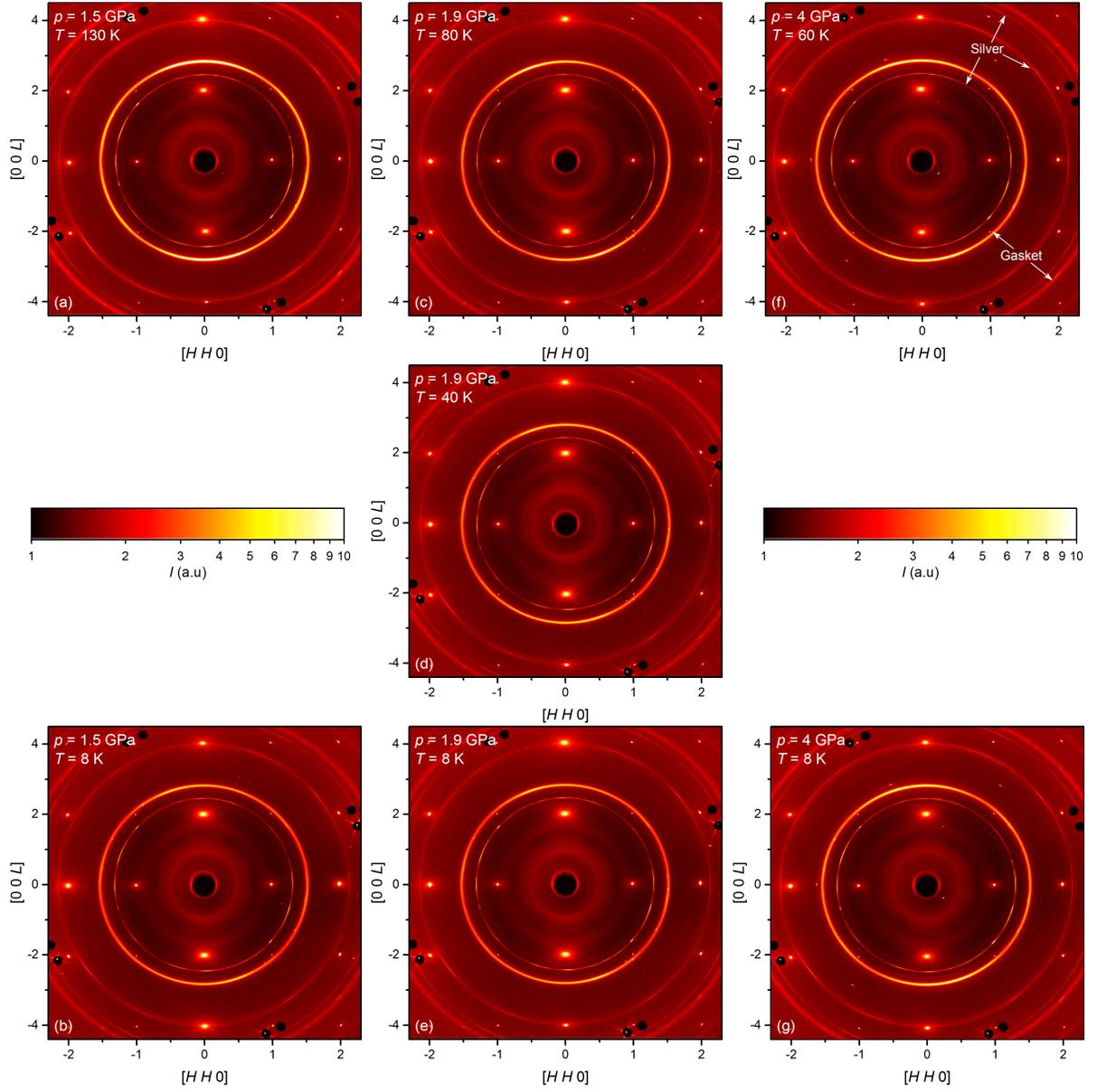

FIG. S11. High-energy x-ray diffraction data on a LaCrGe$_3$ single crystal measured at different temperatures and pressures. Image plots of the $(H\,H\,L)$ plane are shown in each panel with intensities color coded to a log plot as indicated in the color bars. The large, non-central, black circles are from masking the Bragg peaks from the diamond anvils in the DAC, whereas the polycrystalline rings are from the silver foil and the stainless steel gasket, as exemplarily indicated by the white arrows in the top right panel.

## MAGNETISM UNDER PRESSURE FROM NEUTRON AND $\mu$SR STUDIES

### Neutron scattering under pressure

In Fig. S14, we show the integrated intensity data as a function of temperature of the $(1\,0\,0)$ Bragg peak (a) and the $(0\,0\,2)$ Bragg peak (b), which were obtained in neutron diffraction experiments at HB1. For the $(1\,0\,0)$ Bragg peak, we find a clear increase of the intensity upon cooling through the ferromagnetic transition temperature $T_{\mathrm{FM}}(p)$ (see blue arrows). The positions of the arrows correspond to the transition temperatures determined from our thermodynamic measurements and approximately coincide with the points where the behavior in the integrated intensities deviates from the extrapo-

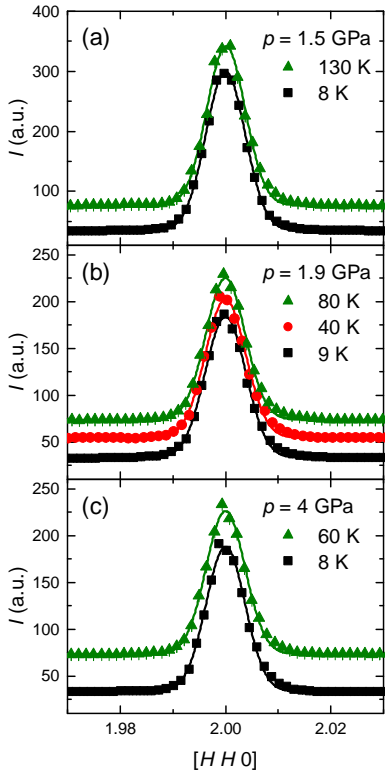

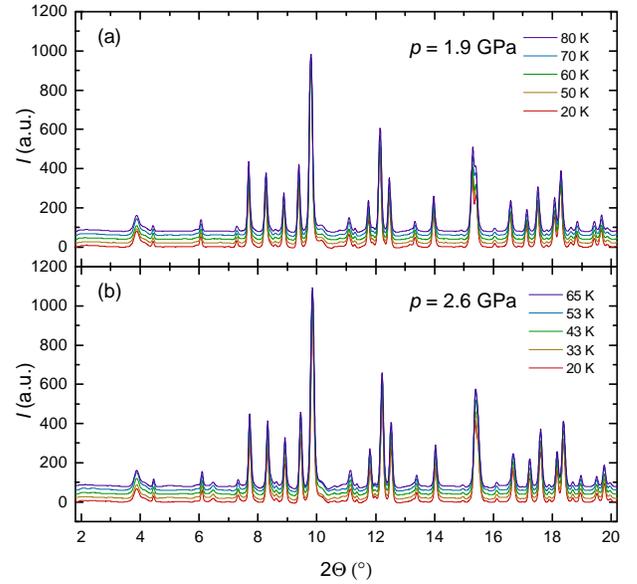

FIG. S13. Powder x-ray diffraction intensities of LaCrGe$_3$ under pressure. The data shows no significant change of peak intensities which would imply the absence of a symmetry-lowering structural component to all salient phase transitions at $p = 1.9$ GPa (a), and 2.6 GPa (b). Data is offset for clarity.

FIG. S12. Cuts along the $(H\,H\,0)$ direction of the $(2\,2\,0)$ peak from high-energy x-ray diffraction data on a LaCrGe$_3$ single crystal under pressure. The data shows no peak splitting or broadening, indicating the absence of a symmetry-lowering structural phase transition. Cuts of the $(2\,2\,0)$ peak are shown at several temperatures for $p = 1.5$ GPa (a), 1.9 GPa (b), and 4 GPa (c). Data are offset for clarity.

lated high-temperature behavior. The increase of the $(1\,0\,0)$ Bragg peak is fully consistent with a ferromagnetic order with moments aligned along the crystallographic $c$ axis[30] and corresponds to 1.5(3) $\mu_B$ at $p = 0$ GPa and $T = 5$ K. The value of magnetic moment was determined from the $(1\,0\,0)$ Bragg peak intensity, $I_{100}$, relative to a set of nuclear Bragg peaks and compared to calculated intensities[31]. For low pressures, the temperature dependence of the $(1\,0\,0)$ Bragg peak shows a typical order-parameter behavior and is thus fully consistent with the second-order nature of the FM transition. Upon increasing pressure, $T_{FM}$ is shifted to lower temperatures. For 1.9 GPa, the ordered moment is 1.4(3) $\mu_B$ at $T = 9$ K.

We also observe an increase of the $(0\,0\,2)$ Bragg peak intensity for almost all pressures at $T_{FM}$, as shown in Fig. S14 (b). This effect, however, is likely not related to the magnetic order itself, but rather due to extinction effects, i.e., a change of the mosaicity of the crystal upon cooling through a transition can lead to a sudden increase of the neutron intensity as a consequence of the suppression of multiple scattering[32]. For example, a magnetic transition can change the mosaicity of a crys-

tal through magnetoelastic effects and the formation of magnetic domains. With x-ray single crystal diffraction, we verified the presence of strong extinction effects by Renninger scans[33] of Bragg peaks by rotating the crystal about the axis of the scattering vector. We found a variation of Bragg peak intensities as the azimuthal angle is changed at ambient pressure, characteristic for strong extinction effects. Extinction effects can be expected to be large for strong nuclear Bragg peaks, such as the $(0\,0\,2)$ Bragg peak, whereas they are negligible for weak nuclear Bragg peaks, such as $(1\,0\,0)$. In addition, if the change in intensity of the $(0\,0\,2)$ Bragg peak was magnetic in nature, then other Bragg peaks, e.g., $(1\,0\,2)$ and $(2\,0\,2)$, corresponding to the same magnetic order, should show a similar increase of magnetic intensity, which was not observed in our experiment. Since extinction effects are dependent on factors such as the size and shape of a specific sample, the strain applied to a specific sample, and the scattering configuration, the effect is expected to be strongly sample-dependent. For our measurements on LaCrGe$_3$, extinction release coincides with $T_{FM}$ for $p \leq 1.9$ GPa (see blue arrows). For higher pressures, the effects of extinction release are weaker. At 2.5 GPa, there is an increase of the $(0\,0\,2)$ intensity upon cooling, but we cannot assign a characteristic temperature solely on the neutron data, whereas for 3.5 GPa no increase can be observed.

Now we return to the survey of potential magnetic Bragg peaks for $p \gtrsim 1.9$ GPa, related to the new magnetic phases associated with $T_1$ and $T_2$. In Figs. S15 and

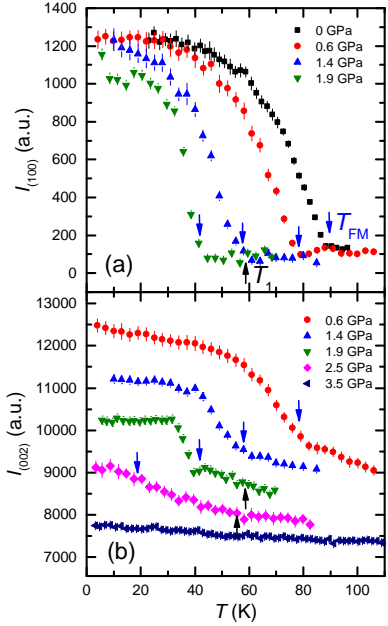

FIG. S14. Integrated intensity of the $(1\,0\,0)$ Bragg peak (a) and the $(0\,0\,2)$ Bragg peak (b) in elastic neutron scattering experiments at HB1 on a single crystal of LaCrGe$_3$ under pressures up to 1.9 GPa (a) and 3.5 GPa (b). Blue arrows indicate the position of the ferromagnetic transition at $T_{\mathrm{FM}}$. Data in (b) was offset for clarity. Data at ambient pressure were taken in experiment N0, data up to 1.9 GPa were taken in N1, data at 2.5 GPa and 3.5 GPa in N2. The positions of the arrows correspond to the transition temperatures determined from our thermodynamic measurements and approximately coincide with the points where the behavior in the integrated intensities deviates from the extrapolated high-temperature behavior.

S16 we show scans along the high-symmetry $H$ and $L$ directions over wide regions of reciprocal space at base temperature for pressures of 1.9 GPa (a-c), 2.5 GPa (d-f), and 3.5 GPa (g-i). Based on our $T$-$p$ phase diagram, LaCrGe$_3$ is ordered ferromagnetically at base temperature for 1.9 GPa and 2.5 GPa, and is in the new magnetic phase for 3.5 GPa, for which a modulated AFM$_q$ was expected[3,4]. We do not observe a significant intensity of the $(1\,0\,0)$ Bragg peak for 2.5 GPa and 3.5 GPa. The nuclear contribution to the Bragg peak is likely reduced due to pressure-induced shifts of atomic positions and changes of lattice parameters. At the same time, the weak $(1\,0\,0)$ Bragg peak at 2.5 GPa implies at maximum only a weak FM contribution at base temperature [see inset in Fig. S15 (d)]. This increase in the intensity is within the significance level of our experiment, but nonetheless consistent with the formation of FM order with small correlation length at base temperature (see also the large $\lambda_{\mathrm{T}}$ in μSR experiments at a similar pressure, discussed in the main text). At 3.5 GPa at low temperatures, a magnetic contribution to the $(1\,0\,0)$ Bragg peak can be ruled out within our experimental limits, discussed be-

low. In addition, we do not observe magnetic superstructure peaks along the high symmetry directions at any pressure, indicating the absence of AFM order, in particular for 3.5 GPa, for which LaCrGe$_3$ was proposed to be in the AFM$_Q$ region down to lowest temperature[3]. The only peaks observed during the experiment are structural Bragg peaks from LaCrGe$_3$, the pressure cell, and the pressure medium, as indicated in Fig. S15 and Fig. S16.

Given the absence of magnetic Bragg peaks, we can calculate a lower boundary for the observable magnetic moment ($\mu$) in our experiment for particular cases of long-range magnetic order. We discuss three cases of magnetic order which were suggested in [3]: (i) long-range FM with $\mu \parallel \mathbf{c}$, (ii) an AFM structure consisting of FM-$ab$ planes with $\mu \parallel \mathbf{c}$, which are stacked along the $c$ axis in a $++++++++++----------$ sequence, and (iii) the intermediate case with FM-$ab$ planes stacked in a $100\times+$ and $100\times-$ sequence along the $c$ axis. The FM order would yield a Bragg peak at $(1\,0\,0)$ with a minimum observable moment of $0.3\,\mu_{\mathrm{B}}$ at 3.5 GPa. The second structure would yield satellite peaks at positions $(1.1\,0\,0)$ and $(0.9\,0\,0)$, which are clearly separate from the $(1\,0\,0)$ Bragg peak. For those satellite peaks we are sensitive to an ordered moment of $0.5\,\mu_{\mathrm{B}}$ at 3.5 GPa. We would observe the small-$q$ peaks for the third structure as a broadened peak at position $(1\,0\,0)$, but the sensitivity for observing a peak at this position is the same as for the FM structure at 3.5 GPa, $\mu = 0.3\,\mu_{\mathrm{B}}$.

Since we found no evidence for AFM order along high-symmetry directions in the crystal, we then searched larger sections of reciprocal space using CORELLI. CORELLI allows for the simultaneous measurement of large sections of three-dimensional reciprocal space by utilizing a white-beam Laue technique with energy discrimination by modulating the incident beam with a statistical chopper[17]. This allows CORELLI to efficiently separate the elastic and diffuse scattering from the sample, and is useful for identifying short- and long-range order. By applying pressure in a DAC at CORELLI we were able to reach pressures from 0.8 GPa to 3.2 GPa at base temperature. In Fig. S17 we show a clear increase of intensity of the $(1\,0\,0)$ Bragg peak when cooling through the FM transition at 0.8 GPa. This observation shows clearly that we are sensitive to the FM transition at CORELLI, and that we can expect to detect AFM order or short-range FM order with a minimum estimated correlation length of 15 nm with a similar magnetic moment for higher pressures. Note that the estimate of correlation length for our CORELLI experiment is 15 nm vs. 12 nm for the HB1 experiment. To search for superstructure peaks we increased the pressure up to 3.2 GPa, at which point we no longer observed the $(1\,0\,0)$ Bragg peak increasing upon cooling and for which the phase diagram in the main text as well as in Ref. [3] indicates the presence of the new magnetic phase below $\approx 50$ K. In Fig. S18 we show 2D images of the $(H\,K\,0)$, $(H\,H\,0)$ and

($H\,0\,L$) reciprocal planes at $T = 5$ K (a, c, e) and 30 K (b, d, f), for which LaCrGe$_3$ is below $T_2$ at those pressures. Bragg peaks from LaCrGe$_3$ are clearly seen alongside rings from the polycrystalline steel gasket. After an exhaustive search of peaks within the 3D reciprocal space at CORELLI we found no evidence of superstructure peaks indicative of AFM from FM order. For the FM and AFM phases, introduced above in the discussion of the HB1 results, our sensitivity at CORELLI amounts to $\mu = 0.4\,\mu_B$ for the FM phase with $\mu \parallel \mathbf{c}$ and $\mu = 0.7\,\mu_B$ for the AFM (+++++++++++---------- ) phase with $\mu \parallel \mathbf{c}$.

In conclusion, we found no indications of long-range magnetic order within the high-pressure phase in careful measurements along the high-symmetry directions on the triple-axis HB1, and within the full 3D reciprocal space measurements done at CORELLI. Summarizing all results from HB1 and CORELLI, our sensitivity for magnetic order would be a lower limit for correlation length of 15 nm for an ordered moment of $1.5\,\mu_B$ like in the FM phase, or a lower limit of $0.7\,\mu_B$ for any long-range AFM order.

## $\mu$SR data under pressure

*General introduction* - During a $\mu$SR experiment, almost 100% spin-polarized muons are implanted into the sample of interest, where they thermalize at interstitial lattice sites. Once stopped, the muon precesses around the direction of the local magnetic field $B$ at the stopping site with the Lamor frequency $\omega_m = \gamma_m \mu_0 B$ with $\gamma_m/(2\pi) = 135.5$ MHz/T being the muon gyromagnetic ratio. The muon is unstable with a life time of $2.2\,\mu$s and decays into a positron and two neutrinos. The time- and the direction-dependence of the positron emission is monitored during a $\mu$SR experiment. From this information on the emitted positron, the muon precession and relaxation can be inferred, and thus, directly the local magnetic field in the sample. The muon therefore is a magnetic micro-probe that allows for tracing of the internal magnetic fields at a local level, and for investigations of the static and dynamic magnetism.

When $\mu$SR experiments are performed on a magnetic sample with simple magnetic order, which implies a well-defined magnetic field $B$ at any of the $n$ inequivalent muon stopping sites ($n \geq 1$, depending on the sample), then the superposition of the signals from all of the muon stopping sites is observed experimentally. The

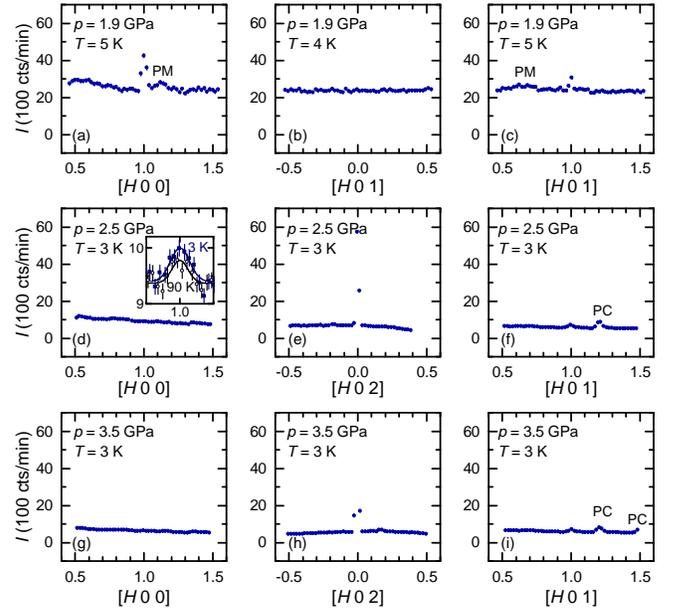

FIG. S15. Selected scans from neutron diffraction experiments on a single crystal of LaCrGe$_3$ at HB1 along the high symmetry directions [$H\,0\,0$] (a, d, g), [$H\,0\,1$] (b, c, f, i), and [$H\,0\,2$] (e, h) at base temperature for pressures of 1.9 GPa (a-c), 2.5 GPa (d-f), and 3.5 GPa (g-i). Different regions of reciprocal space were measured in the data due to the supports in the Palm Cubic Cell blocking incoming and outgoing neutrons for some configurations. The labels Al, PC and PM indicate that the observed peaks are associated with Aluminum (Al), the pressure cell (PC), and the pressure medium (PM). Unlabeled peaks correspond to Bragg peaks associated with the crystal structure of LaCrGe$_3$. Data in (a-c) were taken in experiment N1, (d-i) in N2. Inset in (d) shows the [$H\,0\,0$] scan, measured with 5 minutes per data point, close to the ($1\,0\,0$) Bragg peak at 3 K and 90 K.

measured asymmetry (i.e., the normalized difference between positron counts on the detectors in forward and backward direction) in zero magnetic field for a powder sample, or an aggregate of crystals with random orientation, is given by

$$
A(0)P_{\mathrm{ZF}}(t)
= \sum_{i=1}^{n} A_i(0) \left[ \frac{1}{3} \exp(-\lambda_{\mathrm{L},i}t) + \frac{2}{3} \exp(-\lambda_{\mathrm{T},i}t) \cos(\gamma_m B_{\mathrm{int},i}t) \right],
\tag{S4}
$$

with $A(0)$ ($A_i(0)$) the initial asymmetry of the muon ensemble (of the muon at the $i$-th stopping site) and $P_{\mathrm{ZF}}(t)$ the time-dependent polarization function of the muon ensemble. The spatial averaging due to the random orientation leads to a non-oscillating component with a weight of $1/3$ for muons, whose spins are parallel to the internal field vector at the stopping site, $B_{\mathrm{int},i}$, and therefore show an exponential relaxation with rate $\lambda_{\mathrm{L},i}$, as well as an oscillating component with weight $2/3$, for which

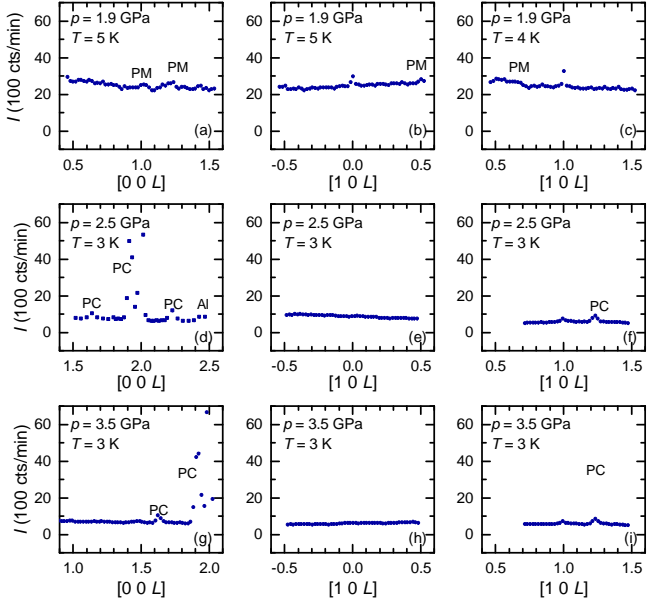

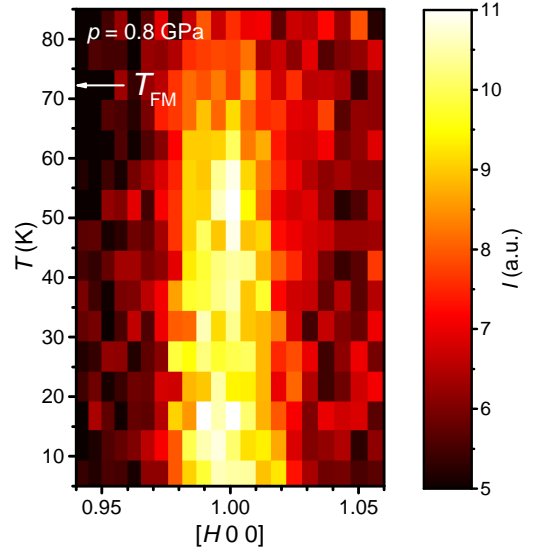

FIG. S16. Selected scans from neutron diffraction experiments on a single crystal of LaCrGe$_3$ at HB1 along the high-symmetry directions $[0\,0\,L]$ (a, d, g), and $[1\,0\,L]$ (b, c, e, f, h, i) at base temperature for pressures of 1.9 GPa (a-c), 2.5 GPa (d-f), and 3.5 GPa (g-i). Different regions of reciprocal space were measured in the data due to the supports in the Palm Cubic Cell blocking incoming and outgoing neutrons for some configurations. The labels Al, PC and PM indicate that the observed peaks are associated with Aluminum (Al), the pressure cell (PC), and the pressure medium (PM). Unlabeled peaks correspond to Bragg peaks associated with the crystal structure of LaCrGe$_3$. Data in (a-c) were taken in experiment N1, (d-i) in N2.

FIG. S17. Neutron diffraction intensity of a LaCrGe$_3$ single crystal as a function of $[H\,0\,0]$ and temperature clearly showing the increase of $(1\,0\,0)$ peak intensity associated with the FM transition. Measurement was taken at CORELLI in a DAC with $p = 0.8$ GPa. The arrow, labeled with $T_{\text{FM}}$, indicates the position of the ferromagnetic transition, as determined from our thermodynamic measurements under pressure. Data were taken in experiment N3.

the muons precess around the internal field vector. The relaxation rate, $\lambda_{\text{T},i}$, which is associated with the oscillating component, is a measure of the width of the static field distribution $\Delta B_{\text{int},i}/\gamma_{\text{m}}$, whereas $\lambda_{\text{L},i}$ is solely related to dynamical magnetic fluctuations. Note that for LaCrGe$_3$, an earlier, ambient-pressure, study[3] showed that the muon time-spectra was best fitted by considering three inequivalent muon stopping sites. However, the three internal field values $B_{\text{int},i}$ were found to be so close to each other, that, for simplicity, we will consider only one muon stopping site for fitting the data inside the pressure cell, given that the higher background contribution in pressure-cell experiments does not allow for taking high-enough statistics to reliably distinguish different muon stopping sites with very similar internal fields.

In addition to zero-field experiments, $\mu$SR measurements can also be performed in external fields. Here, weak-transverse field (wTF) measurements, in which a small external field, $\mu_0 H_{\text{ext}}$, is applied perpendicular to the initial muon-spin direction, is a commonly used method to determine the onset magnetic transition temperature and the magnetic volume fraction. When muons

stop in a non-magnetic sample, the external magnetic field causes a steady precession of the muon spin around its direction, giving rise to long-lived oscillations in the measured $\mu$SR asymmetry. In contrast, when muons stop in a magnetically-ordered sample, then the $\mu$SR signal becomes more complex and reflects the precession around the vector combination of $B_{\text{int}}$ and $\mu_0 H_{\text{ext}}$, which due to the random orientation of the crystallites leads to a broad distribution of precession frequencies. Therefore, the contribution to the muon asymmetry from muons, which do not experience a finite internal fields, can be accurately determined as a function of temperature. For the case of a weak transverse field, i.e., $\mu_0 H_{\text{ex}} \ll B_{\text{int}}$, the fitting function of the $\mu$SR asymmetry becomes simplified such that

$$
\begin{aligned}
&A(0)P_{\text{wTF}}(t) \\
&= A_{\text{nmag}}(0)\cos(\gamma_{\text{m}}\mu_0 H_{\text{ext}}t + \phi)\exp(\frac{-\sigma_{nm}^2 t^2}{2}) \\
&+ A_{\text{mag}}(0)P_{\text{ZF}}(t),
\end{aligned}
\tag{S5}
$$

with $A_{\text{nmag}}(0)$ [$A_{\text{mag}}(0)$] the initial non-magnetic [magnetic] asymmetry, $\phi$ a phase factor, $\sigma_{nm}$ the relaxation rate caused by nuclear moments, and $P_{\text{ZF}}(t)$ the function defined in Eq. S4.

Overall, in pressure-cell experiments, a large fraction of the muons stop in the pressure cell ($\approx 50\%$). This additional contribution has to be included in the data

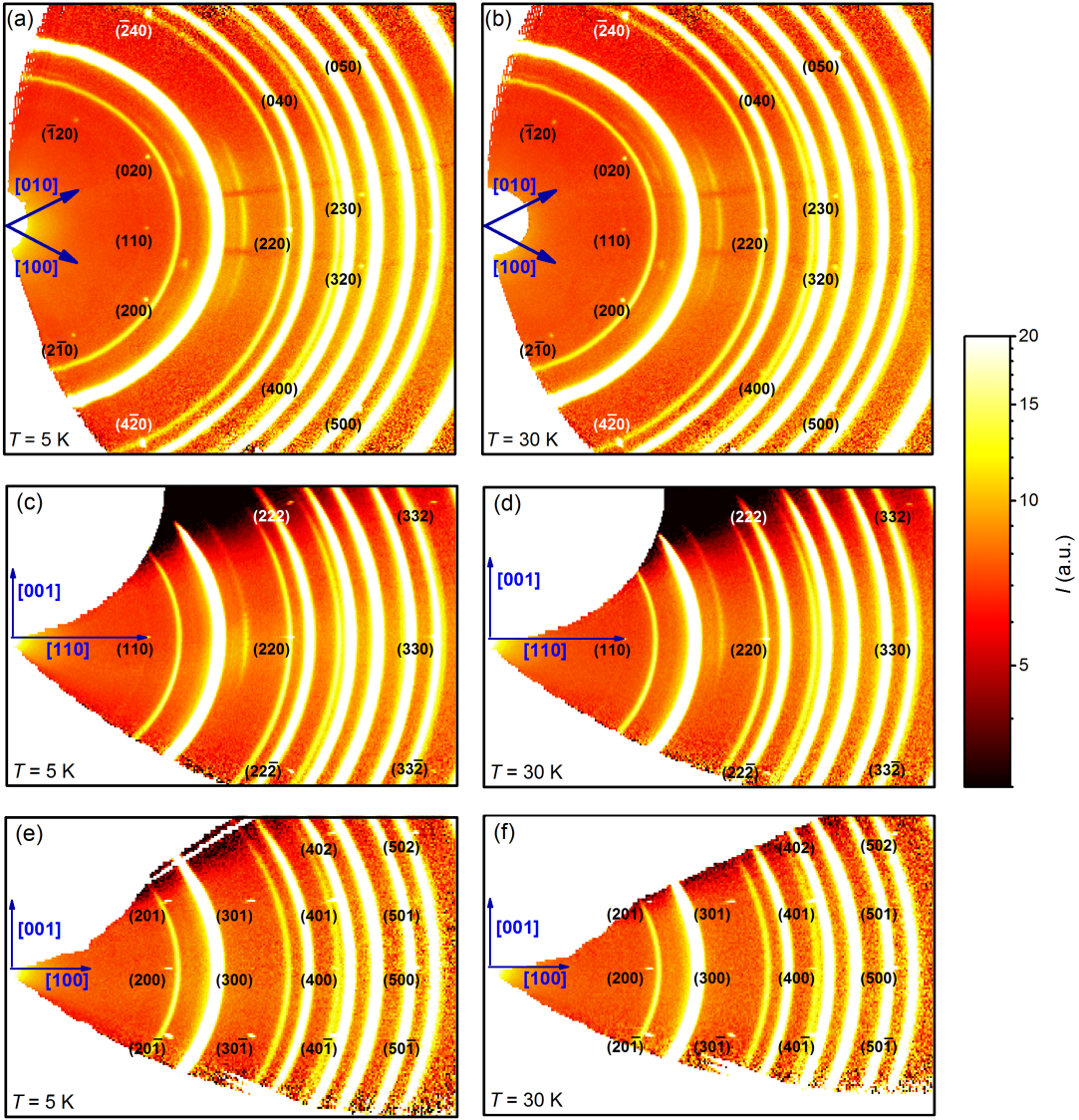

FIG. S18. Several slices of neutron diffraction data taken at CORELLI of a LaCrGe$_3$ single crystal in a DAC with $p = 3.2$ GPa. For each panel crystallographic Bragg peaks are indicated in black or white text, and the reciprocal space direction of the crystal is indicated in blue with arrows. Polycrystalline rings are from the steel gasket with strong intensity modulation arising from the texture and strain in the gasket. Cuts of the neutron data are shown for the $(H\,K\,0)$ plane (a, b), $(H\,H\,L)$ plane (c, d), and $(H\,0\,L)$ plane (e, f) at two temperatures, $T = 5$ K (a, c, e) and 30 K (b, d, f). Data were taken in experiment N3.

analysis, so that the measured asymmetry $A(t)$ reads as

$$A(t) = A_{\mathrm{s}}(0)P_{\mathrm{s}}(t) + A_{\mathrm{pc}}(0)P_{\mathrm{pc}}(t), \quad (S6)$$

with $A_{\mathrm{s}}(0)$ [$A_{\mathrm{pc}}(0)$] being the initial sample [pressure-cell] asymmetry and $P_{\mathrm{s}}(t)$ [$P_{\mathrm{pc}}(t)$] the sample [pressure-cell] polarization function. The sample polarization function either corresponds to $P_{\mathrm{ZF}}(t)$ for the case of zero-field experiments, as defined in Eq. S4, or to $P_{\mathrm{wTF}}(t)$ for the case of weak-transverse field experiments (see Eq. S5).

The background of the pressure cell[21] is typically determined in an independent set of experiments and can then be described by two depolarization channels (one from nuclear moments and one from electronic moments), using a damped Kubo-Toyabe depolarization function,

$$A_{\mathrm{pc}}^{\mathrm{ZF}}(t) =$$
$$A_{\mathrm{pc}}^{\mathrm{ZF}}(0)\left(\frac{1}{3} + \frac{2}{3}(1 - \sigma_{\mathrm{PC}}^2 t^2)\exp(-\sigma_{\mathrm{PC}}^2 t^2)\right)\exp(-\lambda_{\mathrm{PC}} t), \quad (S7)$$

with $\lambda_{\mathrm{PC}}$ the relaxation rate, which is related to electronic moments, and $\sigma_{\mathrm{PC}}$ the relaxation rate, related to the nuclear moments. For the case of LaCrGe$_3$ under pressure, it also needs to be taken into account that samples, which do exhibit a strong macroscopic magnetiza-

tion, will induce a magnetic field in their surrounding, which can be felt by muons that stop in the pressure cell. Typically, this is the case for superconducting or ferro- and ferrimagnetic samples. As a result, the muons stopping regions of the pressure cell closest to the sample undergo a precession around the magnetic field, which is the vector sum of the applied field and the field induced by the sample with strong magnetization (the sum is denoted as $B_{PC}$). This leads to an additional depolarization of the muon spin polarization, the size of which depends on the external field, the field created by the sample as well as the stopping site distribution of the muons in the pressure cell with respect to the spatial distribution of $B_{PC}$. In these cases, the pressure cell contribution cannot be determined in an independent set of experiments or described by the Eq. S7 above, and instead follows in wTF experiments

$$A_{PC}^{wTF}(t) = A(0)\exp(-\lambda_{PC}t)\exp(-\sigma_{PC}^2 t^2/2)\cos(\gamma_m B_{PC}t+\phi)$$
$$(S8)$$

with $\lambda_{PC}$ the relaxation rate, the size of which is determined by the influence of a sample with macroscopic magnetization on the pressure cell as well as the electronic relaxation rate, and $\sigma_{PC}$ the relaxation rate, related to the nuclear moments. The electronic relaxation rate is found to be temperature-independent and was determined to be $\approx 0.05\,\mu s^{-1}$ in the non-magnetic state of LaCrGe$_3$, i.e., for $T > T_{FM}$ at ambient pressure, for the used pressure cell. Therefore, if the muons stopping in the pressure cell do not experience any field, that is created by the sample, then $\lambda_{PC} \simeq 0.05\,\mu s^{-1}$, and the pressure cell asymmetry shows long-lived oscillations. Instead for $\lambda_{PC} \gtrsim 0.05\,\mu s^{-1}$, the signal is damped, reflecting the additional depolarization of the precession of muons, that stop in the pressure cell, as a result of the field created by the sample.

Following the same line of arguments, any sample that exhibits a large, remnant magnetization, will distort the pressure-cell $\mu$SR signal. Thus, $\mu$SR measurements inside the pressure cell also allow for the estimation of whether the sample exhibits a remnant magnetization or not. A remnant magnetization is typical for ferromagnets, however, the remanency can be very small, potentially even beyond the resolution of $\mu$SR experiments.

Experimentally, the test for a remnant magnetization is performed in the following way. First, the sample is cooled below the transition temperature in zero field, and an initial $\mu$SR spectrum is recorded. In a next step, the external field is ramped isothermally to a specific, finite value, held constant for a short period of time, and then removed again. Then, at zero field, the $\mu$SR spectrum is recorded again. The recorded pressure-cell response after the application and subsequent removal of the magnetic field can be described by following function

$$A(t) = A(0)[(1-\zeta)G_{KT}(t)\exp(-\lambda_{PC}t)+\zeta], \quad (S9)$$

with $1-\zeta$ being the spectral weight of the relaxing component, $G_{KT}(t) = 1/3+2/3(1-\sigma^2 t^2)\exp(-\sigma^2 t^2/2)$ being the Gaussian Kubo-Toyabe depolarization function reflecting the field distribution at the muon site created by nuclear moments and $\lambda_{PC}$ is the exponential relaxation describing the influence of the sample on the pressure cell. Again, $\lambda_{PC} \simeq 0.05\,\mu s^{-1}$ implies no remanent magnetization, whereas $\lambda_{PC} \gtrsim 0.05\,\mu s^{-1}$ implies a remanent magnetization and the exact value of $\lambda_{PC}$ is expected to be dependent on the external field that was applied prior to the collection of the spectrum (as long as the applied field is smaller than the saturation field). In the present experiment, we performed a set of these experiments at high pressures, $p = 2.55$ GPa, at three distinct temperatures. At each temperature, in total 5 different fields were applied and a spectrum was recorded each time after decreasing the respective field back to zero.

*$\mu$SR measurements in zero field (ZF) at $p = 0.2$ GPa* - Figure S19 shows selected zero-field $\mu$SR spectra of LaCrGe$_3$ at $p = 0.2$ GPa for temperatures in the range $10\,K \leq T \leq 89.5\,K$, i.e., across $T_{FM}(0.2\,GPa) \simeq 82\,K$. The $T = 10\,K$ data is shown again separately below in Fig. S25. For $T < T_{FM}$, a well-defined muon spin precession is observed, which confirms the presence of a finite internal field $B_{int}$. For temperatures just below $T_{FM}$, i.e., for $T = 80.3\,K$, weak and highly damped oscillations are observed. For $T > T_{FM}$ (see $T = 89.5\,K$ data), no precession of the muon spins is discernible, indicating that $B_{int} = 0$. The solid lines in Fig. S19 correspond to fits to the experimental data to Eqs. S4 and S6. The temperature-dependence of the fit parameters for all investigated temperatures will be discussed below.

*$\mu$SR measurements in weak-transverse field (wTF) at $p = 0.2$ GPa* - Next, we show selected $\mu$SR time-spectra of LaCrGe$_3$ at $p = 0.2$ GPa, which were taken in a weak-transverse field of 30 Oe after zero-field cooling for various temperatures across $T_{FM}$ in Fig. S20 (a) as well as on enlarged scales around $t \approx 2.5\,\mu s$ (b). For $T > T_{FM}$, large and only weakly-damped oscillations with maximum amplitude close to 0.25, i.e., the maximum for the used spectrometer, are observed. This observation corresponds to the expected precession of the spins in the non-magnetic sample and the non-magnetic pressure cell, induced by the wTF. In contrast, for $T < T_{FM}$, the oscillations are damped, since the sample exhibits a strong internal field, but also the pressure cell are exposed to a strong magnetic field, which is created by the ferromagnetic sample

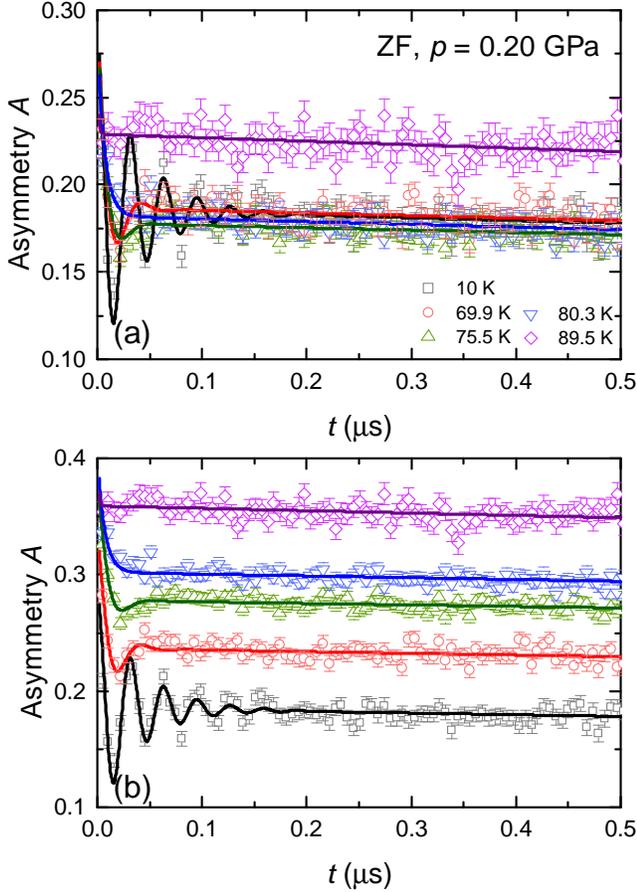

FIG. S19. Short-time μSR spectra (symbols) of LaCrGe₃, taken in a zero field (ZF) and at a pressure $p = 0.20\,\mathrm{GPa}$ for different temperatures. Lines correspond to fits of the data to Eqs. S4 and S6. Data in (b) are the same as in (a), but offset for clarity.

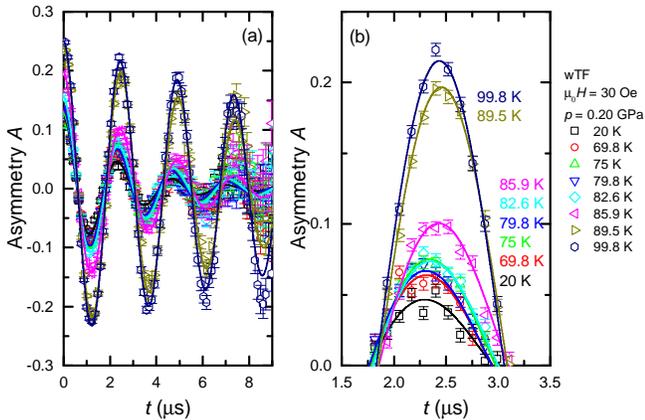

FIG. S20. Muon-time spectra (symbols) of LaCrGe₃, taken in a weak-transverse field (wTF) of 30 Oe and at a pressure $p = 0.20\,\mathrm{GPa}$ for different temperatures (a-b). (b) shows the data, presented in (a), on enlarged scales around the local maximum close to $t \approx 2.5\,\mu\mathrm{s}$. Lines correspond to fits of the data to Eqs. S5 and S6.

inside the pressure cell. The ordering therefore leads to an additional depolarization of the muons, which explains the strongly reduced amplitude of the oscillations. The maximum size of the oscillation amplitude for $T < T_{\mathrm{FM}}$ is fully consistent with full-volume fraction, which will be elucidated below in more detail, when discussing the detailed evolution of the fit parameters with temperature, which are extracted from the fits to Eqs. S5 and S6 (solid lines in Fig. S20), will be discussed in the following.

*Temperature evolution of μSR fitting parameters at $p = 0.2\,GPa$ and comparison with thermodynamic measurements* - In Fig. S21, we show the temperature ($T$) evolution of the μSR fitting parameters (a-d) at a pressure of 0.2 GPa. This includes the evolution of the internal field, $B_{\mathrm{int}}$, (a) and the transverse relaxation rate, $\lambda_{\mathrm{T}}$, (b) which were both extracted from fitting the ZF μSR data, as well as the magnetic asymmetry, $A_{\mathrm{mag}}$, (c) and the relaxation rate of the pressure cell, $\lambda_{\mathrm{PC}}$, (d) which were extracted from fitting the wTF data. We compare this data with data of the specific heat, $\Delta C/T$, (e), the relative length change along the $c$ axis, $(\Delta L/L)_c$ and the thermal expansion coefficient, $\alpha_c$, (f) and the temperature-derivative of the $c$-axis resistance, $\mathrm{d}R_c/\mathrm{d}T$, (g), all taken at similar pressure values.

We find that $B_{\mathrm{int}} \simeq 2200\,\mathrm{Oe}$ at lowest temperatures [see Fig. S21 (a)], which is very similar to the previous μSR data[3]. $B_{\mathrm{int}}$ decreases upon increasing $T$ and extrapolates to zero close to $T_{\mathrm{FM}} \approx 82\,\mathrm{K}$. At lowest temperatures, $\lambda_{\mathrm{T}} \simeq 25\,\mu\mathrm{s}^{-1}$, somewhat larger, but still consistent with previous reports[3], and increases with increasing temperature, until it reaches a peak at $T_{\mathrm{FM}}$, above which $\lambda_{\mathrm{T}}$ decreases rapidly [see Fig. S21 (b)]. This behavior of $\lambda_{\mathrm{T}}$, which quantifies the width of the static field distribution at the muon stopping site, corresponds to the typical behavior for a sample which undergoes a magnetic transition. Only very close to the phase transition, the field distribution becomes wide as the field starts to occur in the sample when magnetic order develops, whereas well below the ordering temperature, $\lambda_{\mathrm{T}}$ is small, reflecting the well-ordered magnetism in LaCrGe₃ at 0.2 GPa (see below for a comparison of $\lambda_{\mathrm{T}}$ at 0.2 GPa and 2.55 GPa). $A_{\mathrm{mag}}$ is almost constant for $T \leq T_{\mathrm{FM}}$ at a value of $\approx 0.12$ [see Fig. S21 (c)], which reflects that approximately 50% of the muons stop in the pressure cell. Therefore this value of $A_{\mathrm{mag}}$ strongly suggests that the magnetic volume fraction reaches 100% at $T_{\mathrm{FM}}$, given that the maximum asymmetry of the setup is close to 0.25, which was determined in a separate experiment. Above $T_{\mathrm{FM}}$, $A_{\mathrm{mag}}$ decreases rapidly to zero, as the sample becomes non-magnetic. $\lambda_{\mathrm{PC}} \simeq 0.43\,\mu\mathrm{s}^{-1}$ at lowest temperatures [see Fig. S21 (d)], the finite size of which reflects the influence of the magnetic field, created by the ferromagnetic sample inside the pressure cell, on the pressure cell. Upon increasing $T$, $\lambda_{\mathrm{PC}}$ initially stays roughly

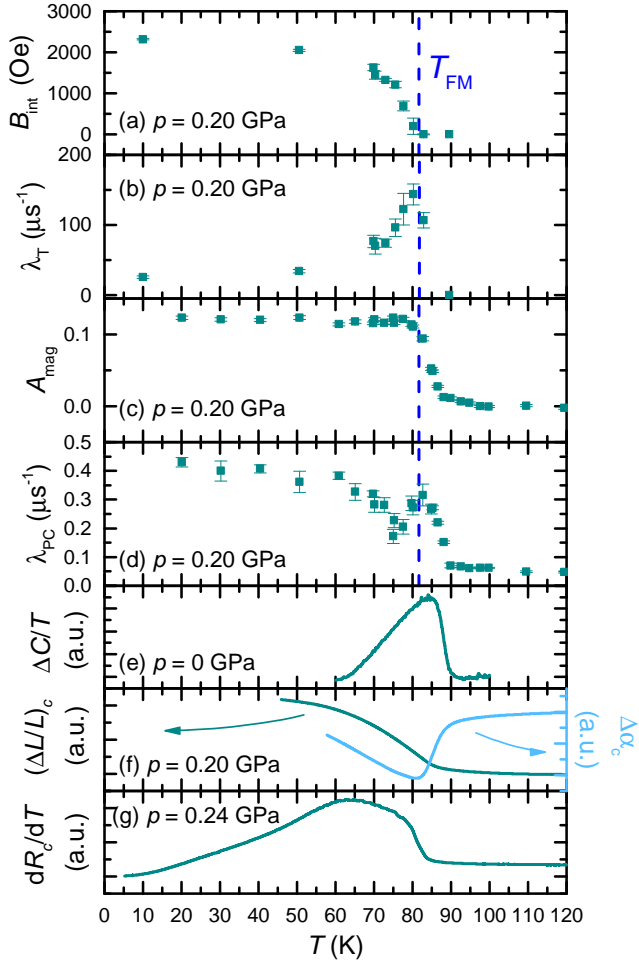

FIG. S21. Comparison of several low-pressure microscopic and thermodynamic data sets close to a pressure, $p$, of 0.2 GPa as a function of temperature, $T$. (a,b,c,d) Internal field, $B_{int}$ (a), transverse relaxation rate $\lambda_T$ (b), magnetic asymmetry, $A_{mag}$ (c) and relaxation rate of the pressure cell, $\lambda_{PC}$, from zero-field (a,b) and weak-transverse field (wTF) (c,d) μSR measurements at $p = 0.2$ GPa; (e) Anomalous contribution to specific heat, $\Delta C/T$ at $p = 0$ GPa; (f) Relative length change and thermal expansion coefficient along the crystallographic $c$ axis, $(\Delta L/L)_c$ (left axis) and $\alpha_c$ (right axis), respectively, at $p = 0.2$ GPa; (g) Temperature-derivative of the resistance along the $c$ axis, $dR_c/dT$, at $p = 0.24$ GPa. In (a)-(d), blue dashed lines indicates the position of the anomalies, associated with the ferromagnetic transition at $T_{FM}$.

constant and then starts to decrease as $T$ is approaching $T_{FM}$. However, instead of $\lambda_{PC}$ just approaching a value close to zero, the behavior of $\lambda_{PC}$ is more complex. In more detail, $\lambda_{PC}$ first goes through a minimum at $T \approx 76$ K, followed by a maximum at $\approx 82$ K and then decreases and saturates at a value close to zero for $T > T_{FM}$. This complex behavior of $\lambda_{PC}$ was not discovered in the previous study[3], likely due to the large data point spacing in temperature. Whereas the maximum in $\lambda_{PC}$ is highly likely related to the FM ordering

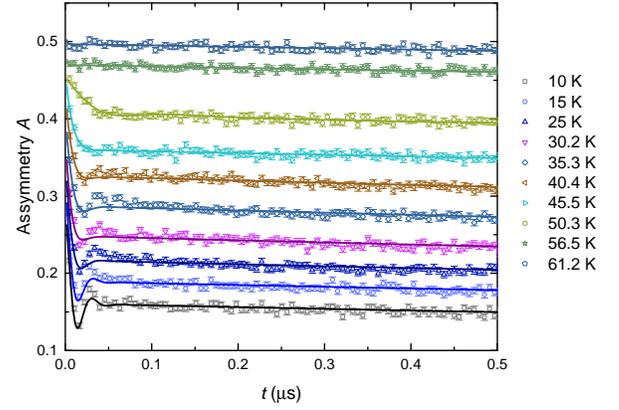

FIG. S22. Muon-time spectra (symbols) of LaCrGe$_3$, taken in zero field (ZF) and at a pressure $p = 2.55$ GPa for different temperatures. Lines correspond to fits of the data to Eqs. S4 and S6. Data were offset for clarity.

at $T_{FM}$, we speculate that the minimum is rather related to the proposed crossover[4] from FM to FM2 in LaCrGe$_3$ upon cooling at low pressures. However, as we will show below, we do not find any corresponding feature in our thermal expansion measurements. The observed features in the μSR fitting parameters at $T_{FM}$ are well consistent with the positions of the FM anomalies in the thermodynamic and transport studies of the present work (Note that the shown specific heat data set was taken at ambient pressure, and thus at a slightly smaller pressure than the other data sets, which were taken at $\approx 0.2$ GPa).

*μSR measurements in zero field (ZF) at $p = 2.55$ GPa* - Fig. S22 shows a larger set of zero-field μSR spectra for $10$ K $\leq T \leq 61$ K. A subset of this data was shown in Fig. 3 of the main text, and the evolution of $B_{int}$ and $\lambda_T$ was shown in Fig. 4 of the main text and discussed there.

*μSR measurements in weak-transverse field (wTF) at $p = 2.55$ GPa* - Figure S23 shows selected μSR time-spectra of LaCrGe$_3$ at $p = 2.55$ GPa, which were taken in a weak-transverse field of 30 Oe after zero-field cooling. The temperature evolution of the fit parameters, extracted from fitting Eqs. S5 and S6 to this data, were already shown and discussed in the main text in Fig. 4 (c) and (d). At this point, we would only like to describe the behavior of these selected data sets, in analogy to the discussion of the wTF data at 0.2 GPa, across the characteristic temperatures $T_1(p = 2.55$ GPa$) \simeq 56$ K and $T_2(p = 2.55$ GPa$) \simeq 49$ K. Figure S23 (a) shows the wTF muon-time spectra for temperatures $T < T_2$ and $T > T_1$. For $T > T_1$, large and only weakly-damped oscillations with amplitude close to 0.25, i.e., the maximum for the used spectrometer are observed, which correspond to the expected precession of the spins in the

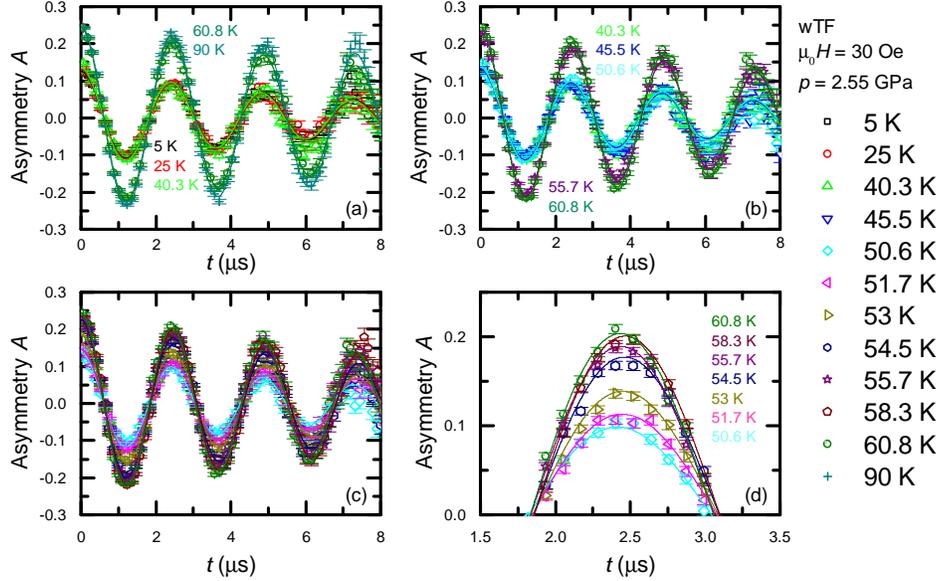

FIG. S23. Muon-time spectra (symbols) of LaCrGe₃, taken in a weak transverse field (wTF) of 30 Oe and at a pressure $p = 2.55$ GPa for different temperatures (a-d). (d) shows the data, presented in (c), on enlarged scales around the local maxmimum close to $t \approx 2.5\,\mu$s. Lines correspond to fits of the data to Eqs. S5 and S6.

non-magnetic sample and the non-magnetic pressure cell, induced by the wTF. In contrast, for $T < T_2$, the muon oscillations are strongly damped, since the sample exhibits a strong internal field, but also the pressure cell is exposed to a strong magnetic field, which is created by a sample with macroscopic magnetization inside the pressure cell. The ordering therefore results in an additional depolarization of the muons, which explains the strongly reduced amplitude of the oscillations. Figure S23 (b) shows wTF spectra for temperatures, which are closer in temperature to $T_1$ and $T_2$. At $T = 60.8$ K and 55.7 K, we find oscillations with large amplitude, whereas for $T \leq 50.6$ K, the oscillations are small in amplitude and heavily damped. A further data set with even finer temperature spacing around $T_1$ and $T_2$ is shown in Fig. S23 (c) and on enlarged scales in (d). These plots show that the oscillations become maximally reduced in amplitude at $T_2$, whereas their amplitude is still large at $T_1$. Note that the magnetic asymmetry, inferred at this pressure at low temperatures, is identical, within the error bars, to the magnetic asymmetry at low pressures and low temperatures [see Fig. 4(c) in the main text and Fig. S21 (c)], consistent with the notion of full magnetic volume fraction. Altogether, this leads to the conclusion full magnetic volume fraction can only be observed for $T \leq T_2$.

*Experimental test for a remanent magnetization of the sample in μSR measurements at $p = 2.55$ GPa* - In this section, we present our experimental test of

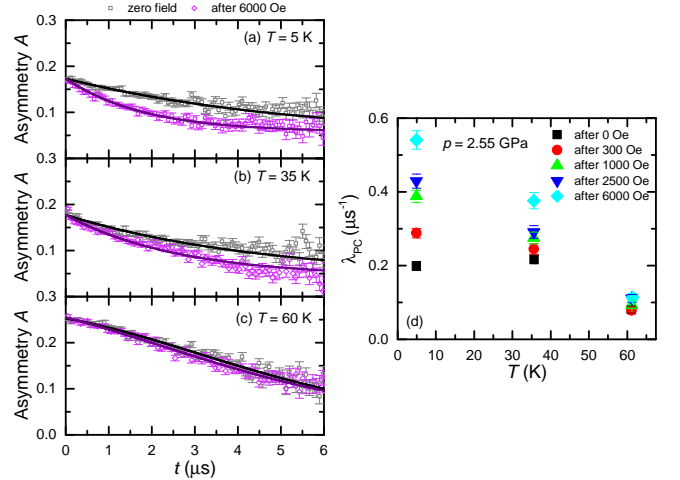

FIG. S24. (a,b,c) Initial zero-field μSR time-spectra and zero-field spectra after increasing and decreasing the magnetic field to 6000 Oe (symbols) at $T = 5$ K (a), 35 K (b) and 60 K (c) at $p = 2.55$ GPa. Solid lines are fits to the experimental data by Eq. S9; (d) Pressure-cell relaxation rate, $\lambda_{\text{PC}}$, as a function of temperature, $T$, at $p = 2.55$ GPa. $\lambda_{\text{PC}}$ was extracted from fitting the experimental data.

whether LaCrGe₃ exhibits a remanent magnetization at $p = 2.55$ GPa by μSR measurements, by searching for a change of the pressure-cell μSR spectrum after magnetizing the sample. We note that similar experiments were already conducted in Ref. [3] at a similar pressure of 2.3 GPa, i.e., clearly in the region of the phase diagram, in which the new phases exist. In this previous report,

it was argued that no significant response of the pressure cell was found at 2.3 GPa, which was interpreted as an absence of an ordering with ferromagnetic component at this pressure. In the present study, we revisited this question by performing measurements with higher statistics. As we will argue below, this new set of data strongly suggests that LaCrGe₃ *does* exhibit a remanent magnetization below $T_2$. As outlined in detail in the main text, this observation together with (i) the finite $\lambda_{PC}$ below $T_2$ in wTF experiments (see main text) and (ii) the absence of a clear magnetic intensity in the vast majority of reciprocal space in neutron scattering experiments (see SI), has led us to redefine the nature of the new magnetic phase that is associated with the avoidance of ferromagnetic criticality.

The detailed experimental procedure, which was used to check for a remanent magnetization in $\mu$SR measurements, was described in the introduction above. Figure S24 shows the initial ZF muon-time spectrum after zero-field cooling, together with the ZF time-spectra after increasing the field to 6000 Oe and subsequently going back to ZF at $T = 5$ K (a), 35 K (b) and 60 K (c). According to our thermodynamic phase diagram [see Fig. 2 of the main text or Fig. S4 (b)], LaCrGe₃ is in the FM phase at 5 K (a), in the $T_2$ phase at 35 K (b) and in the high-temperature paramagnetic phase at 60 K (c). Solid lines are fits to the experimental data by Eq. S9. The raw data already indicate that for $T = 5$ K and 35 K the muon-time spectra are different from the initial time-spectra after the application of any field. In contrast, at $T = 60$ K, the application of any field leaves the ZF time-spectra almost unmodified from the initial ZF time spectrum. This already strongly suggests the presence of a remanent magnetization not only for $T = 5$ K, but also for 35 K. A quantitative analysis of this behavior is obtained from considering the evolution of the fit parameter $\lambda_{PC}$ with temperature and magnetic field, which is shown in Fig. S24 (d). We note that the $\lambda_{PC}$ values obtained here from ZF experiments are usually larger by a factor of $\approx$ 1.8 than those $\lambda_{PC}$ values obtained in wTF experiments[34] (shown in the main text in Fig. 4). Whereas the ZF $\lambda_{PC}$ is small for 60 K and almost independent of field, $\lambda_{PC}$ is clearly larger and stronger field-dependent for 35 K and 5 K. This all suggests that for 5 K and 35 K there exists a remanent magnetization, which creates a field that the muons, which stop in the pressure cell, are exposed to. Thus, this result speaks in strong favor of the fact that the magnetic phase below $T_2$ also has a remanent magnetization. In this regard, we would like to note that the finite $\lambda_{PC}$ below $T_2$, which was inferred from wTF measurements and is shown in the main text in Fig. 4(d), is also fully consistent with that notion.

*Direct comparison of low- and high-pressure $\mu$SR data*

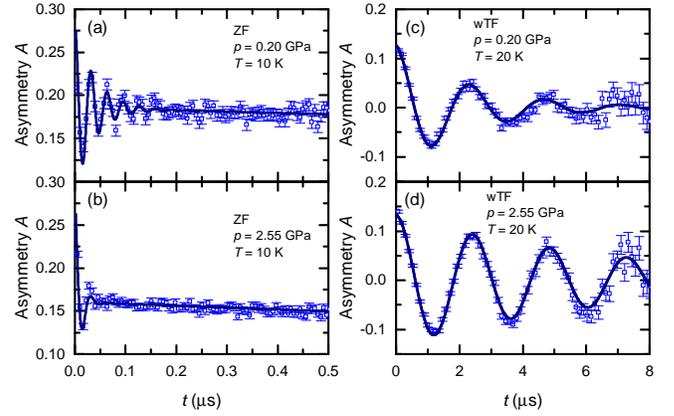

FIG. S25. Comparison of zero-field [ZF; (a-b)] and weak-transverse field [wTF; (c-d)] muon-time spectra (symbols) for low pressure, $p = 0.20$ GPa (a,c) and high pressure, $p = 2.55$ GPa (b,d). ZF data were taken at $T = 10$ K, wTF data were taken at $T = 20$ K. Lines are fits to the experimental data by Eq. S4 for the ZF data and Eq. S5 for the wTF data.

- Finally, we want to explicitly compare the ZF muon-time spectra at low temperatures ($T = 10$ K) for 0.2 GPa (a) and 2.55 GPa (b), as well as the wTF time spectra at $T = 20$ K for the same pressures (c-d). The comparison of the ZF spectra shows that the precession is much stronger damped for high pressures, as also quantified by the respective $\lambda_T$ values, which are depicted in Fig. S21 and Fig. 4 in the main text, respectively. This implies that the static field distribution (i.e., the disorder in field the muon experiences) is three to four times larger than the one at 0.2 GPa, whereas the size of the internal field $B_{int}$ remains similar. Similar observations were also made in Ref. [3]. In addition, the comparison of the respective wTF data shows that the damping of the muon precession is larger for low pressures of 0.2 GPa than for high pressures of 2.55 GPa, i.e., that $\lambda_{PC}(0.2 \, \text{GPa}) > \lambda_{PC}(2.55 \, \text{GPa})$. This result implies that the macroscopic magnetization of the sample assembly is smaller for 2.55 GPa than for 0.2 GPa despite the very similar values of the internal field. Overall, in the main text, these observations have led us to a reinterpretation of the magnetism below $T_2$ for 2.55 GPa in terms of a short-range magnetically ordered state with ferromagnetic component.

## ESTIMATION OF THE POSITION OF THE TRICRITICAL POINT

In the following, we discuss our estimation of the position of the pressure-induced tricritical point at $(p_{tr}, T_{tr})$, at which the character of the ferromagnetic (FM) transition changes from second order to first order. To this end, we will focus on an analysis of the anomalies in the thermal expansion coefficient, given the presence of pro-

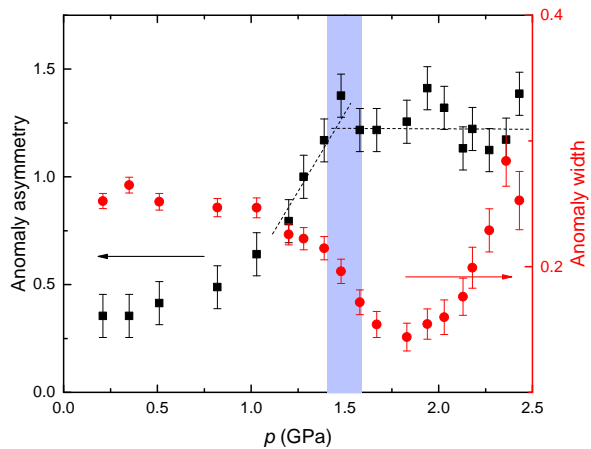

FIG. S26. Asymmetry (left axis) and width (right axis) of the thermal expansion anomalies in LaCrGe$_3$ along the crystallographic $ab$ axis, which was shown in Fig. S6. The asymmetry was determined from $(T_r - T_m)/(T_m - T_l)$, with $T_m$ being the temperatures, at which the peak of the thermal expansion anomaly occurs, and $T_r$ and $T_l$ being the temperatures, at which the thermal expansion reaches 50% of the peak value, respectively. Correspondingly, the width was calculated as $(T_r - T_l)/T_m$. Dashed lines are to guide of the eyes to highlight the change of the behavior of the asymmetry close to 1.5 GPa.

nounced features in this quantity over a wide range of the phase diagram.

*Analysis of the shape of the anomaly* - In thermodynamic quantities, a second-order phase transition often manifests itself in a strongly asymmetric, mean-field type anomaly, whereas a first-order phase transition usually shows up as a symmetric, somewhat broadened peak. As discussed in the main text as well as the SI above, we observe a change from an almost mean-field-type jump in the thermal expansion coefficient $\Delta\alpha_i$ for low pressures to an almost symmetric, sharp peak for high pressures. This signals a pressure-induced change of the character of the transition from second-order to first-order. To quantify this change and to determine the position of the associated tricritical point, we evaluated the asymmetry of the expansion anomaly $\Delta\alpha_a$ by using the following expression $(T_r - T_m)/(T_m - T_l)$ with $T_m$ being the temperature, at which $\Delta\alpha_{ab}$ reaches its maximum value, and $T_r$ ($T_l$) being the higher (lower) temperature, at which $\Delta\alpha_{ab}$ exhibits 50% of the maximum value of $\Delta\alpha_{ab}$. The evolution of the so-determined asymmetry is shown in Fig. S26 (left axis). For low pressures, the asymmetry parameter is less than 0.5, signaling a very asymmetric anomaly. With increasing pressure, the asymmetry parameter increases rapidly to a value close to 1 (corresponding to a perfectly symmetric peak), and flattens off (see dashed line) at a value of $\approx$ 1.2. This behavior

therefore meets the expectation for the above-described change from second order to first order. Thus, we use the pressure, at which the asymmetry parameter levels off, to determine the position of the tricritical point. This results in $p_{tr} = 1.5(1)$ GPa, and the corresponding $T_{cr} = 53(3)$ K was inferred from the thermodynamic phase diagram in Fig. 2 of the main text or Fig. S4 (b) in the SI above. We can also consider the width of the $\Delta\alpha$ feature, which we determine via $(T_r - T_l)/T_m$ and is displayed on the right axis of Fig. S26. The width clearly shows a strong decrease right around $p_{tr}$, consistent with an increase in sharpness of the transition feature, once the transition becomes first-order. We assign the subsequent increase of the width with pressure for higher pressures, which is on the first glance not consistent with the notion of a sharp-first order transition, to an increased slope $dT_{FM}/dp$, which naturally accounts for an increase in broadening, the higher the pressure is.

*Measurements of thermal hysteresis* - As a complementary approach, we can also consider the evolution of the thermal hysteresis at the ferromagnetic transition with pressure. In Figs. S27 (a) and (b), we show two example data sets of the anomalous contribution to the thermal expansion coefficient, $\Delta\alpha_c$, around the ferromagnetic transition upon warming and cooling at $p = 0.35$ GPa (a) and 1.9 GPa (b). Whereas we find only a tiny thermal hysteresis for low pressure, which is probably related to the intrinsic hysteresis of our experimental setup, we observe a clear hysteresis for larger pressures. This confirms clearly that the transition becomes first order for higher pressures. A quantitative analysis of the evolution of the thermal hysteresis, $\Delta T$, defined as the difference between transition temperatures upon warming and cooling, with pressure is shown in Fig. S27. $\Delta T$ starts to increase at $\approx 1.5$ GPa, as visualized by the blue lines, which is consistent with the position of the tricritical point, discussed above.

## PROBING THE PROPOSED FM2 TRANSITION

A previous study on LaCrGe$_3$[4] demonstrated, based on resistance measurements at ambient and finite pressures $p \lesssim 1.8$ GPa and in zero and finite magnetic field, that LaCrGe$_3$ undergoes a crossover from the well-established ferromagnetic (FM) state to another ferromagnetic state, which was correspondingly labeled with FM2. However, no clear feature associated with this crossover was detected in previous specific heat measurements[1] at ambient pressure. In this section, we want to discuss to what extent our present set of thermodynamic, $\mu$SR and neutron scattering measurements under pressure provide further insight into the presence of this crossover.


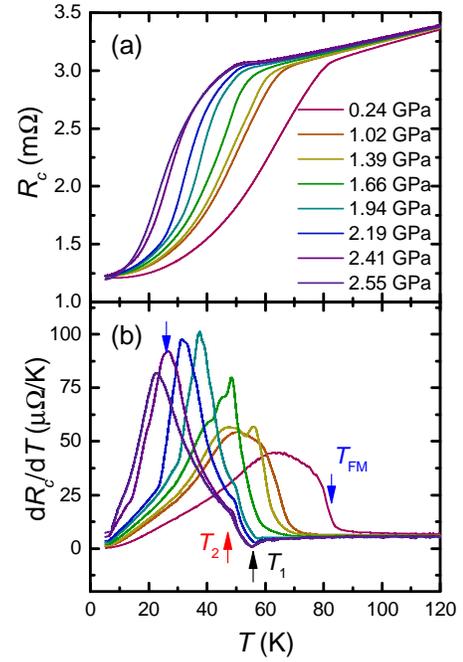

FIG. S29. (a) Resistance of LaCrGe$_3$ along the crystallographic $c$ direction, $R_c$, as a function of temperature, $T$, for different pressures $0.24\,\text{GPa} \leq p \leq 2.55\,\text{GPa}$; (b) Temperature-derivative of out-of-plane resistance, $dR_c/dT$, vs. $T$ for the same pressures as in (a). Blue, black and red arrows indicate the position of the anomalies, that are associated with the phase transitions at $T_{\text{FM}}$, $T_1$ and $T_2$, respectively.

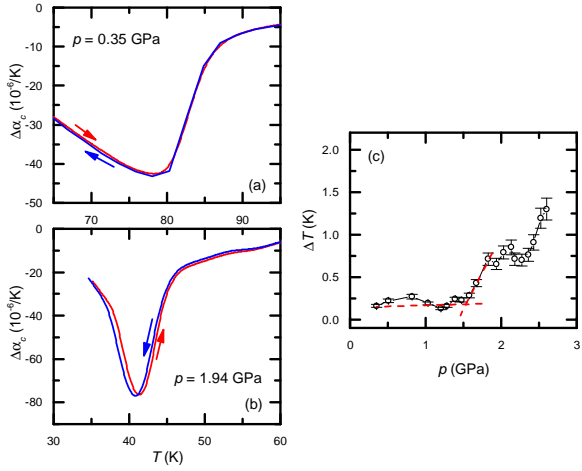

FIG. S27. Thermal hysteresis at the ferromagnetic transition in LaCrGe$_3$; (a) Anomalous contribution to the thermal expansion coefficient along the $c$ axis, $\Delta\alpha_c$, at $p = 0.35\,\text{GPa}$ upon warming (red) and cooling (blue). The rate of temperature change was $\pm 0.25\,\text{K/min}$; (a) $\Delta\alpha_c$ at $p = 1.94\,\text{GPa}$ upon warming (red) and cooling (blue); (c) Thermal hysteresis, $\Delta T$, defined as the difference between transition temperatures upon warming and cooling, as a function of pressure. Red dotted lines indicate the onset of a measurable thermal hysteresis beyond the experimental hysteresis of the setup.

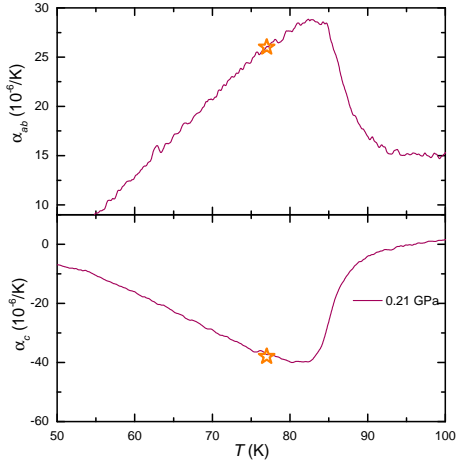

FIG. S28. Thermal expansion coefficients along the crystallographic $ab$ direction, $\alpha_{ab}$ (top), and along the crystallographic $c$ direction, $\alpha_c$ (bottom), vs. $T$ for LaCrGe$_3$ for $p = 0.21\,\text{GPa}$. The orange stars indicate position of the minimum in $\lambda_{\text{PC}}$, inferred from the present $\mu$SR measurements at a pressure of $0.2\,\text{GPa}$.

Figure S28 shows a plot of the thermal expansion anomalies $\alpha_i$ ($i = ab, c$) at $p = 0.21\,\text{GPa}$. The orange stars indicate the position of minimum in $\lambda_{\text{PC}}$, which was observed in our $\mu$SR data at $0.2\,\text{GPa}$ (see Fig. S21) and which might be potentially related to the FM-FM2 crossover. However, our data of the thermal expansion coefficients do not show any discernible feature at

this temperature nor at any other temperature (also the ambient-pressure thermal expansion data does not reveal any signature of the crossover). Thus, we cannot provide any thermodynamic evidence for this crossover from our data. Similarly, we did not find any anomaly in our neutron data of the intensity of the $(1\,0\,0)$ Bragg peak.

## RESISTANCE DATA UNDER PRESSURE

In this section, we want to provide more details of our resistance data set of LaCrGe$_3$ under pressure. We note that, in contrast to the previously-published resistance under pressure data[3], which were performed with current in the $ab$ plane ($R_{ab}$), we performed the present resistance data set with the current along the crystallographic $c$ direction to infer $R_c$. In this way, we explore the directional anisotropy of the resistance in order to demonstrate that the herein-reported phase transition at $T_1$ leaves a clear fingerprint in $R_c(T)$ for high pressures.

Figure S29 shows selected data of $R_c$ as a function of temperature, $T$, for different pressures in the range $0.24\,\text{GPa} \leq p \leq 2.41\,\text{GPa}$ (a), together with the

temperature-derivative of the same data in (b). For low pressures, we find a clear decrease of $R_c$ upon cooling through the ferromagnetic (FM) transition at $T_{FM}$, associated with the loss of spin-disorder scattering. For very high pressures, e.g., for 2.41 GPa, we find first a small increase of $R_c(T)$ upon cooling through $T_1$, before the resistance drops quickly below $T_2$. This behavior becomes more apparent when considering the $T$-derivative of the $R_c$ data. For low pressures, $dR_c/dT$, shows a step-like feature at $T_{FM}$, which is followed by a broad maximum at lower temperatures. The broad maximum was also observed in previous work[4] and was associated with a crossover to another ferromagnetic state at $T_{FM2}$. We have discussed the ambiguity of the thermodynamic evidence for this additional crossover in the previous section. Irrespective of this discussion, the mid-point of the step-like increase of $dR_c/dT$ can be used to infer the transition temperature $T_{FM}$ for low pressures. Upon increasing pressure, the step-like feature in $dR_c/dT$ evolves into a clear peak. At the same time, above a finite pressure close to 1.5 GPa, the broad maximum associated with the potential $T_{FM2}$ becomes indiscernible. Whenever $dR_c/dT$ shows a clear peak rather than a step-like feature, see, e.g., the data sets for $p \geq 1.66$ GPa in Fig. S29 (b), we used the peak position in $dR_c/dT$ to infer $T_{FM}$. Note that we find a signature of the FM transition for $T \geq 5$ K in $R_c(T)$ up to 2.55 GPa, the highest pressure measured in this experiment (the corresponding data is shown in the main text). In addition to the features that are associated with $T_{FM}$, we also find clear anomalies at $T_1$ and $T_2$ in $dR_c/dT$, see all data sets for $p \geq 1.94$ GPa in Fig. S29 (b). The subtle increase in $R_c$ at $T_1$ gives rise to a discernible minimum in $dR_c/dT$, the position of which we use to determine $T_1$ (see black squares). The transition at $T_2$ is associated with a decrease of $R_c(T)$ upon cooling, which gives rise to a clear kink in $dR_c/dT$. The position of this kink is used to infer $T_2$ and is visualized by the red triangles in Fig. S29 (b). Altogether, this data set shows that the phase transition at $T_1$, which was first reported in the present work, leaves a clear fingerprint not only in $C(T)$ and $\alpha_i(T)$, but also in $R_c(T)$ as well.

[1] X. Lin, V. Taufour, S. L. Bud'ko, and P. C. Canfield, Phys. Rev. B **88**, 094405 (2013).

[2] P. C. Canfield, T. Kong, U. S. Kaluarachchi, and N. H. Jo, Philos. Mag. **96**, 84 (2016).

[3] V. Taufour, U. S. Kaluarachchi, R. Khasanov, M. C. Nguyen, Z. Guguchia, P. K. Biswas, P. Bonfà, R. De Renzi, X. Lin, S. K. Kim, et al., Phys. Rev. Lett. **117**, 037207 (2016).

[4] U. S. Kaluarachchi, S. L. Bud'ko, P. C. Canfield, and V. Taufour, Nat. Commun. **8**, 546 (2017).

[5] E. Gati, G. Drachuck, L. Xiang, L.-L. Wang, S. L. Bud'ko, and P. C. Canfield, Rev. Sci. Instrum. **90**, 023911 (2019).

[6] E. Gati, L. Xiang, S. L. Bud'ko, and P. C. Canfield, Ann. Phys. p. 2000248 (2020).

[7] S. L. Bud'ko, A. N. Voronovskii, A. G. Gapotchenko, and E. S. Itskevich, Zh. Eksp. Teor. Fiz. **86**, 778 (1984).

[8] M. S. Torikachvili, S. K. Kim, E. Colombier, S. L. Bud'ko, and P. C. Canfield, Rev. Sci. Instrum. **86**, 123904 (2015).

[9] A. Eiling and J. S. Schilling, Journal of Physics F: Metal Physics **11**, 623 (1981).

[10] L. Xiang, E. Gati, S. L. Bud'ko, R. A. Ribeiro, A. Ata, U. Tutsch, M. Lang, and P. C. Canfield, Rev. Sci. Instrum. **91**, 095103 (2020).

[11] A. D. Krawitz, D. G. Reichel, and R. Hitterman, J. Am. Ceram. Soc. **72**, 515 (1989).

[12] N. Kabeya, K. Imura, K. Deguchi, and N. K. Sato, Journal of the Physical Society of Japan **80**, SA098 (2011).

[13] https://www.almax-easylab.com/ProductDetails.aspx?PID=50&IID=10029.

[14] N. Aso, Y. Uwatoko, T. Fujiwara, G. Motoyama, S. Ban, Y. Homma, Y. Shiokawa, K. Hirota, and N. K. Sato, AIP Conference Proceedings **850**, 705 (2006).

[15] S. Dissanayake, M. Matsuda, K. Munakata, H. Kagi, J. Gouchi, and Y. Uwatoko, Journal of Physics: Condensed Matter **31**, 384001 (2019).

[16] B. Haberl, S. Dissanayake, F. Ye, L. L. Daemen, Y. Cheng, C. W. Li, A.-J. T. Ramirez-Cuesta, M. Matsuda, J. J. Molaison, and R. Boehler, High Pressure Research **37**, 495 (2017).

[17] F. Ye, Y. Liu, R. Whitfield, R. Osborn, and S. Rosenkranz, J. Appl. Crystallogr. **51**, 315 (2018).

[18] B. Haberl, S. Dissanayake, Y. Wu, D. A. Myles, A. M. dos Santos, M. Loguillo, G. M. Rucker, D. P. Armitage, M. Cochran, K. M. Andrews, et al., Rev. Sci. Instrum. **89**, 092902 (2018).

[19] A. Drozd-Rzoska, S. J. Rzoska, M. Paluch, A. R. Imre, and C. M. Roland, The Journal of Chemical Physics **126**, 164504 (2007).

[20] S. Klotz, K. Takemura, T. Strässle, and T. Hansen, Journal of Physics: Condensed Matter **24**, 325103 (2012).

[21] R. Khasanov, Z. Guguchia, A. Maisuradze, D. Andreica, M. Elender, A. Raselli, Z. Shermadini, T. Goko, F. Knecht, E. Morenzoni, et al., High Pressure Research **36**, 140 (2016).

[22] T. F. Smith and C. W. Chu, Phys. Rev. **159**, 353 (1967).

[23] S. Klotz, J.-C. Chervin, P. Munsch, and G. L. Marchand, Journal of Physics D: Applied Physics **42**, 075413 (2009).

[24] N. Tateiwa and Y. Haga, Rev. Sci. Instrum. **80**, 123901 (2009).

[25] T. Barron and G. White, *Heat Capacity and Thermal Expansion at Low Temperatures* (Springer US, 1999).

[26] R. Küchler, T. Bauer, M. Brando, and F. Steglich, Rev. Sci. Instrum. **83**, 095102 (2012).

[27] G. M. Schmiedeshoff, A. W. Lounsbury, D. J. Luna, S. J. Tracy, A. J. Schramm, S. W. Tozer, V. F. Correa, S. T. Hannahs, T. P. Murphy, E. C. Palm, et al., Rev. Sci. Instrum. **77**, 123907 (2006).

[28] A. Lindbaum and M. Rotter, *Handbook of Magnetic Materials* (North Holland, 2002), chap. Spontaneous magnetoelastic effects in gadolinium compounds.

[29] H. Bie, O. Y. Zelinska, A. V. Tkachuk, and A. Mar, Chemistry of Materials **19**, 4613 (2007).

[30] J. Cadogan, P. Lemoine, B. R. Slater, A. Mar, and M. Avdeev, in *Solid Compounds of Transition Elements II* (Trans Tech Publications Ltd, 2013), vol. 194 of *Solid State Phenomena*, pp. 71–74.

[31] J. Rodríguez-Carvajal, Physica B: Condensed Matter **192**, 55 (1993).

[32] S.-L. Chang, *X-ray Multiple-Wave Diffraction* (Springer-Verlag Berlin Heidelberg New York, 2004).

[33] M. Renninger, Acta Cryst. A **24**, 143 (1968).

[34] A. Yaouanc and P. Dalmas De Réotier, *Muon Spin Rotation, Relaxation and Resonance* (Oxford University Press, Oxford, 2011).



\end{document}